\documentclass[%
 aip,
 amsmath,amssymb,
preprint,%
]{revtex4-1}

\usepackage{graphicx}%
\usepackage{dcolumn}%
\usepackage{bm}%
\usepackage{xcolor}
\usepackage{hyperref}

\usepackage[utf8]{inputenc}
\usepackage[T1]{fontenc}
\usepackage{etoolbox}
\usepackage{enumitem}
\usepackage{booktabs}
\usepackage{siunitx}
\usepackage{float}
\usepackage{tabularx}
 \usepackage{multirow}

\makeatletter
\def\@email#1#2{%
 \endgroup
 \patchcmd{\titleblock@produce}
  {\frontmatter@RRAPformat}
  {\frontmatter@RRAPformat{\produce@RRAP{*#1\href{mailto:#2}{#2}}}\frontmatter@RRAPformat}
  {}{}
}%
\makeatother
\usepackage{mathrsfs}

\newcommand{\elec}{\text{e}}
\newcommand{\gas}{\text{g}}
\newcommand{\ion}{\text{i}}
\newcommand{\el}{\text{el}}
\newcommand{\iz}{\text{iz}}

\newcommand{\partialx}[1][]{\frac{\partial{#1}}{\partial x}}
\newcommand{\boltz}{k_\mathrm{B}}
\newcommand{\cx}{\text{cx}}
\newcommand{\ex}{\hat{\vec{e}}_x}
\newcommand{\eperp}{\hat{\vec{e}}_{\perp}}
\newcommand{\eperpgi}[1][1]{\hat{\vec{e}}_{\perp g #1}}
\newcommand{\eperpgOne}{\eperpgi[1]}
\newcommand{\eperpgTwo}{\eperpgi[1]}

\begin{document}

\preprint{AIP/123-QED}

\title[]{One dimensional high-order moment models with realistic collisions for nonequilibrium ion transport in weakly ionized plasmas}

\author{A. Berger}
\author{A. Alvarez Laguna}%
\email{alvarez@lpp.polytechnique.fr}
\affiliation{Laboratoire de Physique des Plasmas, Centre National de la Recherche Scientifique, Sorbonne Université, École polytechnique, Institut Polytechnique de Paris, route de Saclay 91128 Palaiseau, France%
}%

\begin{abstract}
{\small
Ion-neutral collisions are fundamental in the transport of partially ionized plasmas. When the collisional scales are comparable to the system scales or the electric field is strong, nonequilibrium conditions for the ions arise that lie outside the scope of classical transport models due to the large drift, strong heat flux, and temperature anisotropy. In this paper, we propose the resolution of non-linear high-order moment closures for simulating nonequilibrium ion dynamics in one dimensional weakly ionized plasmas. We compare a four-moment model based on an anisotropic Maxwellian (considering the mass, axial momentum, and axial and perpendicular energies), a five-moment model based on a hyperbolic quadrature-based closure (considering the first five axial moments), and a novel six-moment model hyperbolic quadrature-based closure (considering the first five axial moments and the perpendicular energy). We derive the analytical expressions of the collision source terms in the moment equations from the Boltzmann operator for electron-impact ionization and ion-neutral scattering collisions with arbitrary differential cross sections. Our novel formulation generalizes the classical Chapman-Cowling collision theory for arbitrary drift velocity, temperature anisotropy and heat flux, with strictly realizable (positive) distributions. The models are validated via non-linear simulations benchmarked against kinetic solutions for argon plasmas with realistic isotropic scattering and charge-exchange cross sections across a wide pressure range (0.05–500 mTorr), considering a bounded plasma between floating walls and a direct-current discharge. The six-moment model robustly and accurately captures ion dynamics under all studied conditions, particularly under strong nonequilibrium, where temperature anisotropy and heat flux cannot be treated as local transport phenomena. It also reconstructs the distribution function with high fidelity, without noise, and at a computational cost comparable to classical fluid models.}
\end{abstract}

\maketitle

\section{Introduction}

Ion-neutral collisions play a fundamental role in the transport of partially ionized plasmas in a wide variety of scenarios: gas discharges for industrial applications~\cite{Liberman05,Chabert11,Raimbault09}, the plumes of electric thrusters~\cite{Goebel08_8}, the edge region of tokamak plasmas \cite{Rozhansky22} or partially ionized astrophysical plasmas~\cite{Ballester18}. When the scales of physical processes become comparable to those of ion-neutral collision, the motion of ions (influenced by the electromagnetic field) and neutrals can decouple significantly. Certain collisional processes, such as ionization or resonant charge exchange, produce ions with velocities significantly lower than those that were accelerated by the electric field in upstream regions. This velocity disparity creates nonequilibrium conditions that lie outside the scope of classical transport models, remaining a key challenge for fluid models of partially ionized plasmas.

Classical transport models are primarily based on the drift-diffusion approximation. Under highly collisional conditions, the motions of ions and neutrals are ``tied together'' (in Braginskii's own words \cite{Braginskii65}), allowing the plasma to be described with the fluid equations. These equations solve a single momentum equation for ions and neutrals with a drift-diffusion approximation, where the transport fluxes (\textit{e.g.} diffusion velocities, heat flux vector) are function of the local gradients of the fluid variables and the local electric field. The transport coefficients can be computed with the Chapman-Enskog expansion (\textit{e.g.}, Refs.\cite{Chapman70, Braginskii65, Ferziger72, Alexeev94, Ern98, Kustova98, Giovangigli13, Graille09}) or with the linearized high-order moment equations with Grad's closure (\textit{e.g.}, Refs.\cite{Zhdanov02,Zhdanov16,Hunana25}).

However, as the length and time variations become comparable to those of collisions (or the electric force largely exceeds the collisional drag force), the motions of ions and neutrals are no longer ``tied together''. As a result, these conditions, strictly speaking, fall outside the regime of applicability of the fluid equations\cite{Braginskii65}. These nonequilibrium conditions are often referred to as non-local\cite{Wu_2026}, since the nonequilibrium processes cannot be captured by classical transport models based on local gradients of fluid variables and the local electric field. In practice, however, these plasmas are often modeled with multi-fluid equations that solve the momentum equations for ions and neutrals separately (\textit{e.g.}, Refs.\cite{Meier12, Leake13, Khomenko14, AlvarezLaguna16, AlvarezLaguna17, AlvarezLaguna20, Sahu2020, Gangemi25,Poli24}). While the multi-fluid equations can describe the decoupling between ions and neutrals, the collisional terms are often modeled with a simple Bhatnagar-Gross-Krook (BGK) operator\cite{Bhatnagar54} or other models that fail to take into account for the nonequilibrium distribution function of the ions. As it will be shown in this work, such simplistic BGK model can lead to quantitatively erroneous results.

In low pressure gas discharges, the ion drift velocity can far exceed the thermal speed of ions and neutrals, leading to ion nonequilibrium distribution functions with asymmetric heavy tails. These effects significantly impact collisional rates and cannot be captured by simple BGK relaxation models. Similarly, multi-component models or linearized Grad models\cite{Chapman70, Braginskii65, Ferziger72, Alexeev94, Ern98, Kustova98, Giovangigli13, Graille09} are based on the linearized Boltzmann operator under the assumption that interspecies drift velocities are much smaller than their thermal speeds.

Several approaches have aimed to improve the multi-fluid description of ion-neutral collisions in nonequilibrium gas discharges. Benilov~\cite{Benilov97} proposed a model for the momentum and energy exchange between ions and neutrals, assuming Maxwellian distributions at different temperatures and arbitrary drift velocities (allowing for drifts much larger than the thermal speed). Other approaches rely on phenomenological ansätze for nonequilibrium ion distributions. For example, the variable mobility model~\cite{Godyak86,Liberman05,Chabert11}, widely used in low-to-intermediate pressure gas discharges, assumes a BGK operator that depends on the ion drift velocity with the assumption of a half Maxwellian ion distribution. Recent models include Semenov’s work that assumes a rectangular distribution~\cite{Semenov17}, Mun \textit{et al.} with a semi-empirical mobility~\cite{Mun24} (based on Khrapak \textit{et al.}’s work~\cite{Khrapak20}), and Boccelli \textit{et al.} with a triangular distribution~\cite{Boccelli20-HET}, which captures the asymmetric heavy tails due to the ionization. However, these models are challenging to generalize for different gas mixtures or arbitrary nonequilibrium distributions.

A natural extension of multi-fluid models under nonequilibrium conditions is the method of moments (MOM), which expands fluid equations into a truncated hierarchy of moment equations derived from the kinetic equation, using a mathematical approximation for the closing flux. The MOM has been used to study the ion transport in partially-ionized plasmas under spatially uniform conditions\cite{Robson94} or in one dimensional simulations with simplified collision operators\cite{Boccelli20-HET, Kuldinow24_b}. In a recent work\cite{Berger25}, we have compared different five moment (5M) one dimensional closures, \textit{i.e.}, considering moments up to the fourth-order moment in 1D, to study the transport of ions in a bounded plasma. The closures include regularized Grad method\cite{Grad49, Cai14}, maximum-entropy \cite{Dreyer87, Levermore96, Boccelli20,Boccelli22}, the hyperbolic quadrature MOM (HyQMOM)\cite{McGraw97, Fox22}, and Extended QMOM (EQMOM) closure \cite{Chalons10}. In particular, the 5M HyQMOM proved to be a simple and robust method to model the ions in partially ionized bounded plasmas under different pressure regimes, both in the sheath and the bulk of the plasma. However, in the numerical study of Ref.\cite{Berger25}, a simplified BGK collision model was employed, raising the question of whether the 5M HyQMOM method can effectively model plasmas using higher fidelity collision models, that depend on the energy of the impact as well as on the scattering angle of the collision.

In the present paper, we will extend the HyQMOM model to incorporate realistic collisions by a direct integration of the Boltzmann operator for the dominant ion-neutral processes in noble gases plasmas: isotropic elastic scattering, charge exchange, and electron-impact ionization collisions, while considering realistic cross sections, \textit{i.e.}, that depend on the energy of the colliding particles and their scattering angles. This work is complementary to the study of the electron moment models with the multi-species scattering and reactive Boltzmann operator\cite{Furkal00, AlvarezLaguna22, AlvarezLaguna23, AlvarezLaguna25, AlvarezLaguna26}.

As it will be shown in this paper, the main challenges in the case of ions in partially ionized plasmas can be summarized as follows:
\begin{enumerate}[noitemsep, topsep=\baselineskip]
    \item The charge exchange and isotropic scattering collisions have different  angular dependence of the differential cross section, impacting differently to the different moments, beyond a BGK approximation.
    \item The mass of the ions and the neutrals is approximately equal and, therefore, the full Boltzmann operator cannot be simplified (as done in the electron-neutral collisions).
    \item The drift of the ions often becomes larger than the ion thermal speed and needs to be taken into account in the computation of the collisional rates.
    \item The combined effect of electric field and charge exchange collisions produce distributions that have strong temperature anisotropies (with temperatures in the direction of the electric field that are much larger than in the perpendicular direction) and large skewness (heat flux) in the direction of the electric field.
\end{enumerate}
In the present paper, we will derive analytical expressions that take into account these difficulties in the MOM for 1D gas discharges.

We will validate our models by comparing with particle-in-cell Monte Carlo collisions (PIC-MCC) simulations. In particular, we will study two cases of interest for weakly ionized gas discharges: a plasma between two floating walls and a direct-current (DC) discharge with a cathode sheath with a large electric field. In this study, we will propose three different 1D high-order moment closures, a 4M model (considering balance equations of mass, axial momentum and two energies), 5M model (considering balance equations of mass, axial momentum, axial energy, axial heat flux and axial fourth-order moment), and 6M model (considering the perpendicular energy in addition to the 5M model).  Comparison between the different models with a 3M model (considering mass, momentum, and isotropic energy balance equations) as well as to common BGK operators and kinetic models based on the particle-in-cell (PIC) method will be provided in a wide range of pressures. {To better assess the errors associated with the choice of closure and collisional terms in the high-order moment equations for the ions, we study the ions standalone. As a result, the ionization rate and electric field in the moment simulations will be taken from the kinetic simulations, so the differences in the comparison arise solely from the choice of closure and collisional terms, not from the electron dynamics which will be studied in a separate work.}

The paper is organized as follows. In Section~\ref{sec:One}, we describe the one dimensional geometry under study. We also present the kinetic equation for ions in these setups, including a review of the collision operators, and introduce the moment closure equations analyzed in this work. In Section~\ref{sec:DIBO-computation}, we detail the computation of the collision terms in the moment equations, derived through direct integration of the Boltzmann operator. {Section~\ref{sec:Four} explains how the collisional integrals are numerically integrated and they are validated in 0D cases against MCC kinetic simulations. We study the case of a relaxation toward equilibrium of a distribution of ions initially drifting at supersonic speeds. In a second case, we study ions in presence of a homogenous electric field in a neutral background in a range of reduced fields (from $100$ to $5000$ Td)}. Section~\ref{sec:results} validates our derivation using one dimensional simulations of the moment equations, comparing the results to kinetic simulations for the two aforementioned cases. Particular attention is paid to both the representation of the moments of the distribution and the reconstruction of the distribution function from the moments. Finally, in Section~\ref{sec:Conclusions}, we summarize our findings and discuss the implications of the results.%

\section{Ion kinetic and moment equations}\label{sec:One}

\subsection{Description of one dimensional kinetic simulations of a bounded plasma}\label{sec:set-ups}

\begin{figure}[h]
    \centering
    \includegraphics[width=0.49\linewidth]{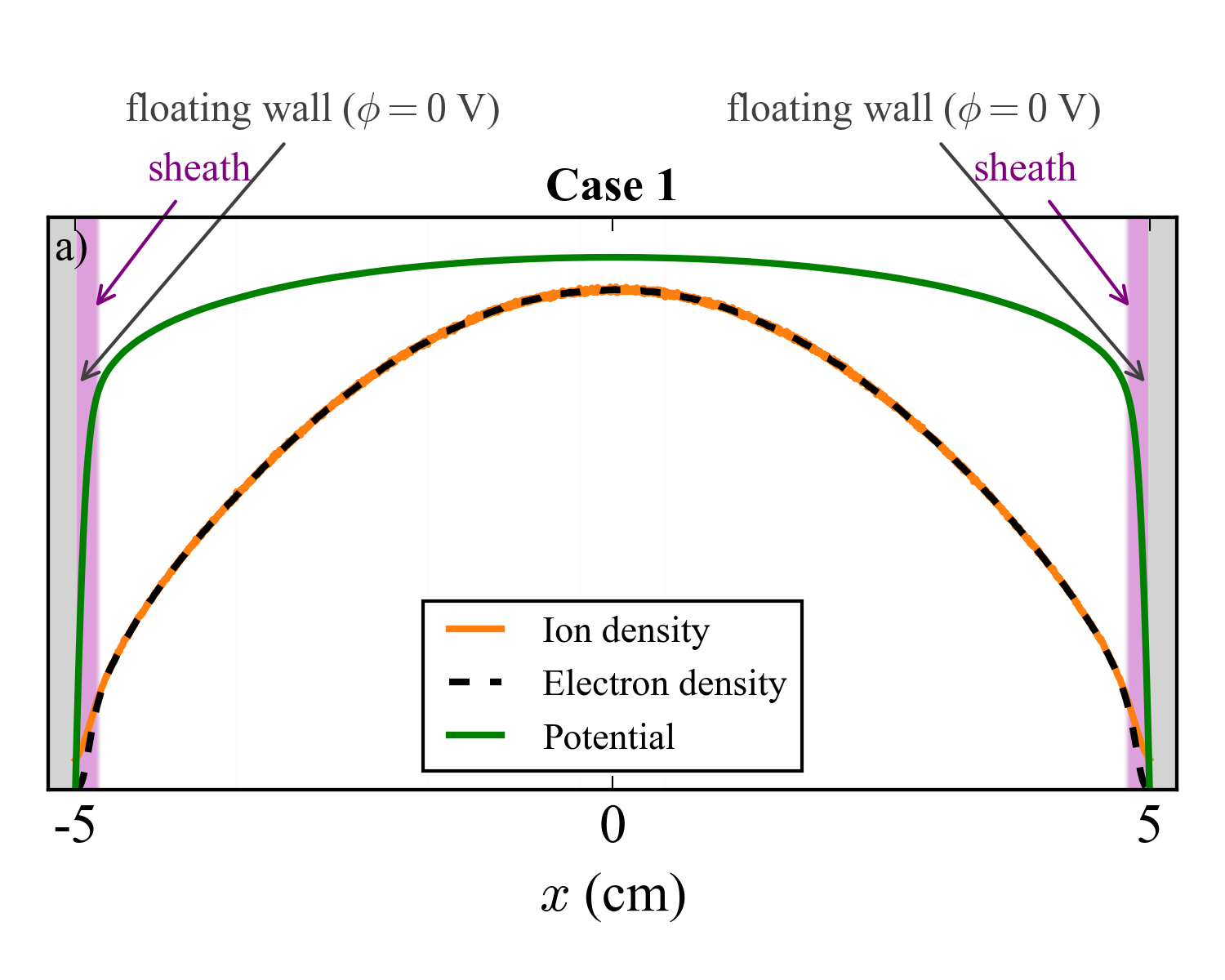}
    \includegraphics[width=0.49\linewidth]{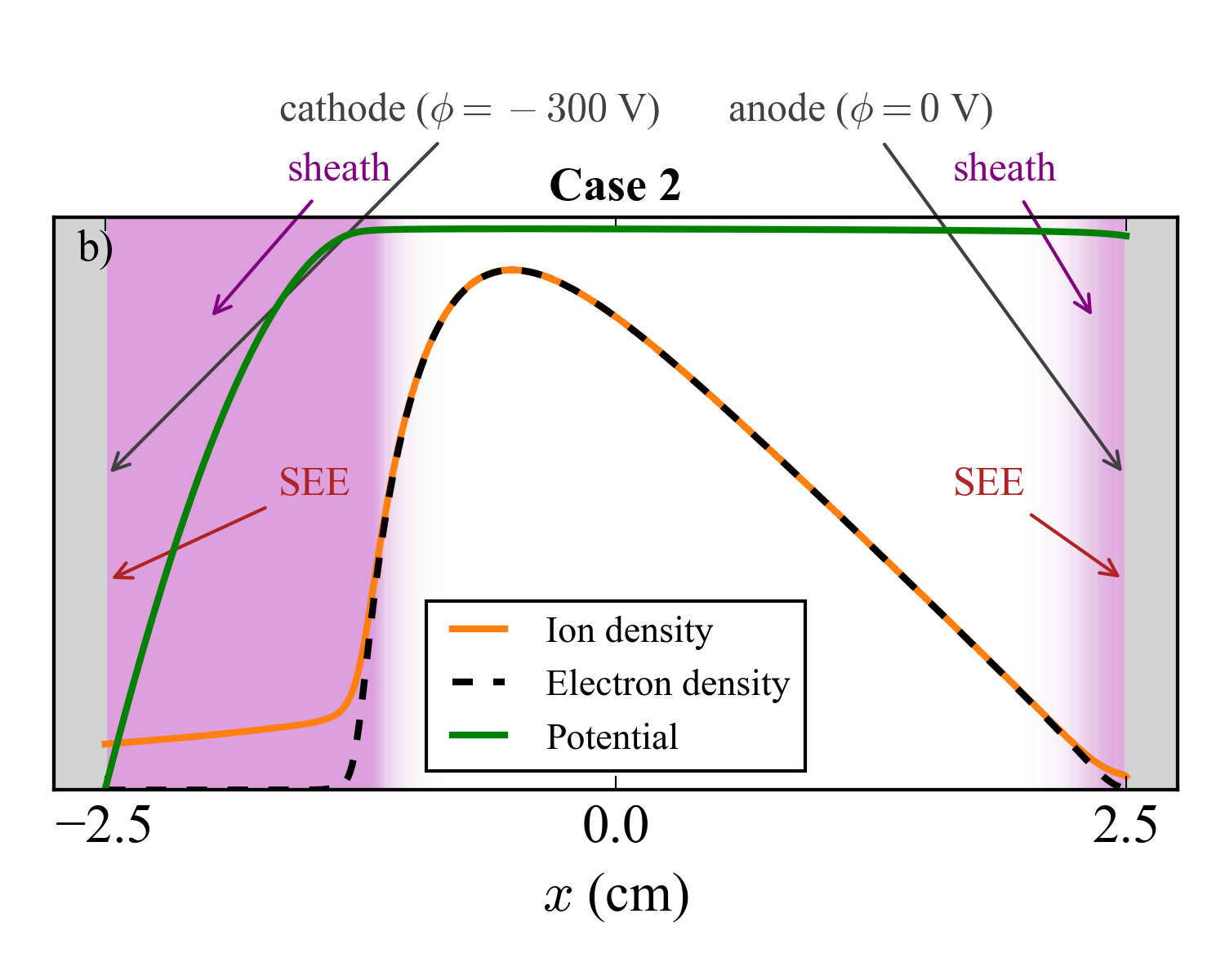}
    \caption{Sketch of the argon numerical setups studied in this paper. a) Discharge between two floating walls with injection of particles proportional to electron density and b) a DC discharge with secondary electron emission and self-consistent ionization. Both configurations are investigated across a range of pressures.
    }
    \label{fig:set-ups}
\end{figure}

In this paper, we will consider the ion dynamics in a one dimensional weakly ionized argon plasma across a range of pressures. In order to be used for comparison with the moment models, we carry out kinetic simulations based on the 1D-3V (\textit{i.e.}, one dimension in physical space and three dimensions in velocity space) PIC-MCC, based on the null collision method\cite{Vahedi95}. We will consider two cases in an argon plasma (See Fig.~\ref{fig:set-ups}): (1) a plasma between two-floating walls and (2) a direct-current (DC) discharge. In both cases, the argon ion-atom collision cross sections are taken from LXCat (Phelps database\cite{PhelpsLX}), including isotropic and charge exchange (with a backscattering approximation){\cite{Phelps94}}. In addition, the plasma density is considered to be much smaller than that of the gas, so the gas is assumed to be a constant uniform background, not affected by the plasma, and the Coulomb collisions are neglected.

Case 1 (Fig.~\ref{fig:set-ups} a) reproduces a plasma between two floating walls, with the same set-up that was studied in a previous work\cite{Berger25} but with realistic ion-neutral collisions instead of a BGK collision operator. The domain extends from $x\in[-L,L]$ with $L=5$~cm. The neutral gas is treated as a spatially homogeneous background at $T_\gas=300$~K at four different gas pressures ranging from $p_\gas = 5\cdot 10^{-2},\,5\cdot 10^{-1},\,\,5,$ and $ 50$ mTorr. These correspond to characteristic Knudsen numbers $\text{Kn} = \lambda_{\ion \gas}/{(2L)} = ({2}n_\gas\sigma_0L)^{-1} = 6.2\cdot[1,~10^{-1},~10^{-2},~10^{-3}]$, which goes from nearly collisionless to collisionally-dominated regimes (for this estimate, we take the characteristic cross section as $\sigma_0 = 10^{-18}$ m$^2$). Electrons and singly-ionized argon ions are simulated with the PIC-MCC method. The steady state solution is obtained as follows. Initially, electrons and ions are uniformly distributed with densities $n_\elec(t=0)=n_\ion(t=0)=10^{15}$~m$^{-3}$ and temperatures $T_\elec=5$~eV and $T_\ion=T_\gas$. At the boundaries, charged species are absorbed, and the electric potential is set to zero: $\phi(x=\pm L)=0$ (\textit{i.e.}, floating walls). To compensate for particle losses, at each time step, electrons and ions are injected in the domain with a probability that is proportional to the local electron density and the number equal to the ions lost at the walls. As a result, the only collisional process considered for the electrons is elastic collisions with the gas, with the cross section from Ref.~\cite{PhelpsLX}. The electrons are injected with a Maxwellian distribution at $T_{inj} = 5$ eV and the ions following the neutral distribution (which is a Maxwellian at $T_\gas$). This injection method mimics ionization and the electron heating processes that sustains the discharge and, for the ions, is equivalent to the electron-impact ionization MCC algorithm. The simulations are run with $2000$ cells (except the case at $50$ mTorr that uses $3000$ cells in order to better resolve the sheath). The time step is $\Delta t = 0.37\cdot 10^{-10}$ s and the average number of particles per cell is $N_\mathrm{PPC} = 100$.

Case 2 (Fig.~\ref{fig:set-ups} b) reproduces a one dimensional DC discharge. The domain extends from $x\in[-L,L]$ with $L=2.5$~cm at two different gas pressures $p_\gas = 200$ and $500$ mTorr, with the gas at $T_\gas = 300$ K. We impose a difference in electrical potential at the left (cathode) and right (anode) boundaries, with $\phi(x={-L})=-300$ V and $\phi(x=L)=0$. The plasma is sustained by a self-consistent electron-impact ionization collisions with the electron elastic and inelastic collisional processes from Ref.\cite{PhelpsLX} and the current in the plasma is sustained by a secondary electron emission at the cathode with an effective secondary electron coefficient model $\gamma_{eff}=0.03$, implemented as described in Ref.~\cite{Schulze22}. The simulations are initialized with a uniform constant plasma density and evolved until convergence with a grid that uses $1200$ cells and the time step is $\Delta t = 0.5\cdot 10^{-11}$ s, chosen to resolve both the collision frequency as well as the CFL condition of the particle pusher. In the converged solution, the average number of particles per cell is $N_\mathrm{PPC} \approx 80$ and $300$, for $p_\gas = 200$ and $500$ mTorr, respectively.

\subsection{Ion kinetic equation}

The ion kinetics can be described with the Boltzmann equation in a 1D-3V space, as follows,
\begin{eqnarray}\label{eq:kinetic}
    \frac{\partial{f}}{\partial t} + v_x  \frac{\partial{f}}{\partial x} + \frac{eE_x}{m} \frac{\partial{f}}{\partial v_x} &=& \left. \frac{\delta f}{\delta t}\right\rvert^\el_\text{c} + \left. \frac{\delta f}{\delta t}\right\rvert^\iz_\text{c},
\end{eqnarray}
where $f(t, x, \vec{v})$ is the ion distribution function, $m$ is the ion mass, $e$ is the elementary charge, $E_x$ is the electric field along the $x$ direction, and right-hand side term accounts for the rate of change due to collisions with the gas atoms, \textit{i.e.}, ion-gas elastic (isotropic and charge exchange) collisions and ionization collisions. For simplicity, we will omit the ion index from the ion properties in the following, as this paper focuses on the ion dynamics, retaining it only where necessary to avoid ambiguity.

In both Cases 1 and 2, the 1D spatial geometry and the geometry of the collisions ensures symmetry of the ion distribution function in the velocities perpendicular to \(x\). {Thus, the ion velocity is \(\vec{v} = v_x \ex + \vec{v}_\perp\), where \(\ex\) is the unit vector along the \(x\)-direction and $\vec{v}_\perp = v_y \hat{\vec{e}}_y + v_z\hat{\vec{e}}_z \equiv v_\perp \eperp$, where and \(\eperp\) is the unit vectors along the perpendicular perpendicular to $x$. With these definitions,} the distribution function then simplifies to \(f(t, x, v_x, v_\perp)\).

\subsubsection{Generalized Boltzmann operators for the ion kinetic equation}

We can divide the ion-neutral collisions into two groups\cite{Giovangigli13}: scattering collisions (where the species do not change during the collisions) and reactive collisions (where the species are different before and after the collisions). In noble gases, the dominant scattering cross sections are often divided into two types: isotropic elastic collisions and charge exchange collisions. In both collision types, the kinetic energy of the particles is conserved in the collision, thus we refer to as elastic, $\left. \frac{\delta f}{\delta t}\right\rvert^\el_\text{c} = \left. \frac{\delta f}{\delta t}\right\rvert^{\text{iso}}_\text{c}+\left. \frac{\delta f}{\delta t}\right\rvert^{\cx}_\text{c}$. Finally, the only reactive collision considered in this paper is the electron-impact ionization.

\begin{figure}[h]
    \centering
    \includegraphics[width=0.5\linewidth]{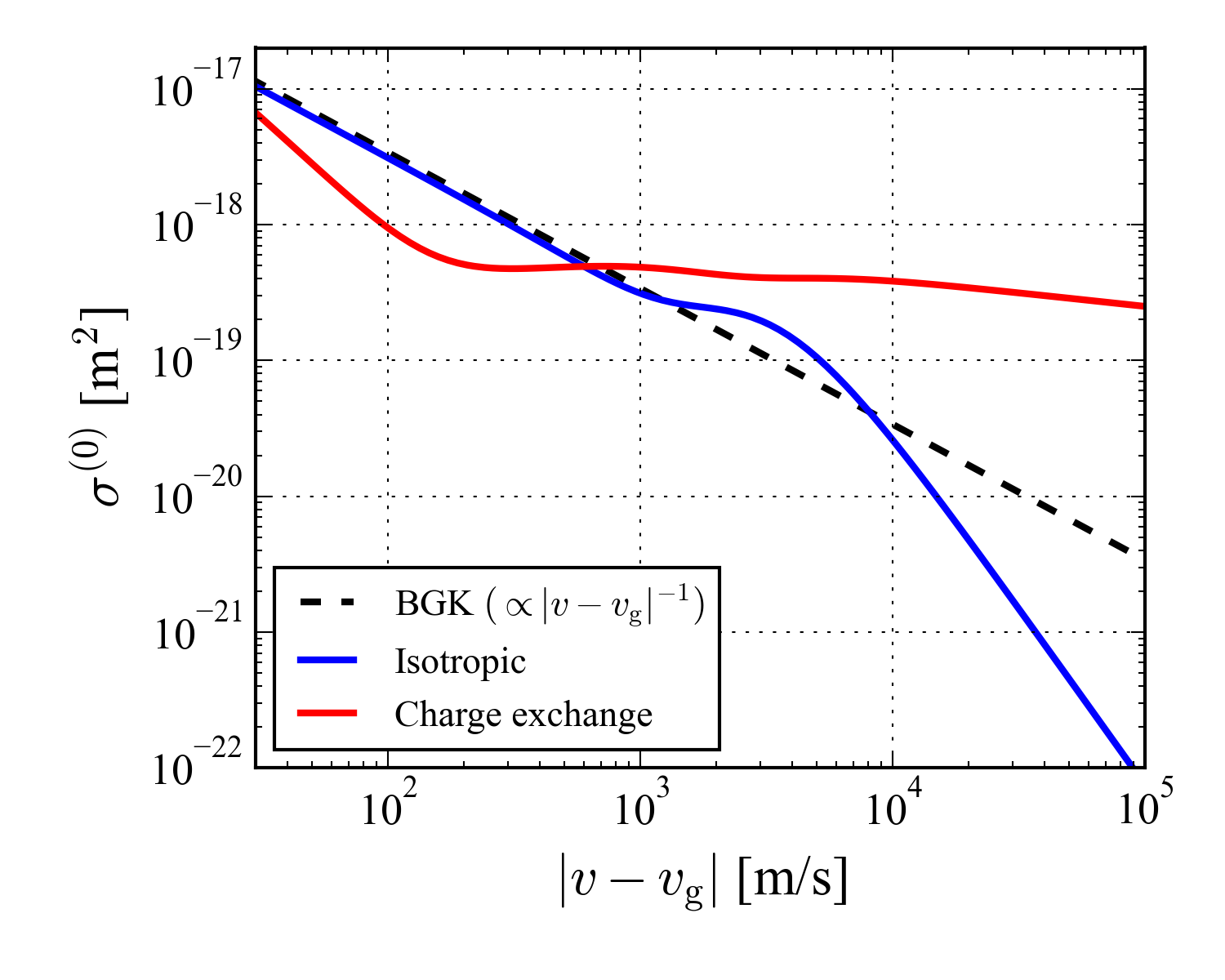}
    \caption{Isotropic scattering and charge exchange ion-atom total cross sections in an argon plasma from Ref.\cite{PhelpsLX}. For comparison, a collision cross section with constant frequency is shown, which in the case of charge exchange can be modeled with a BGK operator.}
    \label{fig:crossSection}
\end{figure}

The ion-neutral elastic collisions are modeled with the multi-species version of the Boltzmann collision operator, as follows,
\begin{eqnarray}\label{eq:Boltzmann}
    \left. \frac{\delta f}{\delta t}\right\rvert^{\el}_\text{c} &=& \int_{\mathbb{R}^3}\int_{\mathbb{S}^2} \left[f(\vec{v}') f_{\gas}(\vec{v}'_{\gas}) - f(\vec{v}) f_{\gas}(\vec{v}_{\gas})\right]\,\lvert \vec{v}-\vec{v}_{\gas}\rvert\,\sigma(\lvert \vec{v}-\vec{v}_{\gas}\rvert, \chi)  \,\mathrm{d}^2\Omega \mathrm{d}^3\vec{v}_{\gas},
\end{eqnarray}
 the subscript $\gas$ refers to the neutral gas atom properties, $\sigma$ is the differential cross section, $\chi$ is the scattering angle, and $\mathrm{d}^2\Omega$ is the unit sphere element of the collision angles, and the primed velocities refer to the velocities of the restitution collision (which conserves momentum and kinetic energy), as follows,
\begin{eqnarray}
    m \vec{v} + m_\gas \vec{v}_\gas &=& m \vec{v}' + m_\gas \vec{v}'_\gas,\\
    \frac{1}{2}mv^2 + \frac{1}{2}m_\gas {v}_\gas^2 &=& \frac{1}{2}mv'^2 + \frac{1}{2} m_\gas {v}'^2_\gas.
\end{eqnarray}
The multi-species Boltzmann operator (Eq.~\eqref{eq:Boltzmann}) models both isotropic and charge exchange collisions. The key difference lies in the angular and energy dependence of their differential cross sections: isotropic scattering treats collisions as independent of the angle (as an energy-dependent hard-sphere interaction), while charge exchange is modeled with a backscattering approximation (\textit{i.e.}, with a Dirac-delta angular dependence, as detailed in Refs.~\cite{Berger25, Robson17}), as follows,
\begin{multline}
    \sigma^{\text{iso}}(\lvert \vec{v}-\vec{v}_{\gas}\rvert, \chi) = \frac{\sigma^{(0),\,\text{iso}}(\lvert \vec{v}-\vec{v}_{\gas}\rvert)}{4\pi} \quad\text{and}\quad \\
    \sigma^{\cx}(\lvert \vec{v}-\vec{v}_{\gas}\rvert, \chi)=\frac{\sigma^{(0),\,\cx}(\lvert \vec{v}-\vec{v}_{\gas}\rvert)}{2\pi}\left[\delta(\cos\chi-\cos\chi_0)\right]_{\chi_0\rightarrow\pi},
\end{multline}
where $\sigma^{(0)}(\lvert \vec{v}-\vec{v}_{\gas}\rvert) = \int_{\mathbb{S}^2} \sigma(\lvert \vec{v}-\vec{v}_{\gas}\rvert, \chi) \mathrm{d}^2\Omega$ is the  total cross section. Both cross sections are presented in Fig.~\ref{fig:crossSection} and compared to a Langevin cross section, i.e., that scales as $\sigma^{(0)}\propto \lvert \vec{v}-\vec{v}_{\gas}\rvert^{-1} $.

Alternatively, the effect of ionization collisions in the ion kinetic equation is written with a generalized Boltzmann operator for reactive collisions, as derived by Alexeev-Giovangigli\cite{Giovangigli13,Alexeev94},
\begin{equation}\label{eq:BoltzmannReact}
    \left. \frac{\delta f}{\delta t}\right\rvert^\iz_\text{c} = \int \left( f_\gas f_{\elec1} - f_\ion f_{\elec2} f_{\elec3} \frac{\beta_\ion\beta_\elec}{\beta_\gas}\right) \mathcal{W}^{\ion\elec\elec}_{\gas \elec} \, \mathrm{d}^3\vec{v}_\gas \, \mathrm{d}^3\vec{v}_{\elec1} \, \mathrm{d}^3\vec{v}_{\elec2} \, \mathrm{d}^3\vec{v}_{\elec3},
\end{equation}
where $\beta_\alpha = (h_\mathrm{p}/m_\alpha)^3$ is the statistical weight of the species $\alpha$ with Planck's constant $h_\mathrm{p}$ and $\mathcal{W}^{\ion\elec\elec}_{\gas \elec}$ is the transition probability of the collision $\gas + \elec \rightarrow \ion + \elec + \elec $. The velocities of the colliding particles are related by the following conservation laws
\begin{eqnarray}
    m_\elec \vec{v}_{\elec1} + m_\gas \vec{v}_\gas &=& m_\elec \vec{v}_{\elec2} + m_\elec \vec{v}_{\elec3} + m_\ion \vec{v}_\ion,\\
    \frac{1}{2}m_\elec v_{\elec1}^2 + \frac{1}{2}m_\gas {v}_\gas^2 &=& \frac{1}{2}m_{\elec}v_{\elec2}^2 + \frac{1}{2}m_{\elec}v_{\elec3}^2 + \frac{1}{2} m_\ion {v}^2_\ion.
\end{eqnarray}

The transition probability can be expressed as an effective cross section,
\begin{equation}\label{eq:transProbability}
    \mathcal{W}^{\ion\elec\elec}_{\gas \elec} \, \mathrm{d}^3\vec{v}_\ion \, \mathrm{d}^3\vec{v}_{\elec2} \, \mathrm{d}^3\vec{v}_{\elec3} = |\vec{v}_{\elec1} - \vec{v}_\gas| \mathrm{d}(\sigma^{\ion\elec\elec}_{\gas \elec}),
\end{equation}
where the differential cross section depends, in general, on the relative velocity between the impacting electron and neutral and their angle, as well as on the relative velocities of the two resulting electrons with the ion and their scattering angles (\textit{cf.} Eq.~(184) of Alexeev \textit{et al.}\cite{Alexeev94}).

However, Eq.~\eqref{eq:BoltzmannReact} can be simplified under the nonequilibrium weakly ionized conditions of most low temperature plasmas. First, the cross section is assumed to depend only on the relative velocity between the electron and the neutral. Due to the large mass disparity, this relative velocity is approximated by the electron velocity, \textit{i.e.}, $|\vec{v}_{\elec1} - \vec{v}_\gas|\approx|\vec{v}_{\elec1}|$. Also, the cross section is assumed isotropic in the scattering angles, \textit{i.e.}, $\sigma^{(0),\,\iz}(|\vec{v}_{\elec}|) = \int \mathrm{d}(\sigma^{\ion\elec\elec}_{\gas \elec})$. Second, due to the mass disparity between the electron and the atom, the velocity of the ion after the ionization is assumed to be that of the neutral, \textit{i.e.}, $\vec{v}_\ion \approx \vec{v}_\gas$, and, therefore, the Jacobian $|\mathrm{d}^3\vec{v}_\ion/\mathrm{d}^3\vec{v}_\gas|=1$ (that can be injected when substituting Eq.~\eqref{eq:transProbability} into Eq.~\eqref{eq:BoltzmannReact}). Third, note that Eq.~\eqref{eq:BoltzmannReact} takes into account the micro-reversibility of the collisions, \textit{i.e.}, the three-body recombination. However, far from chemical equilibrium, the three-body recombination can be neglected, so the term proportional to $f_\ion f_{\elec2} f_{\elec3}$ in Eq.~\eqref{eq:BoltzmannReact} can be neglected.

As a result, the ionization collision operator largely simplifies to
\begin{equation}\label{eq:BoltzmannReact_simple}
    \left. \frac{\delta f}{\delta t}\right\rvert^\iz_\text{c} \approx \int_{\mathbb{R}^3} f_\gas(\vec{v}_\gas)\, f_{\elec1}(\vec{v}_{\elec1})\,|\vec{v}_{\elec1}|\,\sigma^{(0),\,\iz}(|\vec{v}_{\elec1}|)\,\mathrm{d}^3\vec{v}_{\elec1}  = n_\gas\,n_\elec\,K^{(0)}_\iz\,w_\gas(\vec{v}_\ion),
\end{equation}
where we have used the approximation (due to the small electron mass) $\vec{v}_\ion \approx \vec{v}_\gas$ and we define $w_\gas(\vec{v}_\ion) = f_\gas(\vec{v}_\ion)/n_\gas$ with the neutral number density $n_\gas$, the electron number density $n_\elec$, and the ionization rate constant is defined as $n_\elec K^{(0)}_\iz = \int f_{\elec}\, |\vec{v}_{\elec}|\,\sigma^{(0),\,\iz}(|\vec{v}_{\elec}|)\,\mathrm{d}^3\vec{v}_{\elec}$.

\subsubsection{Common ion-neutral BGK operators in gas discharges}\label{sec:BGK}
In this paper, we will compare our results to the BGK operator approximation\cite{Bhatnagar54}. This model assumes that the collisions relax the distribution to equilibrium at a constant rate, generalized in multi-species\cite{Andries02} as follows,
\begin{eqnarray}
    \left.\frac{\delta f}{\delta t}\right\vert^{\el}_\mathrm{BGK} &=& {-}\nu^\mathrm{BGK}_{\ion\gas} \left(f(\vec{v}) - n w_\gas(\vec{v})\right) \,~~\text{with}~~w_\gas(\vec{v}) = \left(\frac{\gamma_\gas}{2\pi}\right)^{3/2}e^{-\frac{\gamma_\gas v^2}{2}}~~\text{and}~~\gamma_\gas=\frac{m_\gas}{\boltz T_\gas}.
\end{eqnarray}
Note that the BGK operator can be an exact integration of the Boltzmann operator in the case of a Langevin interaction, \textit{i.e.}, $\sigma^{(0)} \propto 1/|v - v_\gas|$ (Refs.\cite{Robson17,Berger25}). However, as shown Fig.~\ref{fig:crossSection} this velocity dependence is not a realistic approximation for argon.

The BGK operator is often used in the theory of transport of charged species in a gas discharge (\textit{e.g.}, Refs.~\cite{Liberman05, Chabert11}). At high pressure, the drift-diffusion model proposed by Schottky~\cite{Schottky24} corresponds to the following frequency,
\begin{equation}\label{eq:Schottky}
    \nu^{\mathrm{Schottky}} = \frac{\bar{v}_\ion}{\lambda_{\ion\gas}} \,,
\end{equation}
with $\bar{v}_\ion = \sqrt{8 \boltz T_\ion / (\pi m_\ion)}$ and the mean free path is estimated with a characteristic cross section as $\lambda_{\ion\gas} = (n_\gas \bar{\sigma}_{\ion\gas}^{(0)})^{-1}$, which effectively assumes a hard sphere interaction (\textit{i.e.}, constant mean-free path). Note that proposed collision frequency should, in general, depend on the ion and neutral temperatures, as the cross section is a function of their relative velocities (which can be justified with the assumption that the ion temperature is equal to the gas temperature locally everywhere).

As noted by Godyak~\cite{Godyak86}, the Schottky expression fails at low pressure, where the ion drift velocity exceeds the thermal velocity ($u_\ion \gg v_{\mathrm{th},\ion}$). Godyak derived an alternative expression by solving for ions in a uniform plasma with a constant electric field and Langevin interaction. The resulting steady-state ion distribution is a half-Maxwellian due to a stationary neutral background with a Dirac distribution. This ion distribution yields a drift velocity from its asymmetry. The mobility computed from this distribution gives an effective momentum transfer frequency proportional to the drift velocity rather than the thermal velocity. as follows,
\begin{equation}
    \nu^{\mathrm{Godyak}} = \frac{\pi}{2} \frac{u_\ion}{\lambda_{\ion\gas}} \,.
\end{equation}

Finally, Chabert et al.~\cite{Chabert11} has proposed an expression that matches both asymptotic regimes, as follows,
\begin{equation}\label{eq:Chabert}
    \nu^{\mathrm{Chabert}}  = \frac{\bar{v}_\ion}{\lambda_{\ion\gas}} \left(1 + \left(\frac{\pi}{2}\frac{u_\ion}{\bar{v}_\ion}\right)^2\right)^{1/2} \,.
\end{equation}

It is to be noted that some of the hypotheses to derive this model are contradictory (\textit{e.g.}, the cross section is a hard sphere in Schottky's model whereas Godyak's considers a Langevin's cross section). Other hypotheses do not agree with the actual dynamics of the discharge at low pressure (in particular, the assumption of the variable mobility model that considers a spatially homogeneous solution for the ion distribution function). Finally, the characteristic mean free path $\lambda_{\ion\gas}$ is an input parameter for the model which can be challenging to assess in the case of gas mixtures and the model does not depend on the gas temperature.

\subsection{Moment equations and closure models}

\subsubsection{General definitions and moment transport equation}

\paragraph{Definition of moments:}

The moment equations are derived by integrating the kinetic equation over velocity space with velocity-dependent weights. {For distribution functions that depend on three velocity dimensions, the velocity moment of order $N=i+j+k$ reads,
\begin{eqnarray}\label{eq:momentConsCart}
    \tilde{M}_{ijk} &=& \int_{\mathbb{R}^3} m v_x^i v^j_y v^k_z f(t, x, v_x, v_y, v_z) \,\mathrm{d}v_x\mathrm{d}v_y \mathrm{d}v_z \,.
\end{eqnarray}
Due to the symmetry in the perpendicular velocity, all moments with odd $j$ and $k$ vanish. For instance, the first moments are:
$\tilde{M}_{000} = \rho = mn$ (mass density, with $n$ the number density),
$\tilde{M}_{100} = \rho u$ (mass density flux, with $u$ the drift velocity),
 $\tilde{M}_{010} = \rho u_y = 0$, and $\tilde{M}_{001} = \rho u_z = 0$.
}

{Given the symmetry $f(t, x, \vec{v}) = f(t, x, v_x, v_\perp)$, we propose a different definition, based on cylindrical coordinates with the polar axis aligned with $v_x$ and integrate over the azimuthal angles in the velocity space to define the moments of order $N = i + 2j$, as follows,
\begin{eqnarray}\label{eq:momentCons}
    M_{i\,2j} &=& \alpha_j\int_{-\infty}^\infty\int_{0}^\infty m v_x^i v^{2j+1}_\perp f(t, x, v_x, v_\perp) \,\mathrm{d}v_x\mathrm{d}v_\perp \,~~~\text{with}~~~\alpha_j=2\pi \frac{(2j-1)!!}{(2j)!!}.
\end{eqnarray}
The first moments are:
$M_{00} = \rho = mn$ (mass density, with $n$ the number density),
$M_{10} = \rho u$ (mass density flux, with $u$ the drift velocity)}. {The conversion from the Cartesian to the cylindrical moments, can be performed with the following formula
\begin{eqnarray}\label{eq:momentCyl}
    \tilde{M}_{i\,2\ell\, 2m} &=& I_{\ell m} M_{i\,2(\ell+m)}~~~\text{with}~~~I_{\ell m} = \frac{(2\ell-1)!!(2m - 1)!!}{(2m + 2\ell-1)!!}\,.
\end{eqnarray}
The normalization factors $\alpha_j$ and $I_{mn}$ appear due to the integration over the azimuthal angle and are chosen in a way that the terms in the main diagonal of the tensors in Cartesian moments read the same in cylindrical moments, i.e.,
\begin{eqnarray}\label{eq:momentCyltoCons}
    M_{i\,2j} = \tilde{M}_{i\,2j\, 0} =\tilde{M}_{i\,0\, 2j}.
\end{eqnarray}
For instance, $\tilde{M}_{020} = \tilde{M}_{002} = M_{02}$ or $\tilde{M}_{040} = \tilde{M}_{004} = M_{04}$. Note that this is not the case of terms outside the diagonal, e.g., $\tilde{M}_{022} = \frac{1}{3}M_{04}$. Nevertheless, the proposed notation allows to compute all the Cartesian moments in the case of cylindrical symmetry.}

{For simplicity, we exploit the property of Eq.~\eqref{eq:momentCyltoCons} to express the cylindrical moments with Cartesian coordinates, where $M_{i\,2j}$ is now defined as,
\begin{eqnarray}\label{eq:momentCylwithy}
    M_{i\,2j} = \int_{-\infty}^\infty\int_{\mathbb{R}^2} m v_x^i v^{2j}_y f(t, x, v_x, v_y, v_z) \,\mathrm{d}v_x\mathrm{d}^2\vec{v}_\perp \,.
\end{eqnarray}
Note that $v_y$ is chosen by convention and, due to the symmetry in the perpendicular plane, $v_z$ would yield the same result. We recall that $\mathrm{d}^2\vec{v}_\perp = \mathrm{d}v_y\mathrm{d}v_z.$
}

The centered moments are defined as,
{\begin{eqnarray}
    P_{i\,2j} &=& \int_{-\infty}^\infty\int_{\mathbb{R}^2} m c^i_x v^{2j}_y f(c_x, v_y, v_z) \,\mathrm{d}c_x \mathrm{d}^2\vec{v}_\perp \quad  \text{with} \quad c_x = v_x-u.
\end{eqnarray}}
Note that as there average velocity along the perpendicular direction is zero and hence $\vec{c}_\perp = \vec{v}_\perp$.

In this study, we will use the following centered moments:
{\begin{equation}\label{eq:Centered6M}
    P_{20} \equiv p_{xx} \equiv p_{x} \,, \quad P_{02} = p_{yy} = p_{zz} \equiv p_{\perp\perp} \equiv p_{\perp} \,, \quad P_{30} \equiv q \,, \quad  P_{40} \equiv r \quad \text{and} \quad P_{50} \equiv s,
\end{equation}}
respectively, the pressure in the $x$ direction, the pressure in the $\perp$ direction, and the heat flux, the kurtosis, and the hyper-skewness in the $x$ direction. Note that due to the geometry, $p_{x\perp} = 0 $.

Finally, the normalized (or standardized) moments are defined as follows,
{\begin{eqnarray}\label{eq:standardMoms}
    P^\star_{i\,2j} &=& \frac{P_{i\,2j}}{\rho v_{Tx}^i v_{T\perp}^{2j}},
\end{eqnarray}}
with the thermal velocities defined as $v_{Tx} = \sqrt{p_{x}/\rho}$ and $v_{T\perp} = \sqrt{p_{\perp}/\rho}$.

\paragraph{Maxwell's transfer equation:}

The general 1D moment transport equation (Maxwell's transfer equation) is obtained, by multiplying {by $m v_{x}^i v_{y}^{2j}$ Eq.~\eqref{eq:kinetic} and integrating over the velocity space}. This yields,
{\begin{equation}\label{eq:MaxwellTransfer}
    \frac{\partial{M_{i\,2j}}}{\partial t} + \frac{\partial{M_{(i+1) \,2j}}}{\partial x} = i \frac{eE_x}{m} M_{(i-1)\,2j} + \mathcal{C}_{i\,2j}.
\end{equation}}

We define the $n$-th order moment of the collision operator as follows,
{\begin{equation}\label{eq:general_coll_term}
    \mathcal{C}_{i\,2j} = \mathcal{C}^{\el}_{i\,2j} + \mathcal{C}^{\iz}_{i\,2j}
    = \int_{\mathbb{R}^3} \!\!\!\!  m v_{x}^i v_{y}^{2j}\left.\frac{\delta f}{\delta t}\right\vert_\text{c} \mathrm{d}v_x\mathrm{d}^2\vec{v}_\perp \,.
\end{equation}}

In particular, the moments of the ionization collisions, can be easily computed from Eq.~\eqref{eq:BoltzmannReact_simple} and they only depend on the distribution function of the gas species, which reads:
{\begin{equation}\label{eq:general_coll_iz}
    \mathcal{C}^{\iz}_{i\,2j}  =
     \int_{\mathbb{R}^3} \!\!\!\!  m v_{x}^i v_{y}^{2j}\left.\frac{\delta f}{\delta t}\right\vert^\iz_\text{c} \mathrm{d}v_x\mathrm{d}^2\vec{v}_\perp = S^\text{iz} \frac{M_{\gas_{i\,2j}}}{m_\gas n_\gas}\,,
\end{equation}}
where $S^\text{iz}=m n_\gas n_\elec K_\text{\iz}^{(0)}$ and $M_{\gas_{i,\,2j}}$ is the $n$-th moment (with $n=i+2j$) of the background gas distribution. In the following, we will consider that the gas distribution is a Maxwellian with no drift velocity and, therefore, the moments can be easily expressed as function of the gas density and temperature, \textit{e.g.}, $M_{\gas_{00}} = m_\gas n_\gas$,  $M_{\gas_{10}} = 0$, $M_{\gas_{20}} = M_{\gas_{02}} = n_\gas\boltz T_\gas$, etc.

\subsubsection{Closure models}
In the following, we describe the equations of three different one dimensional high-order moment hierarchies: a four moment (4M) model considering anisotropic pressure tensor, a 5M HyQMOM model (as described in Ref.\cite{Berger25}), and a 6M HyQMOM model extending the previous model with perpendicular energy. For the sake of completeness, we show as well the 3M equations.

\paragraph{3M model:}

The set of equations solve the balance equations of mass density, momentum, and the contracted energy tensor, i.e., $M_{20} + 2M_{02}$, as follows,
\begin{subequations}\label{eq:system3M}
\begin{eqnarray}
    \frac{\partial \rho }{\partial t}+ \partialx[(\rho u)] &=& S^\text{iz}, \label{eq:3M-0} \\
    \frac{\partial (\rho u)}{\partial t} + \partialx[]\left(\rho u^2 + p\right) &=&\frac{e E_x}{m} \rho+ \mathcal{C}^{\el}_{10}, \\
    \frac{\partial}{\partial t}\left(\rho u^2 + 3p\right) + \partialx[]\left(\rho u^3 + 5pu\right) &=& 2 \frac{e E_x}{m} \rho u + \mathcal{C}^{\el}_{20} + 2\mathcal{C}^{\el}_{02} + 3S^\text{iz} \frac{\boltz T_{\gas}}{m_\gas}.
\end{eqnarray}
\end{subequations}
The closure flux assumes that the two temperatures are equal, i.e., $p_x = p_\perp \equiv p$, and the heat flux to be zero, i.e., $P^{(3)}_{30} = P^{(3)}_{12} = 0$.

\paragraph{4M model with anisotropic pressure:}

The set of equations solve the balance equations of mass density, momentum, energy along the $x$ direction and energy in the perpendicular direction, as follows,
\begin{subequations}\label{eq:system4M}
\begin{eqnarray}
    \frac{\partial \rho }{\partial t}+ \partialx[(\rho u)] &=& S^\text{iz}, \label{eq:4M-0} \\
    \frac{\partial (\rho u)}{\partial t} + \partialx[M_{20}] &=&\frac{e E_x}{m} \rho+ \mathcal{C}^{\el}_{10}, \\
    \frac{\partial M_{20}}{\partial t} + \partialx[M_{30}] &=& 2 \frac{e E_x}{m} \rho u + \mathcal{C}^{\el}_{20} + S^\text{iz} \frac{\boltz T_{\gas}}{m_\gas},   \label{eq:4M-2}\\
    \frac{\partial M_{02}}{\partial t} + \partialx[M_{12}] &=& \mathcal{C}^{\el}_{02} + S^\text{iz} \frac{\boltz T_{\gas}}{m_\gas}.
\end{eqnarray}
\end{subequations}
The closure flux assumes that the heat flux is zero, i.e., $P_{30} = P_{12} = 0$. As a result, the moments computed as a function of the centered moments are:
\begin{subequations}\label{eq:consMoments5M}
\begin{eqnarray}
M_{20} &=& \rho u^2 + p_{x},  \\
M_{02} &=& p_{\perp},  \\
M_{30} &=&\rho u^3 + 3 u p_{x},  \\
M_{12} &=& u p_{\perp}.
\end{eqnarray}
\end{subequations}

\paragraph{5M HyQMOM model:}

This closure is the same as used in Ref.~\cite{Berger25}, which is purely 1D-1V. The set of equations consider the moments up to the fourth-order moment in the $x$ direction without considering the perpendicular direction, as follows:
\begin{subequations}\label{eq:system5M}
\begin{eqnarray}
    \frac{\partial \rho }{\partial t}+ \partialx[(\rho u)] &=& S^\text{iz}, \label{eq:5M-0} \\
    \frac{\partial (\rho u)}{\partial t} + \partialx[M_{20}] &=&\frac{e E_x}{m} \rho+ \mathcal{C}^{\el}_{10},  \label{eq:5M-1} \\
    \frac{\partial M_{20}}{\partial t} + \partialx[M_{30}] &=& 2 \frac{e E_x}{m} \rho u + \mathcal{C}^{\el}_{20} + S^\text{iz} \frac{k_\text{B} T_{\gas}}{m_\gas},   \label{eq:5M-2}\\
    \frac{\partial M_{30}}{\partial t} + \partialx[M_{40}] &=& 3 \frac{e E_x}{m} M_{20} + \mathcal{C}^{\el}_{30},  \label{eq:5M-3} \\
    \frac{\partial M_{40}}{\partial t} + \partialx[M_{50}] &=& 4 \frac{e E_x}{m} M_{30} + \mathcal{C}^{\el}_{40} + 3 S^\text{iz} \left(\frac{k_\text{B} T_{\gas}}{m_\gas}\right)^2.\label{eq:5M-4}
\end{eqnarray}
\end{subequations}
As described in Ref.\cite{Berger25}, the closure can be computed as follows:
\begin{subequations}\label{eq:consMoments6M}
\begin{eqnarray}
M_{20} &=& \rho u^2 + p_{x},  \\
M_{30} &=&\rho u^3 + 3 u p_{x} + q,  \\
M_{40} &=&\rho u^4 + 6 u^2 p_{x} + 4 u q + r, \\
M_{50} &=&\rho u^5 + 10 u^3 p_{x}+ 10 u^2 q + 5ru + s.
\end{eqnarray}
\end{subequations}
where the closing flux can be computed from the moments by the following relation in standardized moments, as defined in Eq.~\eqref{eq:standardMoms}, (see Ref.\cite{Berger25} for more details),
\begin{equation}\label{eq:5MClosure}
    s_\star = 2 r_\star q_\star - q_\star^3.
\end{equation}

\paragraph{6M HyQMOM model with anisotropic pressure:}

In this paper, we propose to improve the 5M HyQMOM, by including the effect of the energy in the perpendicular direction. The system of moment equations reads:
\begin{subequations}\label{eq:system6M}
\begin{eqnarray}
    \frac{\partial \rho }{\partial t}+ \partialx[(\rho u)] &=& S^\text{iz}, \label{eq:6M-0} \\
    \frac{\partial (\rho u)}{\partial t} + \partialx[M_{20}] &=&\frac{e E_x}{m} \rho+ \mathcal{C}^{\el}_{10},  \label{eq:6M-1} \\
    \frac{\partial M_{20}}{\partial t} + \partialx[M_{30}] &=& 2 \frac{e E_x}{m} \rho u + \mathcal{C}^{\el}_{20} + S^\text{iz} \frac{k_\text{B} T_{\gas}}{m_\gas},   \label{eq:6M-2}\\
    \frac{\partial M_{02}}{\partial t} + \partialx[M_{12}] &=& \mathcal{C}^{\el}_{02} + S^\text{iz} \frac{\boltz T_{\gas}}{m_\gas},  \\
    \frac{\partial M_{30}}{\partial t} + \partialx[M_{40}] &=& 3 \frac{e E_x}{m} M_{20} + \mathcal{C}^{\el}_{30},  \label{eq:6M-3} \\
    \frac{\partial M_{40}}{\partial t} + \partialx[M_{50}] &=& 4 \frac{e E_x}{m} M_{30} + \mathcal{C}^{\el}_{40} + 3 S^\text{iz} \left(\frac{k_\text{B} T_{\gas}}{m_\gas}\right)^2. \label{eq:6M-4}
\end{eqnarray}
\end{subequations}
The closure is defined as follows:
\begin{subequations}\label{eq:consMoments4M}
\begin{eqnarray}
M_{20} &=& \rho u^2 + p_{x},  \\
M_{02} &=& p_{\perp},  \\
M_{30} &=&\rho u^3 + 3 u p_{x} + q,  \\
M_{12} &=& u p_{\perp}, \label{eq:heatfluxperp}\\
M_{40} &=&\rho u^4 + 6 u^2 p_{x} + 4 u q + r, \\
M_{50} &=&\rho u^5 + 10 u^3 p_{x}+ 10 u^2 q + 5ru + s.
\end{eqnarray}
\end{subequations}
Here, the closure is computed from Eq.~\eqref{eq:5MClosure} for $s$ and we have the perpendicular heat flux in Eq.~\eqref{eq:heatfluxperp}, i.e., $P_{12} = 0$.

\section{Collisional terms in the moment equations via the direct Integration of the Boltzmann Operator}
\label{sec:DIBO-computation}

In this section we will derive the elastic collision terms $\mathcal{C}^{\el}_{i\,2j}$, defined in Eq.~\eqref{eq:general_coll_term}, with the Boltzmann operator of Eq.~\eqref{eq:Boltzmann}, as follows,
{\begin{multline}\label{eq:integral0}
    \mathcal{C}^{\el}_{i\,2j} = \int_{\mathbb{R}^3}\int_{\mathbb{R}^3}\int_{\mathbb{S}^2}  m v_{x}^i v_{y}^{2j} \left(f' f_{\gas}' - f f_{\gas}\right)\,\lvert \vec{v}-\vec{v}_{\gas}\rvert\,\sigma  \,\mathrm{d}^2\Omega \mathrm{d}^3\vec{v}_{\gas}\mathrm{d}^3\vec{v} \\
    = \int_{\mathbb{R}^3}\int_{\mathbb{R}^3}\int_{\mathbb{S}^2}  m \left(v_{x}'^i v_{y}'^{2j} - v_{x}^i v_{y}^{2j}\right) f f_{\gas}\,\lvert \vec{v}-\vec{v}_{\gas}\rvert\,\sigma  \,\mathrm{d}^2\Omega \mathrm{d}^3\vec{v}_{\gas}\mathrm{d}^3\vec{v},
\end{multline}}
where $f' = f(t,x,\vec{v}')$. The second equality is obtained by using the reciprocity relations of the integration of the multi-species Boltzmann operator (see, \textit{e.g.}, Ref.~\cite{Zhdanov02}). In particular, in the moment equations \eqref{eq:system3M}, \eqref{eq:system4M}, \eqref{eq:system5M} and \eqref{eq:system6M}, we require collisional terms $\mathcal{C}^{\el}_{i0}$ for $i\in(1,4)$ in the $x$ direction and $\mathcal{C}^{\el}_{02}$ for the perpendicular direction.

There are two fundamental steps in the integration of Eq.~\eqref{eq:integral0}. The first step, is the integration over the scattering angles, i.e., the integration over the solid angle $\Omega$, which can be performed analytically for arbitrary distribution functions. The second step is the integration over the velocity spaces of ions and gas atoms, which requires a nonequilibrium distribution function that is compatible with the closure used in the moment equations.  We detail the analytical derivations in the following sections.

\subsection{Integration over the scattering angles}

As detailed in previous works\cite{AlvarezLaguna22,AlvarezLaguna26}, in Eq.~\eqref{eq:integral0}, the part that depends on the scattering angles reduces to,
\begin{equation}\label{eq:angularI_ij}
    {\mathcal{I}_{i\,2j} = \int_{\mathbb{S}^2}  m\left(v_{x}'^i v_{y}'^{2j} - v_{x}^i v_{y}^{2j}\right) \sigma(\lvert \vec{v}-\vec{v}_{\gas}\rvert, \Omega)  \,\mathrm{d}^2\Omega.}
\end{equation}
In particular, we require the integrals for $\mathcal{I}_{i0}$ for $i\in(1-4)$ and $\mathcal{I}_{02}$. In order to perform the integration, we change the velocities to the center of mass velocity and relative velocity, respectively,
\begin{equation}
    \vec{G} = \frac{m \vec{v} + m_\gas \vec{v}_\gas}{m + m_\gas}=\vec{G}' \quad\quad \vec{g} = \vec{v} - \vec{v}_\gas\quad\text{and}\quad\vec{g}' = \vec{v}' - \vec{v}_\gas'
\end{equation}
where we have used the conservation of momentum in the center of mass velocity. It is to be noted that the Jacobian of this transformation is unity\cite{Zhdanov02}.

The relevant terms for our study of the integral Eq.~\eqref{eq:angularI_ij} in the new set of variables reads
\begin{subequations}
\begin{align}
    & m \left[(v'_{x})^k - v_{x}^k\right] = \mu \sum_{n=1}^k \binom{k}{n} G_x^{k-n} \left(\frac{\mu}{m}\right)^{n-1} \left[ (g'_x)^n - g_x^n\right], \label{eq:difference} \\
    & {m \left[(v'_{y})^2 - v_{y}^2\right] = \mu \left\{2 G_{y} \left(g_{y}' - g_{y}\right) + \tfrac{\mu}{m} \left[(g_{y}')^2 - (g_{y})^2\right]\right\}.}
\end{align}
\end{subequations}
where the reduced mass is $\mu = m m_\gas/(m + m_\gas)$.

In order to perform the integration over the scattering angles, we recall that the differential cross section depends on the scattering angle between the relative velocities $\vec{g}$ and $\vec{g}'$, \textit{i.e.}, $\sigma(|g|, \chi)$ where, $\vec{g}\cdot\vec{g}' = |g|^2\cos{\chi}$ (using the conservation of energy $|g'| = |g|$ ). As a result, we perform the integral of Eq.~\eqref{eq:angularI_ij} in spherical coordinates by choosing $\vec{g}$ as the polar direction, \textit{i.e.}, $ \vec{g}' = \vec{g} \cos\chi + g\sin\chi \left( \cos\varphi \,\eperpgOne + \sin\varphi \,\eperpgTwo \right)$ where $\eperpgi[1,\,2]$ are the unit vectors in the perpendicular directions of the vector $\vec{g}$.

Injecting Eq.~\eqref{eq:difference} into Eq.~\eqref{eq:angularI_ij}, we obtain the following expression. The integration over the angles $\varphi$ and $\chi$ is detailed in Appendix \ref{app:IntegrationUnitSphere}. The final result reads,
\begin{multline}\label{eq:integralIk0Final}
    \mathcal{I}_{k0} = \int_{\mathbb{S}^2} \!\!\!\! m \left[(v'_{x})^k - v_{x}^k\right] \sigma(g, \Omega) \text{d}^2\Omega\ = \quad \mu \sum_{j=1}^k \binom{k}{j} G_x^{k-j} \left(\frac{\mu}{m}\right)^{j-1} \int_{\mathbb{S}^2} \!\!\!\!  \left[ (g'_x)^j - g_x^j\right] \sigma(g, \Omega) \text{d}^2\Omega\\=
    \mu \sum_{j=1}^k \binom{k}{j} G_x^{k-j} \left(\frac{\mu}{m}\right)^{j-1} \sum_{i=0}^{\lfloor j/2 \rfloor} \binom{2i}{i} \binom{j}{2i} g_x^{j-2i} \left(\tfrac{1}{4}\left(g^2 - g_x^2\right)\right)^{i} \sum_{l=0}^i \binom{i}{l} (-1)^{i-l+1} Q^{(j-2l)}(g),
\end{multline}
where the transport cross section\cite{Chapman70} is defined as,
\begin{equation}
    Q^{(l)}(|\vec{g}|) = 2\pi\int_0^{\pi} (1 - \cos^l\chi)\,\sigma(|\vec{g}|,\,\chi)\sin\chi\,d\chi.
\end{equation}
Note that due to Galilean invariance of collisions\cite{gallagher13}, the integral $\mathcal{I}_{02}$ can be computed from the previous expression, as follows,
{\begin{equation}
    \mathcal{I}_{20} + 2\mathcal{I}_{02} = \int_{\mathbb{S}^2} \!\!\!\! \, m \left[(\vec{v}')^2 - \vec{v}^2\right] \sigma(g, \Omega) \text{d}^2\Omega = \int_{\mathbb{S}^2} \!\!\!\! \, \mu \vec{G}\cdot\left[\vec{g}' - \vec{g}\right] \sigma(g, \Omega) \text{d}^2\Omega = -\mu \vec{G}\cdot\vec{g} Q^{(1)}(g) \,,
\end{equation}
which gives}
\begin{eqnarray}\label{eq:integralI02Final}
    {\mathcal{I}_{02} = \int_{\mathbb{S}^2} \!\!\!\! \, m \left[(v'_{y})^2 - v_{y}^2\right] \sigma(g, \Omega) \text{d}^2\Omega = -\mu \left[ 2 G_y g_y Q^{(1)}(g) + \tfrac{1}{2}\tfrac{\mu}{m}  \left(3 g_y^2 - g^2\right) Q^{(2)}(g) \right].}
\end{eqnarray}

\subsection{Integration over ion and neutral velocities with different distributions}

In this section, we detail the integration of the moment of the collision terms (Eq.~\eqref{eq:integral0}) with the results of the integration over the scattering angles (Eqs.~\eqref{eq:integralIk0Final} and \eqref{eq:integralI02Final}), as follows,
\begin{equation}\label{eq:integral1}
    \mathcal{C}^{\el}_{i\,2j} = \int_{\mathbb{R}^3}\int_{\mathbb{R}^3}\mathcal{I}_{i\,2j}(\vec{g},\, \vec{G}) f(\vec{g},\vec{G}) f_{\gas}(\vec{g},\vec{G})\,\lvert \vec{g}\rvert\, \mathrm{d}^3\vec{G}\mathrm{d}^3\vec{g}.
\end{equation}
In order to perform the integration over the velocity spaces of ions and neutrals, we need to consider a mathematical expression for the ion distribution function that is consistent with the moments that are known from the resolution of the moment equations, including the closure flux. We stress that in the case of nonequilibrium there are an infinite number of distribution functions that satisfy a finite set of moments. We will choose distribution functions that allow for analytical integrations of Eq.~\eqref{eq:integral1}, and that guarantee a strictly positive distribution (as opposed to Grad's method that can create negative tails far from thermodynamics equilibrium).

\begin{figure}
    \centering
    \includegraphics[width=\linewidth]{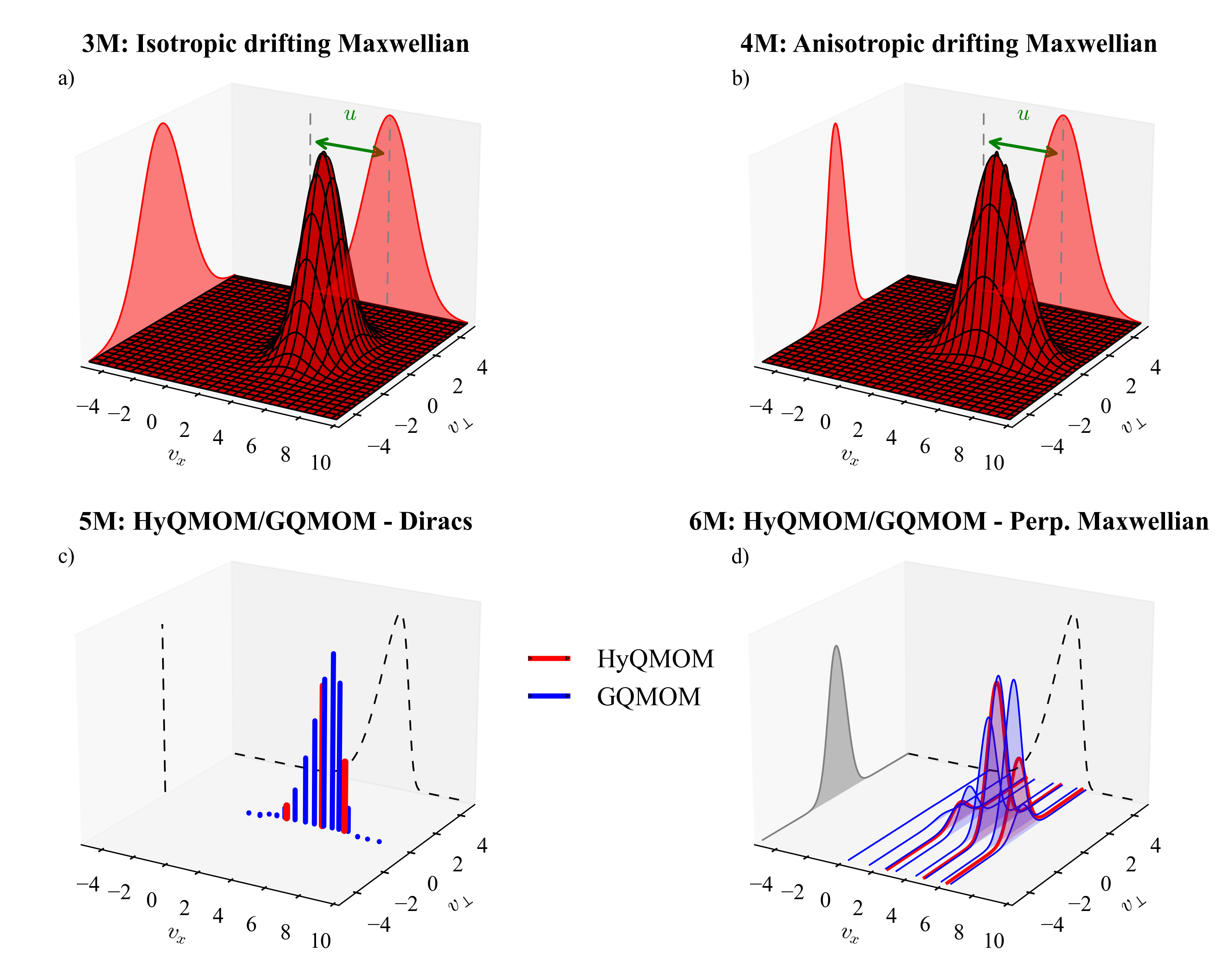}
    \caption{Schematic of the ansatz for the distribution functions used by the different moment models: a) Drifting isotropic Maxwellian for the 3M model, b) drifting anisotropic Maxwellian for the 4M model, c) Dirac distributions for the 5M HyQMOM model, and d) Dirac distributions with Maxwellian distribution in the perpendicular direction for the 6M model.}
    \label{fig:AnsatzVDF}
\end{figure}

The ion distributions considered in this work are schematically depicted in Fig.~\ref{fig:AnsatzVDF}. They are a drifting (with arbitrary Mach) isotropic Maxwellian for the 3M model, a drifting anisotropic Maxwellian (with arbitrary Mach and temperature anisotropies) for the 4M model, Dirac distributions for the 5M HyQMOM model, and  Dirac distributions with Maxwellian distribution in the perpendicular direction for the 6M model.

For simplicity, we will henceforth drop the superscript and use the notation \(\mathcal{C}_{i,\,2j}\) instead of \(\mathcal{C}^{\el}_{i,\,2j}\).

\subsubsection{3M model: Isotropic drifting Maxwellian}

In this section, we reformulate the results of Benilov\cite{Benilov97} that will be used to compare to high-order models. To simplify the notation, and without any loss of generality, we will consider a distribution for the gas without drift, which is equivalent to changing referential to the one moving at the gas velocity.

The ion and neutral VDFs are assumed to be two isotropic Maxwellian distributions at different temperature, with the ions drifting at arbitrarily large velocities, as follows,
\begin{subequations}\label{eq:Maxwellians}
\begin{align}
    f^\mathrm{(3M)}(\vec{v}) &= n \left(\frac{\gamma_\ion}{2\pi}\right)^{3/2} \exp\left( -\tfrac{\gamma_\ion}{2}\left(\vec{v} - \vec{u}\right)^2 \right) \,, \\
    f_\gas(\vec{v}_\gas) &= n_\gas \left(\frac{\gamma_\gas}{2\pi}\right)^{3/2} \exp\left( -\tfrac{\gamma_\gas}{2}v_\gas^2 \right) \,~~~\text{with}~~~\gamma_s = \frac{m_s}{k_\mathrm{B}T_s} ~~~\text{for}~s\in\{\ion,\gas\}.
\end{align}
\end{subequations}
The details of the derivation are provided in Appendix \ref{app:3M-4Mmodel}.

The collision terms for the 3M model read:
\begin{subequations}\label{eq:collTerm1_isoMaxw}
    \begin{align}
        \mathcal{C}_{10} =& -\tfrac{16}{3}\mu n n_\gas \Omega^{(1,1,1)}_{\mathrm{iso}} u\,, \label{eq:collTerm1_isoMaxw1}\\
        \mathcal{C}_{20} =& \tfrac{\mu}{m + m_\gas} n n_\gas\left\{\tfrac{32}{3}\Omega^{(1,1,2)}_{\mathrm{iso}}\boltz(T_\gas - T) - \tfrac{32}{3}\Omega^{(1,1,1)}_{\mathrm{iso}}\tfrac{T_\gas}{T_{\ion,\gas}} m u^2 \right\} + 8 \tfrac{\mu}{m} n n_\gas\left(\Omega^{(2,1,0)}_{\mathrm{iso}} - \Omega^{(2,1,2)}_{\mathrm{iso}}\right) \boltz T_{\ion,\gas}\,\label{eq:collTerm1_isoMaxw2}\\
        \mathcal{C}_{02} =& \tfrac{16}{3}\tfrac{\mu}{m + m_\gas} n n_\gas\left(3\Omega^{(1,1,0)}_{\mathrm{iso}} - \Omega^{(1,1,2)}_{\mathrm{iso}}\right)\boltz(T_\gas - T) + 4 \tfrac{\mu}{m} n n_\gas\left(\Omega^{(2,1,2)}_{\mathrm{iso}} - \Omega^{(2,1,0)}_{\mathrm{iso}}\right) \boltz T_{\ion,\gas}.\label{eq:collTerm1_isoMaxw3}
    \end{align}
\end{subequations}
Here, the reduced temperature is $T_{\ion, \gas} = (m_\ion T_\gas + m_\gas T_\ion)/(m_\ion + m_\gas) $.

We introduce a new definition of the rate coefficients $\Omega^{(l,r,s)}_{\mathrm{iso}}$ that are a generalization of the Chapman-Cowling integrals, that depend on the normalized drift velocity and the temperatures, as follows,
\begin{equation}\label{eq:OmegaIso}
    \Omega^{(l,r,s)}_{\mathrm{iso}}(T_{\ion, \gas}, \bar{u}) = \frac{1}{2^{2+r}}\left(\frac{1}{2\pi\gamma_{\ion,\gas}}\right)^{1/2}\int_0^\infty \bar{g}^{3+2r}\, Q^{(l)}(\bar{g}\gamma_{\ion,\gas}^{-1/2})\, \mathbb{I}^{(s)}_{\mathrm{iso}} (\bar{u}\bar{g})\, e^{-\frac{\bar{g}^2}{2} -\frac{\bar{u}^2}{2}} d\bar{g},
\end{equation}
where the integral over the angles between $\vec{g}$ and $\vec{u}$ reads,
\begin{equation}\label{eq:angularIntegral}
    \mathbb{I}^{(s)}_{\mathrm{iso}} (\bar{u}\bar{g}) = \frac{1}{\left(\frac{\bar{u}\bar{g} + (-1)^{s+1}\bar{u}\bar{g}}{s+2} + \frac{1 + (-1)^{s}}{s+1}\right)}\int_{-1}^{1}\zeta^s \exp(\bar{u}\bar{g}\zeta)d\zeta,
\end{equation}
and the normalized drift velocity (or pseudo-Mach number) and normalized relative velocity are defined as
\begin{equation}
    \bar{u} = u\gamma_{\ion,\gas}^{1/2} = \sqrt{\frac{\mu u^2}{\boltz T_{i,g}}}~~~\text{and}~~~\bar{g} = g\gamma_{\ion,\gas}^{1/2}.
\end{equation}
An analytical solution for the angular integral of Eq.~\eqref{eq:angularIntegral} is given in table \ref{tab:IsIso}.

The definition of $\Omega^{(l,r,s)}_{\mathrm{iso}}$ is chosen such that the rate coefficient tends to the classical Chapman-Cowling integrals as the drift velocity tends to zero (for low-Mach number of the relative velocity):
\begin{equation}\label{eq:lowMachLimIso}
    \lim_{\bar{u}\rightarrow 0} \Omega^{(l,r,s)}_{\mathrm{iso}}(T_{\ion, \gas}, \bar{u}) = \Omega^{(l,r)}(T_{\ion, \gas}) ~~~\text{with}~~~\Omega^{(l,r)}(T_{\ion, \gas}) = \frac{1}{2^{2+r}}\left(\frac{1}{2\pi\gamma_{\ion\gas}}\right)^{1/2}\int_0^\infty \bar{g}^{3+2r}Q^{(l)} e^{-\frac{\bar{g}^2}{2}} d\bar{g}.
\end{equation}
This is because the angular integral satisfies $\lim_{\bar{u}\rightarrow 0} \mathbb{I}^{(s)}_{\mathrm{iso}} (\bar{u}\bar{g}) = 1$.

\begin{table}[h]
    \centering
    \begin{tabular}{l @{\hspace{3em}} c}
        $s$ & $\mathbb{I}^{(s)}_{\mathrm{iso}} (x)$ \\
        \hline
        \hline
        $0$ & $\tfrac{\sinh(x)}{x}$ \\
        $1$ & $\tfrac{3x\cosh(x) - 3\sinh(x)}{x^3}$\\
        $2$ & $\tfrac{3x^2\sinh(x) - 6x\cosh(x) + 6\sinh(x)}{x^3}$\\
        $s$ & $\frac{1}{D_s} \left( \frac{e^{x} - (-1)^s e^{-x}}{x} - \frac{s D_{s-1}}{x} \mathbb{I}^{(s-1)}_{\mathrm{iso}}(x) \right)$\\
        \hline
    \end{tabular}
    \caption{Analytical values of $\mathbb{I}^{(s)}_{\mathrm{iso}}$} with $D_s(x) = \frac{x + (-1)^{s+1} x}{s+2} + \frac{1 + (-1)^s}{s+1}$.
    \label{tab:IsIso}
\end{table}

Note that the collision term for the contracted energy reads,
\begin{equation}
    \mathcal{C}_{20} + 2\mathcal{C}_{02} = \tfrac{\mu}{m + m_\gas} n n_\gas\left\{32\Omega^{(1,1,0)}_{\mathrm{iso}}\boltz(T_\gas - T) - \tfrac{32}{3}\Omega^{(1,1,1)}_{\mathrm{iso}}\tfrac{T_\gas}{T_{\ion,\gas}} m u^2 \right\}.
\end{equation}
{Here, we have used the property of the definition Eq.~\eqref{eq:momentCyltoCons}, so, in Cartesian moments, $\mathcal{C}_{02} = \tilde{\mathcal{C}}_{020} = \tilde{\mathcal{C}}_{002}$, hence, $\tilde{\mathcal{C}}_{200} + \tilde{\mathcal{C}}_{020} + \tilde{\mathcal{C}}_{002} = \mathcal{C}_{20} + 2\mathcal{C}_{02}$}. This expression (divided by two due to the different definition of the energy) is equivalent to Ref.~\cite{Benilov97}. Additionally, by introducing the low-Mach limit of Eq.~\eqref{eq:lowMachLimIso}, the source terms are equivalent to the expressions of Chapter 3 of Ref.~\cite{Zhdanov02} (the so-called quasi-hydrodynamic approach).

\subsubsection{4M model: Anisotropic drifting Maxwellian}

 We will assume that the ion distribution is an anisotropic Gaussian with arbitrary axial drift velocity and axial and perpendicular temperatures whereas the neutral distribution is an isotropic Maxwellian with no drift (as in the previous case, this can be generalized to arbitrary neutral drift velocities), as follows,
\begin{equation}\label{eq:anisotropicGaussian}
    f^\mathrm{(4M)}(v_x,\,v_\perp) = n \frac{\gamma^{1/2}_x\gamma_\perp}{(2\pi)^{3/2}} e^{-\frac{\gamma_x}{2}\left(v_x - u \right)^2-\frac{\gamma_\perp}{2}v^2_\perp}~~~\text{with}~~~\gamma_x = \frac{m_\ion}{k_\mathrm{B}T_x},~\gamma_\perp = \frac{m_\ion}{k_\mathrm{B}T_\perp},
\end{equation}
where the axial and perpendicular temperatures are defined as $T_x = p_x/(n \boltz)$ and $T_\perp = p_\perp/(n \boltz)$, respectively.

The details of the derivation are presented in Appendix \ref{app:3M-4Mmodel}. The collision terms for the 4M model read:
\begin{subequations}\label{eq:collTerm1_AnisoMaxw}
    \begin{align}
        \mathcal{C}_{10} =& -\tfrac{16}{3}\mu n n_\gas \Omega^{(1,1,1)}_{\mathrm{an}} u\,,\label{eq:collTerm1_AnisoMaxw1}\\
        \mathcal{C}_{20} =& \tfrac{32}{3}\tfrac{\mu}{m + m_\gas} n n_\gas\left\{\Omega^{(1,1,2)}_{\mathrm{an}}\boltz(T_\gas - T_x) - \Omega^{(1,1,1)}_{\mathrm{an}}\tfrac{T_\gas}{T_{x,\gas}} m u^2 \right\} + 8 \tfrac{\mu}{m_\ion} n n_\gas\left(\Omega^{(2,1,0)}_{\mathrm{an}} - \Omega^{(2,1,2)}_{\mathrm{an}}\right) \boltz T_{x,\gas},\label{eq:collTerm1_AnisoMaxw2}\\
        \mathcal{C}_{02} =& \frac{16}{3}\tfrac{\mu \kappa_{x,\perp} }{m + m_\gas} n n_\gas\left(3\Omega^{(1,1,0)}_{\mathrm{an}} - \Omega^{(1,1,2)}_{\mathrm{an}}\right)\boltz(T_\gas - T_\perp) + 4 \tfrac{\mu}{m} n n_\gas\left(\Omega^{(2,1,2)}_{\mathrm{an}} - \Omega^{(2,1,0)}_{\mathrm{an}}\right) \boltz T_{x,\gas}.\label{eq:collTerm1_AnisoMaxw3}
    \end{align}
\end{subequations}
Here, we define,
\begin{equation}\label{eq:kappa_xperp}
    \kappa_{x,\perp} = \frac{m_\ion T_\gas + m_\gas T_x}{m_\ion T_\gas + m_\gas T_\perp},~~~T_{x,\gas} = \frac{m_\ion T_\gas + m_\gas T_x}{m_\ion + m_\gas }~~~\text{and}~~~ \bar{u} = u_\ion\gamma_{x,\gas}^{1/2}~~~\text{with}~~\gamma_{x,\gas} = \frac{\gamma_x\gamma_\gas}{\gamma_x + \gamma_\gas}.
\end{equation}
As done before in Eq.~\eqref{eq:OmegaIso}, we introduce a new collision rate that is a generalization of the Chapman-Cowling integrals that depends on the temperature anisotropy and the drift velocity, defined as
\begin{equation}\label{eq:OmegaAn}
    \Omega^{(l,r,s)}_{\mathrm{an}}(\kappa_{x,\perp}, T_{x, \gas}, \bar{u}) = \frac{\kappa_{x,\perp}}{2^{2+r}}\left(\frac{1}{2\pi\gamma_{x,\gas}}\right)^{1/2}\int_0^\infty \bar{g}^{3+2r}\, Q^{(l)}(\bar{g}\gamma_{x,\gas}^{-1/2})\, \mathbb{I}^{(s)}_{\mathrm{an}} (\kappa_{x,\perp}, \bar{u}, \bar{g})\, e^{-\frac{\kappa_{x,\perp}\bar{g}^2}{2} -\frac{\bar{u}^2}{2}} d\bar{g},
\end{equation}
where normalized relative velocity reads $\bar{g} = g\gamma_{x,\gas}^{1/2}$
and the integral over the angles between $\vec{g}$ and $\vec{u}$ reads,
{\begin{equation}\label{eq:angularIntegralAn}
    \mathbb{I}^{(s)}_{\mathrm{an}} \left(\bar{u}\bar{g}, \tfrac{(\kappa_{x,\perp} - 1)\bar{g}^2}{2}\right) = \frac{1}{\left(\frac{\bar{u}\bar{g} + (-1)^{s+1}\bar{u}\bar{g}}{s+2} + \frac{1 + (-1)^{s}}{s+1}\right)}\int_{-1}^{1}\zeta^s \exp\left(\bar{u}\bar{g}\zeta + (\kappa_{x,\perp} - 1)\frac{\bar{g}^2\zeta^2}{2}\right)d\zeta,
\end{equation}}
{For completeness, the analytical solutions of this integral are provided in Table~\ref{tab:I_an}, though they are of limited practical use because they are indeterminate in the limit $y \rightarrow 0$ ($\kappa_{x,\,\perp} \to 1$, i.e., isotropic) and require taking the limit. Nevertheless, the solution remains real and continuous for all $y$. Similarly, the exponentials and $\sinh$, $\cosh$ terms can become very large, producing large numerical errors. Therefore, in the following section, we describe how to compute them numerically to avoid these problems.}

We find the equivalence with the 3M isotropic Maxwellian collisional source terms for $T_x = T_\perp = T_\ion$, hence, $T_{x,\gas} = T_{\ion,\gas}$ and $\kappa_{x,\perp} = 1$, which yields,
\begin{equation}
    \Omega^{(l,r,s)}_{\mathrm{an}}(\kappa_{x,\perp} = 1, T_{x, \gas} = T_{\ion, \gas}, \bar{u}) = \Omega^{(l,r,s)}_{\mathrm{iso}}(T_{\ion, \gas}, \bar{u}).
\end{equation}

It can be easily seen that Eq.~\eqref{eq:OmegaAn} for isotropic case ($\kappa_{x,\perp} = 1$) is equivalent to Eq.~\eqref{eq:OmegaIso}. As a result, these definitions satisfy the following relations with respect to the isotropic case,
\begin{equation}
    \lim_{\bar{u}\rightarrow 0} \Omega^{(l,r,s)}_{\mathrm{an}}(\kappa_{x,\perp} = 1, T_{x, \gas}  = T_{\ion, \gas}, \bar{u}) = \Omega^{(l,r)}(T_{\ion, \gas}) ~~~\text{and}~~~\lim_{\bar{u}\rightarrow 0}\mathbb{I}^{(s)}_{\mathrm{an}} (\kappa_{x,\perp} = 1,\bar{u},\bar{g}) = 1.
\end{equation}
\begin{table}[h]
    \centering
    \begin{tabular}{l @{\hspace{1em}} c}
        $s$ & $\mathbb{I}^{(s)}_{\mathrm{an}}(x, y)$ \\
        \hline
        \hline
        $0$ & $\frac{\sqrt{\pi} \exp\left(-\frac{x^2}{4y}\right)}{4\sqrt{y}} \left[ \text{erfi}\left( \frac{x}{2\sqrt{y}}+\sqrt{y}\right) - \text{erfi}\left( \frac{x}{2\sqrt{y}}-\sqrt{y}\right) \right]$ \\
        $1$ & $\frac{3 \exp(y)}{2xy} \sinh(x) - \frac{3}{2xy} \mathbb{I}^{(0)}_{\mathrm{an}}(x, y)$ \\
        $2$ & $\frac{3 e^y (2y \cosh(x) - x \sinh(x))}{4 y^2} + \frac{3 (x^2 - 2y)}{4 y^2} \mathbb{I}^{(0)}_{\mathrm{an}}(x, y)$ \\
        $s$ & $\frac{1}{D_s} \left( \frac{\exp(x + y) - (-1)^{s-1} \exp(-x + y)}{2y} - \frac{s-1}{2y} D_{s-2} \mathbb{I}^{(s-2)}_{\mathrm{an}}(x, y) - \frac{x}{2y} D_{s-1} \mathbb{I}^{(s-1)}_{\mathrm{an}}(x, y) \right)$ \\
        \hline
    \end{tabular}
    \caption{Analytical values of $\mathbb{I}^{(s)}_{\mathrm{an}}$ with $D_s = \frac{x + (-1)^{s+1}x}{s+2} + \frac{1 + (-1)^s}{s+1}$. The expressions are valid for both $y>0$ and $y<0$, as erfi can handle imaginary numbers and the solution remains continuous and real for all y.}
    \label{tab:I_an}
\end{table}

\subsubsection{5M HyQMOM: Dirac distributions}\label{sec:5Mcollisions}

The quadrature method of moments (QMOM)\cite{McGraw97} is a widely used closure for determining the lower-order moments of the velocity distribution function (VDF), exploiting its deep connection to Gaussian quadratures in 1D distribution functions. QMOM approximates the VDF as a sum of Dirac delta functions, where the weights and nodes are determined by the moments, effectively transforming the moment problem into a Gaussian quadrature problem\cite{Marchisio13}. This approach is particularly accurate in computing the collisional terms when the underlying kernels of the Boltzmann operator are smooth (See Ref.\cite{Shizgal81} for an application of the Gaussian quadratures in plasma collisional terms), a condition often met in ion-neutral collisions as the cross sections do not have a threshold (unlike electron inelastic collisions). While the hyperbolic QMOM (HyQMOM)\cite{Fox22} extends this framework to ensure global hyperbolicity without requiring explicit VDF reconstruction, the Generalized QMOM (GQMOM)\cite{Fox23} retains the ability to reconstruct the VDF using the standard QMOM method, therefore retaining the link with Gaussian Quadratures, which is very useful for the computation of collision terms (as shown by Ref.\cite{Shizgal81}).

A distribution function that satisfies the 5M HyQMOM closure is a sum of three Dirac distributions, as follows,
\begin{equation}\label{eq:5MDist}
    f^\mathrm{(5M)}(v_x) = \frac{\rho}{m v_{Tx}}\left[w_0 \delta(v_{x} - u_0) + w_1 \delta(v_{x} - u_1) + w_2 \delta(v_{x} - u_2)\right] \,,
\end{equation}
where an analytical expressions for the weights $w_{0,1,2}$ and abscissae $u_{1,2}$ are given in Ref.\cite{Fox18, Berger25}, as follows,
\begin{eqnarray} \label{eq:HyQMOMinv}
    &w_0 = 1 - (w_1 + w_2) \,,\quad\quad w_1 = \alpha w_2 \,,\quad\quad
    w_2 = \frac{\alpha}{(1+\alpha) \left(\tfrac{{u_1-u}}{v_{Tx}}\right)^2}\,,\quad \quad
    u_0 = u\,, \nonumber \\
    & u_1 = u + {\text{sign}(q)} v_{Tx}\sqrt{\frac{r_\star}{1 - \alpha + \alpha^2}} \,,\quad\quad
    u_2 = u - \alpha u_1 ,\quad\quad
    \alpha  = \frac{2r_\star - q_\star^2 - \lvert q_\star \rvert \sqrt{4r_\star - 3q_\star^2}}{2 (r_\star - q_\star^2)},
\end{eqnarray}
where the star quantities refer to the standardized moments (see a full discussion on the closure in Ref.\cite{Berger25}).%

If we consider a Diracs VDF for the gas (without any drift) and the ions (HyQMOM distribution), the computation of the collision term with an HyQMOM ion VDF is straightforward from the angular integral of Eq.~\eqref{eq:integralIk0Final}. Because of the bilinearity of the collision operator, we can write the collision terms as a sum of the individual Dirac distributions, as follows,
\begin{equation}\label{eq:sumDiracs}
    \mathcal{C}_{k0}
    =  \sum_{i=0}^{2}w_i \mathcal{C}^{(i)}_{k0} \,,
\end{equation}
where $\mathcal{C}^{(i)}_{k0}$ is computed with $u_i$, \textit{i.e.}, the velocity of the Diracs in Eq.~\eqref{eq:HyQMOMinv}. Its expression is derived in Appendix \ref{app:5Mmodel}.

{Note that to derive this expression, we consider the neutral species distribution as a Dirac distribution, which differs from the other models. The rationale for this choice is that, at equilibrium, only the axial moments are considered, resulting in an ion perpendicular temperature of $T_\perp = 0$. From a thermodynamic perspective, this introduces a inconsistency, as an equilibrium distribution should satisfy $T_x = T_\perp$. Note that, with in the derivation shown in Appendix~\ref{app:5Mmodel}, the case where the neutral distribution is approximated with multiple quadrature points is provided and, hence, this work can be used to other neutral axial distributions (approximated with quadrature points).}

The expression reads:
\begin{equation}\label{eq:DiracVDF-simple}
    \mathcal{C}^{(i)}_{k0}
    = -\mu n n_\gas \lvert u_i\rvert \sum_{j=1}^k \binom{k}{j} \frac{\mu^{k-1}}{m_\gas^{k-j}m^{j-1}} u_i^{k} Q^{(j)}\left(\lvert u_i\rvert\right) \,.
\end{equation}
In this case, we can see this solution as a limiting case of the previous ones when $T_{\ion,\gas} = 0$. As we consider the full collision term as the sum of the contributions several Diracs in the $v_x$ direction at different drifting velocities, the model captures effectively, the impact of $T_x$ in the collisions. However, as shown in Fig.~\ref{fig:AnsatzVDF}, the perpendicular distribution remains a Dirac, which will impact the accuracy of the results, in particular, at high pressure (where the drift is often smaller than the thermal speed).

\subsection{6M HyQMOM: Dirac distributions with perpendicular temperature}

In order to take into account the perpendicular temperature, we propose to extend the 5M HyQMOM by solving the energy conservation in the perpendicular direction. The ansatz ion VDF is of the form
\begin{equation}\label{eq:6MDist}
    f^\text{(6M)}(v_x, v_\perp) = n \left(\frac{\gamma_\perp}{2\pi}\right) \mathrm{e}^{-\frac{\gamma_\perp}{2}v_\perp^2} f^\mathrm{(5M)}(v_x) = \sum_{i=0}^2 w_i n \left(\frac{\gamma_\perp}{2\pi}\right) \mathrm{e}^{-\frac{\gamma_\perp}{2}v_\perp^2} \delta(v_{x} - u_i) \,,
\end{equation}
where the weights $w_{0,1,2}$ and abscissae $u_{1,2}$ are given in Eq. \eqref{eq:HyQMOMinv}.

By linearity of the collision term with the ion VDF, similarly to what we did in Sec.~\ref{sec:5Mcollisions}, we will write the collision term as
\begin{equation}
    {\mathcal{C}_{k\,2l}} = \sum_{i=0}^{2} w_i \mathcal{C}^{(i)}_{k\,2l}
\end{equation}

The collision source terms for the 6M model are derived in Appendix \ref{app:6Mmodel} and read:
\begin{subequations}\label{eq:collTerm1_4_AniHyQMOM}
\begin{align}
    \mathcal{C}^{(i)}_{10}
    =&  -\tfrac{16}{3}\mu n n_\gas \tilde{\Omega}^{(1,1,1)}_{\mathrm{an}} u_{i} \,, \\
    \mathcal{C}^{(i)}_{20}
    =& -\tfrac{32}{3}\mu n n_\gas \tilde{\Omega}^{(1,1,1)}_{\mathrm{an}} u_{i}^2 +\tfrac{\mu}{m + m_\gas}n n_\gas\boltz T_\gas \left\{\tfrac{32}{3}\tilde{\Omega}^{(1,1,2)}_{\mathrm{an}} + 8\left(\tilde{\Omega}^{(2,1,0)}_{\mathrm{an}} - \tilde{\Omega}^{(2,1,2)}_{\mathrm{an}}\right)\right\}\,,\label{eq:collTerm1_4_AniHyQMOM_x}\\
    \mathcal{C}^{(i)}_{02} =& \tfrac{\mu }{m + m_\gas} n n_\gas\left\{\tfrac{16}{3}\kappa_{\perp}\left(3\tilde{\Omega}^{(1,1,0)}_{\mathrm{an}} - \tilde{\Omega}^{(1,1,2)}_{\mathrm{an}}\right)\boltz(T_\gas - T_\perp) + 4 \left(\tilde{\Omega}^{(2,1,2)}_{\mathrm{an}} - \tilde{\Omega}^{(2,1,0)}_{\mathrm{an}}\right) \boltz T_{\gas}\right\},\\
    \mathcal{C}^{(i)}_{30} =& -16\mu n n_\gas \tilde{\Omega}^{(1,1,1)}_{\mathrm{an}} u_{i}^3 + \tfrac{16}{5}\tfrac{\mu^2}{(m + m_\gas)m}n n_\gas\boltz T_\gas u_{i} \times \nonumber\\
    &\left\{\tfrac{m}{\mu}\left(10\tilde{\Omega}^{(1,1,2)} + \tfrac{15}{2}\left(\tilde{\Omega}^{(2,1,0)}_{\mathrm{an}} - \tilde{\Omega}^{(2,1,2)}_{\mathrm{an}}\right)\right) - \left(5\tilde{\Omega}^{(1,2,1)}_{\mathrm{an}} + 3\tilde{\Omega}^{(1,2,3)}_{\mathrm{an}}\right) \right. \nonumber\\
    &\left.+ \left(9 \tilde{\Omega}^{(2,2,3)}_{\mathrm{an}}- 5\tilde{\Omega}^{(2,2,1)}_{\mathrm{an}} \right) + 5\left(\tilde{\Omega}^{(3,2,1)}_{\mathrm{an}} - \tilde{\Omega}^{(3,2,3)}_{\mathrm{an}}\right)\right\}\,,\label{eq:collTerm1_4_AniHyQMOM_3}\\
    \mathcal{C}^{(i)}_{40}
    =& -\tfrac{64}{3}\mu n n_\gas \tilde{\Omega}^{(1,1,1)}_{\mathrm{an}} u_{i}^4 + \tfrac{\mu^2}{(m + m_\gas)m}n n_\gas\boltz T_\gas u_{i}^2  \left\{\tfrac{m}{\mu}\left(64\tilde{\Omega}^{(1,1,2)}_{\mathrm{an}} + 48\left(\tilde{\Omega}^{(2,1,0)}_{\mathrm{an}} - \tilde{\Omega}^{(2,1,2)}_{\mathrm{an}} \right) \right)\right.\nonumber\\
    & \left.
    -\tfrac{64}{5}\left(5\tilde{\Omega}^{(1,2,1)}_{\mathrm{an}} + 3\tilde{\Omega}^{(1,2,3)}_{\mathrm{an}} + 5\tilde{\Omega}^{(2,2,1)}_{\mathrm{an}} - 9\tilde{\Omega}^{(2,2,3)}_{\mathrm{an}} - 5\tilde{\Omega}^{(3,2,1)}_{\mathrm{an}} + 5\tilde{\Omega}^{(3,2,3)}_{\mathrm{an}}\right)\right\}\nonumber\\
    & + \left(\tfrac{\mu}{m + m_\gas}\right)^2 n n_\gas\tfrac{\left(\boltz T_\gas\right)^2}{m} \left\{
    \tfrac{64}{5}\left(5\tilde{\Omega}^{(1,2,2)}_{\mathrm{an}} - \tilde{\Omega}^{(1,2,4)}_{\mathrm{an}}\right)
    -\tfrac{8}{5}\left(10\tilde{\Omega}^{(2,2,2)}_{\mathrm{an}}+21\tilde{\Omega}^{(2,2,4)}_{\mathrm{an}}-15\tilde{\Omega}^{(2,2,0)}_{\mathrm{an}}\right)
    \right.\nonumber\\
    &\left.-64\left(\tilde{\Omega}^{(3,2,2)}_{\mathrm{an}} - \tilde{\Omega}^{(3,2,4)}_{\mathrm{an}}\right)
    +4\left(10\tilde{\Omega}^{(4,2,2)}_{\mathrm{an}} - 3\tilde{\Omega}^{(4,2,0)}_{\mathrm{an}} - 7\tilde{\Omega}^{(4,2,4)}_{\mathrm{an}}\right)
    \right\}.\label{eq:collTerm1_4_AniHyQMOM_4}
\end{align}
\end{subequations}
Here, $\kappa_\perp = mT_\gas/(mT_\gas + m_\gas T_\perp)$. The collision integral is obtained from Eq.~\eqref{eq:OmegaAn}, in the case of vanishing axial temperature, as follows,
\begin{equation}\label{eq:OmegaAnTilde}
    \tilde{\Omega}^{(l,r,s)}_{\mathrm{an}}(\kappa_{\perp}, T_{\gas}, \bar{u}) \equiv \Omega^{(l,r,s)}_{\mathrm{an}}\left(\kappa_{x, \perp} = \frac{mT_\gas}{mT_\gas + m_\gas T_\perp},\,T_{x, \gas} = \frac{\mu}{m_\gas}T_\gas,\,\bar{u} = u\gamma_\gas^{1/2}\right).
\end{equation}

\section{Validation tests of collisional integrals}\label{sec:Four}

\subsection{Numerical evaluation of the collision integrals}

The collision integrals $\Omega^{(l,r,s)}_{\mathrm{iso}}$, $\Omega^{(l,r,s)}_{\mathrm{an}}$, and $\tilde{\Omega}^{(l,r,s)}_{\mathrm{an}}$ of Eqs.~\eqref{eq:OmegaIso}, \eqref{eq:OmegaAn}, and \eqref{eq:OmegaAnTilde} depend on the transport cross section. In the case of isotropic and backscattering these can be computed easily from the total cross section as follows:
\begin{equation}
    Q^{(l), \mathrm{iso}}(g) = \sigma^{(0), \mathrm{iso}}(g)\left(1 - \frac{1}{2}\frac{1+(-1)^l}{1+l}\right)~~\text{and}~~
    Q^{(l), \mathrm{cx}}(g) = \sigma^{(0), \mathrm{cx}}(g)\left(1 + (-1)^{l+1}\right).
\end{equation}
Note that the charge exchange $Q^{(l), \mathrm{cx}}$ vanish for even values of $l$.

The collision integrals Eqs.~\eqref{eq:OmegaIso}, \eqref{eq:OmegaAn}, \eqref{eq:OmegaAnTilde} involve two integrals (over the angles and the velocities) that are computed numerically, as follows. First, the angular integrals $\mathbb{I}^{(s)}_{\mathrm{iso}}$ and $\mathbb{I}^{(s)}_{\mathrm{an}}$ of Eqs.~\eqref{eq:angularIntegral} and \eqref{eq:angularIntegralAn} are integrated over $\zeta$ for each value of the velocity $\bar{g}$ with a Simpson rule. In order to avoid having indeterminate values, the value that is numerically integrated is
\begin{equation}
    \mathbb{I}^{(s)}\times\left(\frac{\bar{u}\bar{g} + (-1)^{s+1}\bar{u}\bar{g}}{s+2} + \frac{1 + (-1)^{s}}{s+1}\right) {\times \exp\left(-\frac{\kappa_{x,\perp} \bar{g}^2}{2} - \frac{\bar{u}^2}{2}\right)}
\end{equation}
{(where $\kappa_{x,\perp} = 1$ in the isotropic case)}. The number of points used for the integration over $\zeta$ depends on the Mach regime and are chosen from the following formula  $N_\zeta = \min\left(\max\left(\bar{u}_{\mathrm{max}}\left(\bar{g} - \bar{u}/2\right),\,200\right),\,3000\right)$. In the numerical tests, we have used $\bar{u}_{\mathrm{max}}=20$. Alternatively, the integration over the velocity is performed also with a Simpson rule by interpolating the cross section in a uniform grid of $N_{\bar{g}} = 100$ points over the $\bar{g}$. In order to avoid overflows in the exponential functions, the integration (that has integration limits $\bar{g}\in[0, \infty)$) is performed in the computational domain $\bar{g}\in[\max(0,\,\bar{u}-\bar{u}_{\mathrm{max}}),\,\max(\bar{u},\,\Delta\bar{u})+\bar{u}_{\mathrm{max}}]$ where, in our numerical tests, we choose $\Delta\bar{u} = 2$.

\begin{figure}[t]%
    \centering
    \includegraphics[width=\linewidth]{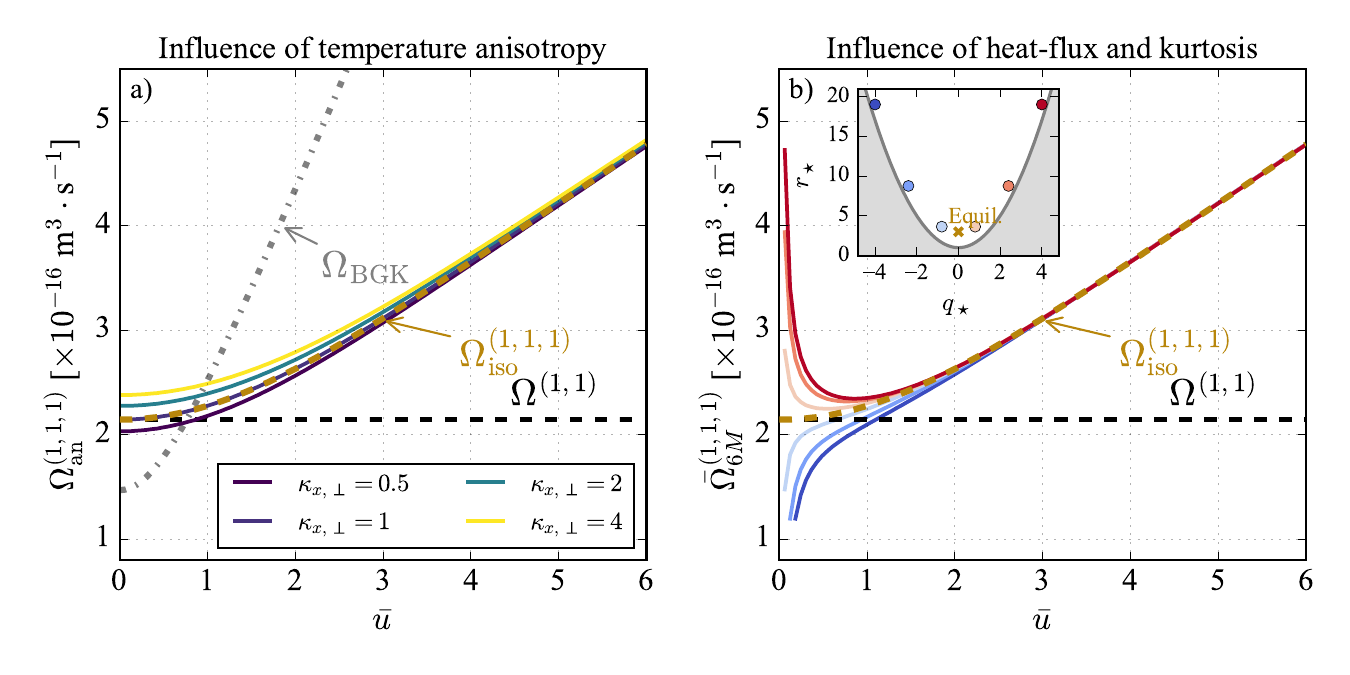}
    \caption{Example of collision rate $\Omega^{(1,1,1)}$ as function of ion drift for $T_{\ion,\gas} = 0.025$ eV. (a) Influence of temperature anisotropy on $\Omega^{(1,1,1)}_{\mathrm{an}}$, compared to $\Omega^{(1,1,1)}_{\mathrm{iso}}$ and the classical Chapman-Cowling integral and the BGK rate compute with Eq.~\eqref{eq:Chabert} $\Omega^{(1,1)}$. (b) Influence of normalized heat flux and kurtosis on $\Omega^{(1,1,1)}$ for $T_\perp = T_{\ion,\gas}$. The chosen cases for normalized heat flux $q_\star$ and kurtosis $r_\star$ are shown in the embedded plot in the realizability space, representing values typical of those observed in the numerical simulations. The result $q_\star=0$, $r_\star = 3$ coincides with $\Omega^{(1,1,1)}_{\mathrm{iso}}$, which coincides with the classical $\Omega^{(1,1)}$ at $\bar{u}=0$.}
    \label{fig:Omega111}
\end{figure}

One example of collision rate is presented in Fig.~\ref{fig:Omega111}. In Fig.~\ref{fig:Omega111} a (left panel), we present the influence of the temperature anisotropy in $\Omega^{(1,1,1)}_{\mathrm{an}}$ as a function of the ion Mach number $\bar{u}$, for a fixed value of $T_{\ion,\gas} = 0.025$ eV. $\Omega^{(1,1,1)}_{\mathrm{an}}$ is computed as the sum of the isotropic and backscattering contributions with the cross sections of Fig.~\ref{fig:crossSection}. The case $\kappa_{x,\perp} = 1$ (defined in Eq.~\eqref{eq:kappa_xperp}) coincides with $\Omega^{(1,1,1)}_{\mathrm{iso}}$ and it tends to the classical Chapman-Cowling collision $\Omega^{(1,1)}$ when $\bar{u}\rightarrow 0$. One can observe that the influence of the pressure anisotropy in the given case is only important for $\bar{u}\lesssim 3$. The case $\kappa_{x,\perp}<1$ corresponds to $T_\perp> T_x$ and it tends to lower the value as compared to the isotropic one, where as  $\kappa_{x,\perp}>1$ has the inverse effect.

In Fig.~\ref{fig:Omega111} a, we have added the value that would correspond to the BGK of Eq.~\eqref{eq:Chabert}. The corresponding value is computed as follows:
\begin{equation}
    m \nu^{Chabert} = \frac{16}{3}\mu \Omega_{\mathrm{BGK}}.
\end{equation}
We note that the value at $\bar{u}\rightarrow0$ is approximately $0.75\Omega^{(1,1)}$ which can lead to error in the high-pressure limit in the bulk. Alternatively, the asymptotic behavior at large velocity is wrong, with a slope that is much larger than the one predicted by the theory based on the integration of the Boltzmann operator.

In Fig.~\ref{fig:Omega111} b (right panel), we present the influence of the normalized heat flux and kurtosis in $\Omega^{(1,1,1)}$ as a function of the ion Mach number $\bar{u}$, for a fixed value of $T_{\ion,\gas} = 0.025$ eV and with $T_\perp = T_{\ion,\gas}$. We compute the rate as an average of the contribution of the three HyQMOM Diracs, as follows, $\bar{\Omega}^{(1,1,1)}_{6M} = \frac{1}{u}\sum_{i = 0}^{2} w_i \tilde{\Omega}^{(1,1,1)}_{\mathrm{an}} u_i$. The values of normalized $q_\star$ and $r_\star$ are shown in the realizability domain and are characteristic values seen in the simulations. We can see that the value of $\bar{\Omega}^{(1,1,1)}_{6M}$ coincides with $\Omega^{(1,1,1)}_{\mathrm{iso}}$ for $q_\star = 0$ and $r_\star=3$, which corresponds to the equilibrium. The impact of the heat flux and kurtosis in this case is limited to Mach numbers $\bar{u}\lesssim2$. We see that the impact in the low Mach region can be significant. The analysis of the impact of the different parameters in the relaxation rates will be done in a future work. In this work, we will focus on the numerical validation of these models against kinetic simulations.

In the numerical tests shown in this paper, the collision source terms are precomputed and stored in tables as functions of $\bar{u}$, $T_x/T_\gas$, and $T_\perp/T_\gas$. Note that in our simulations, $T_\gas = 0.025$\,eV is a constant value. The collision terms are computed for Mach numbers $\bar{u} \in [0, 40]$ using $400$ points. The temperature anisotropies are computed on a grid defined by $\kappa_x = 1/(1 + T_x/T_\gas)$ and $\kappa_\perp = 1/(1 + T_\perp/T_\gas) \in [10^{-3.5}, 1]$, with $30$ points in logarithmic scale. Additionally, the point $T_x = T_\perp = T_\gas$ is included in the table to improve accuracy near thermal equilibrium.

\subsection{0D validation problems}

{\paragraph{0D Relaxation of an ion drifting Maxwellian in a neutral background}

We begin by considering a 0D relaxation problem where argon ions, initially distributed as a drifting Maxwellian with mean velocity $\bar{u} = 15$ and temperature $T_g = 0.025$ eV, relax to thermal equilibrium in an argon neutral background at rest with the same temperature $T_g = 0.025$ eV. This scenario serves as a benchmark to compare all the DIBO moment models introduced in the previous sections, alongside a BGK approximation and the classical Chapman--Cowling collision.

\begin{figure}[htbp]%
    \centering
    \includegraphics[width=\linewidth]{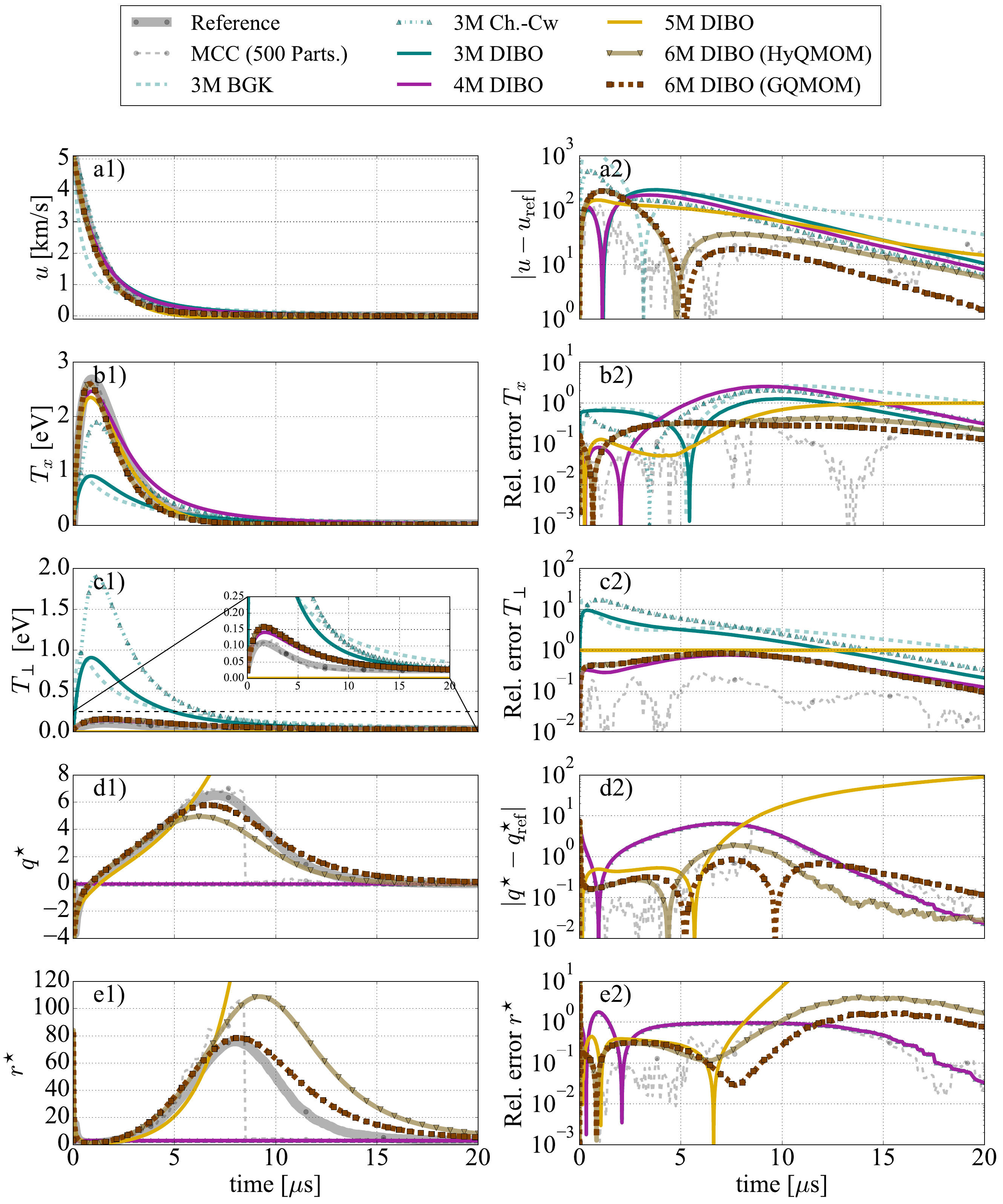}
    \caption{0D relaxation of a Maxwellian ion distribution at $0.025$ eV with a drift of $\bar{u} = 15$ to thermal equilibrium with a neutral background at $0.025$ eV at rest. a1,2) $u_{x}$ evolution and absolute error; b1,2) $T_{xx}$ evolution and relative error; c1,2) $T_\perp$ evolution and relative error; d1,2) $q^\star$ evolution and absolute error; e1,2) $r^\star$ evolution and relative error. The reference solution is computed with an MCC algorithm with $10^6$ particles. The GQMOM is performed with $5$ Diracs.}
    \label{fig:Temperature relaxation}
\end{figure}

Figure~\ref{fig:Temperature relaxation} illustrates the temporal evolution of the axial velocity $u$, axial temperature $T_{x}$, perpendicular temperature $T_\perp$, normalized heat flux $q^\star$ and normalized fourth-order moment $r^\star$, together with their respective errors (we use absolute errors, when the steady state solution of the moment is zero). The reference solution is obtained using a Monte Carlo Collision (MCC) algorithm with $10^6$ particles. In the $6M$ model, we include both the HyQMOM solution (with $3$ Diracs) and the GQMOM\cite{Fox23} performed with $5$ Diracs. Among the models tested, the $6M$ HyQMOM and $6M$ GQMOM exhibit the lowest errors across all studied moments, with errors comparable to those of a low-resolution MCC (with $500$ particles) in the axial moments. The largest relative error is observed in $T_\perp$, which stems from the Maxwellian approximation in the perpendicular direction. However, as demonstrated in subsequent cases, this error does not significantly impact the axial dynamics, as $T_\perp \ll T_x$ in most of the cases.

We would like to draw the attention that, in the DIBO 6M QMOM (both HyQMOM and GQMOM), Eq.~\eqref{eq:collTerm1_4_AniHyQMOM_x} contains only $T_g$ while $T_x$ does not appear explicitly, but is instead carried entirely by the node spacing of the reconstructed VDF. Axial thermal equilibrium can therefore only arise as a numerical cancellation, exact only if three nodes integrate the collision kernel exactly which can only happen if the collision kernel is a polynomial of order $\leq5$. However, this is not the case as the cross section are not polynomials. Nevertheless, the relative errors at steady state remain small: for the $6M$ HyQMOM, the errors are $1.28\%$ for $T_{xx}$ and $1.47\%$ for $T_\perp$, while for the $6M$ GQMOM (with 5 Diracs), they are $0.62\%$ for $T_{xx}$ and $1.42\%$ for $T_\perp$. This error is shown to be independent from the computation of the collisional integrals or the interpolation tables and they are due to the quadrature integration of the collisional terms and therefore due to the closure. Similarly, it can be noted that the $5M$ model relaxes to zero velocity, which is a consequence of choosing a Dirac for the neutral distribution. This choice might be improved by considering a different distribution for the neutral species. However, in this paper, we choose it to be a Dirac to be in thermodynamic equilibrium with the perpendicular direction (which has zero temperature). Nevertheless, this is a choice that can be changed in the future.}

{\paragraph{Ions in a homogeneous electric field}

We also consider a case where ions are initialized as a Maxwellian at rest with $T_{\text{ion}} = 0.025$ eV in the presence of an electric field and a stationary neutral background. Under these conditions, an equilibrium is found between the drag force of the collisions and the electric field force, resulting in a distribution characterized by a drift velocity and anisotropic temperatures (due to charge exchange collisions), as well as a strong heat flux that increases with the electric field strength.

\begin{figure}[htbp]%
    \centering
    \includegraphics[width=\linewidth]{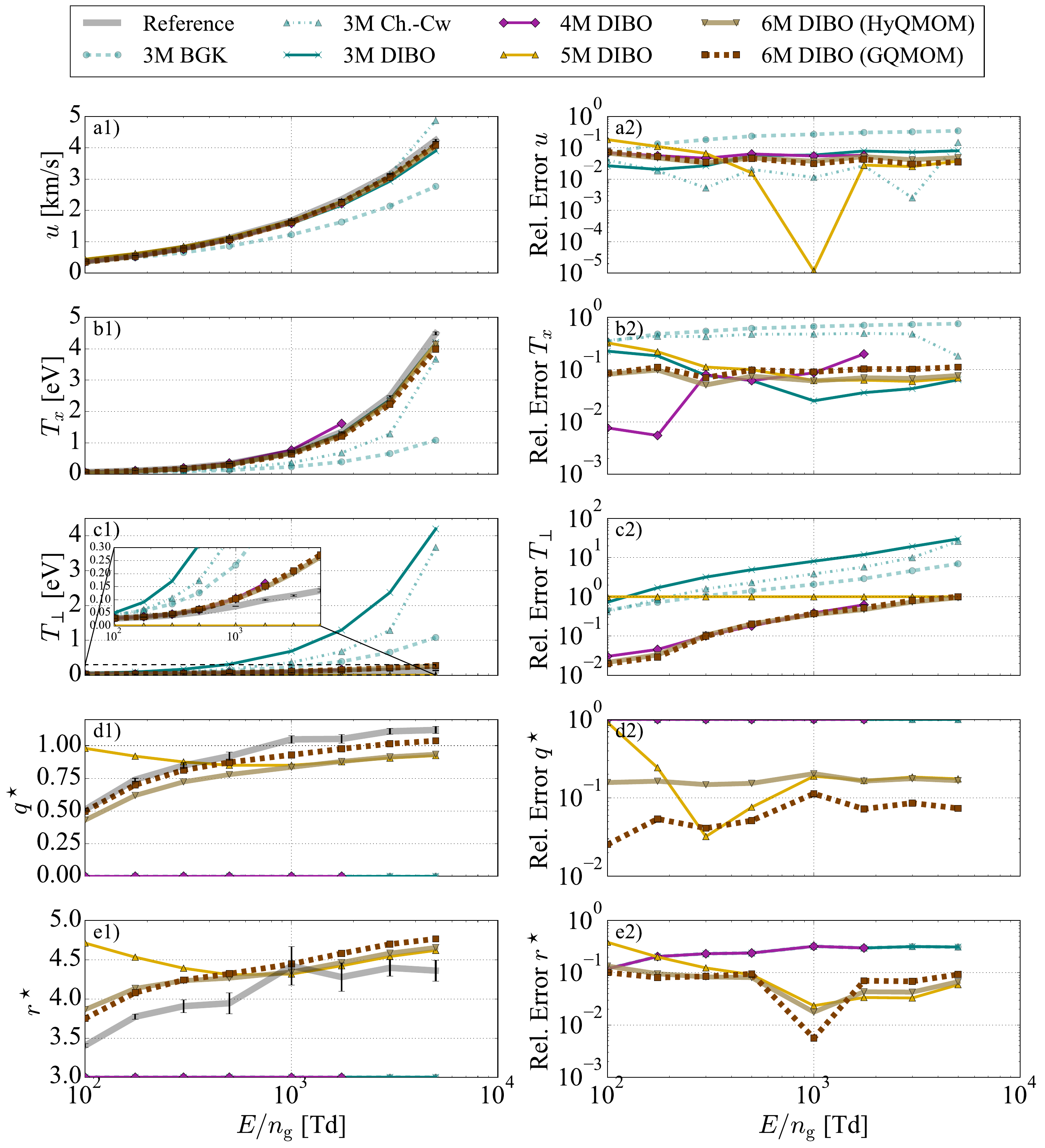}
    \caption{Converged results of ions in a homogeneous electric field in a gas background at different reduced fields. a1,2) $u_{x}$ and relative error; b1,2) $T_{xx}$ and relative error; c1,2) $T_\perp$ and relative error; d1,2) $q^\star$ and relative error; e1,2) $r^\star$ and relative error. The reference solution is computed with an MCC algorithm with $10^4$ particles and averaged over $10~\mu$s, with error bars indicating the standard deviation. The GQMOM is performed with $5$ Diracs.}
    \label{fig:EovN relaxation}
\end{figure}

Figure~\ref{fig:EovN relaxation} presents the converged results of ions in a homogeneous electric field at different reduced fields ranging from $100$ to $5000$ Td. The converged moments $u$, $T_{x}$, $T_\perp$, $q^\star$, and $r^\star$ is shown alongside their respective relative errors. The reference solution is computed with an MCC algorithm with $10^4$ species with $\Delta t = 3.83\times 10^{-11}$ s and averaged over $10~\mu$s, while the GQMOM is performed with $5$ Dirac delta functions. The moment equations are solved with the LSODA method.

\begin{figure}[htbp]
    \centering
    \begin{minipage}{0.49\textwidth}
        \centering
        \includegraphics[width=\linewidth]{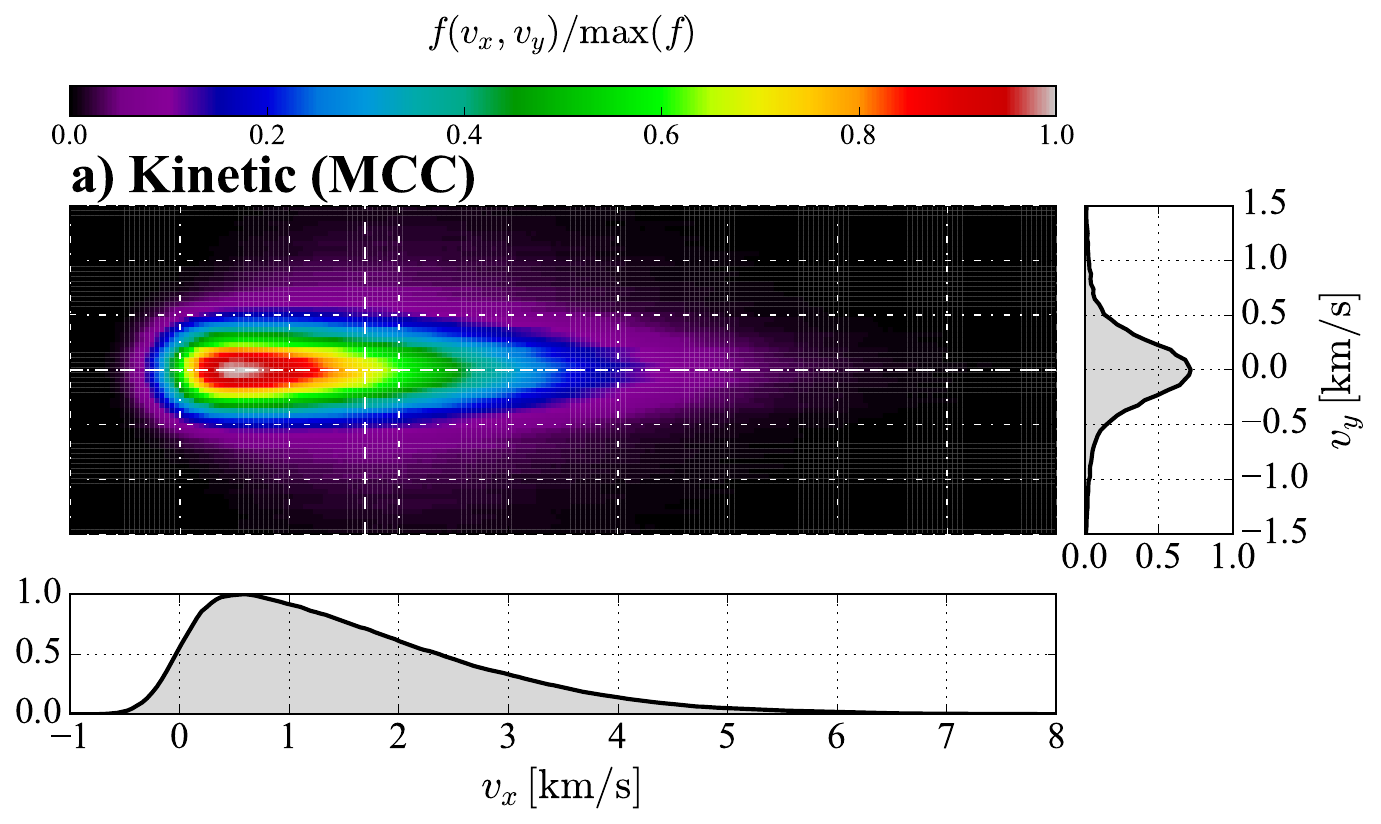}
    \end{minipage}

    \vspace{0.2cm}%

    \begin{minipage}{0.49\textwidth}
        \centering
        \includegraphics[width=\linewidth]{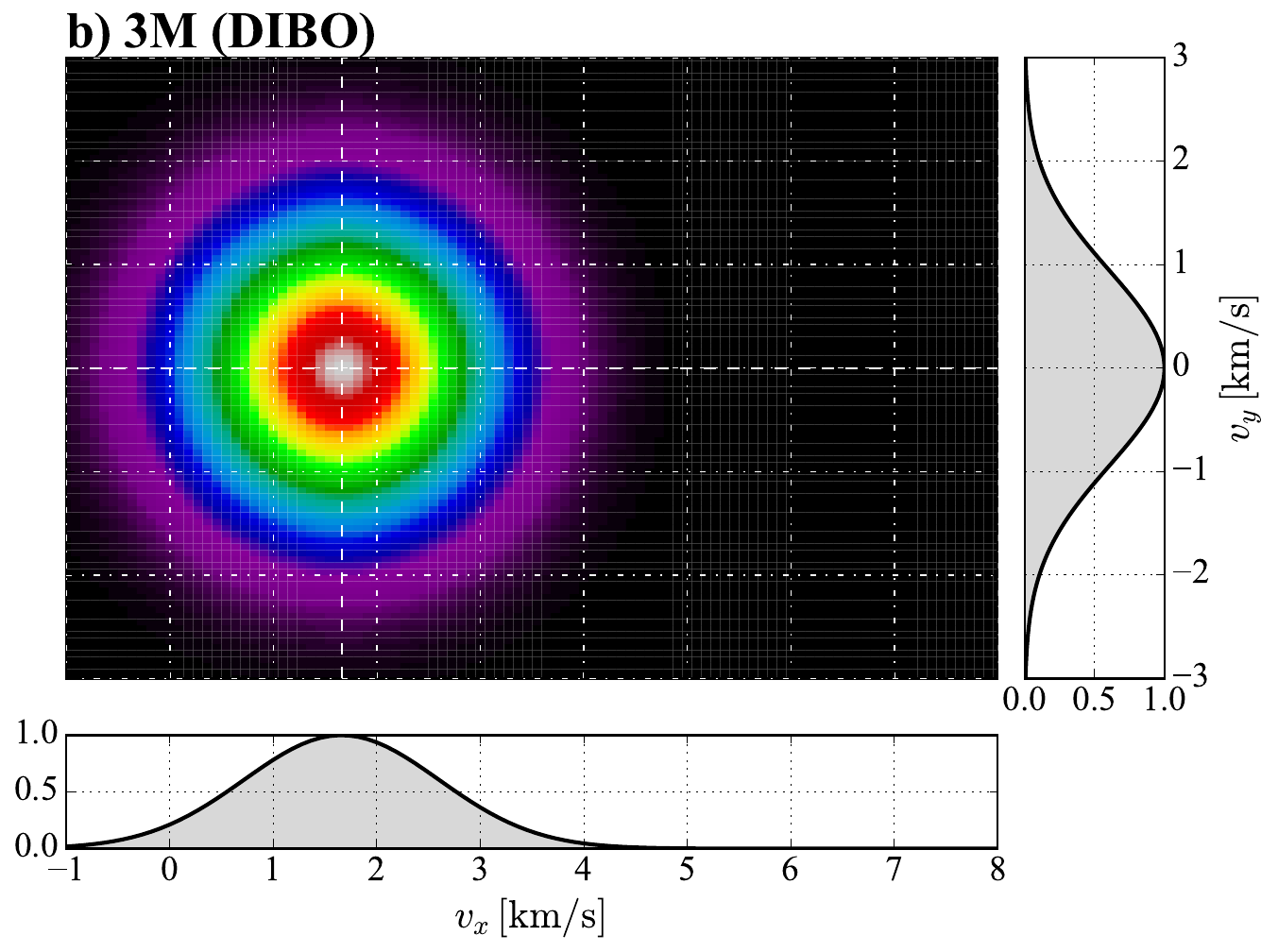}
    \end{minipage}
    \hfill%
    \begin{minipage}{0.49\textwidth}
        \centering
        \includegraphics[width=\linewidth]{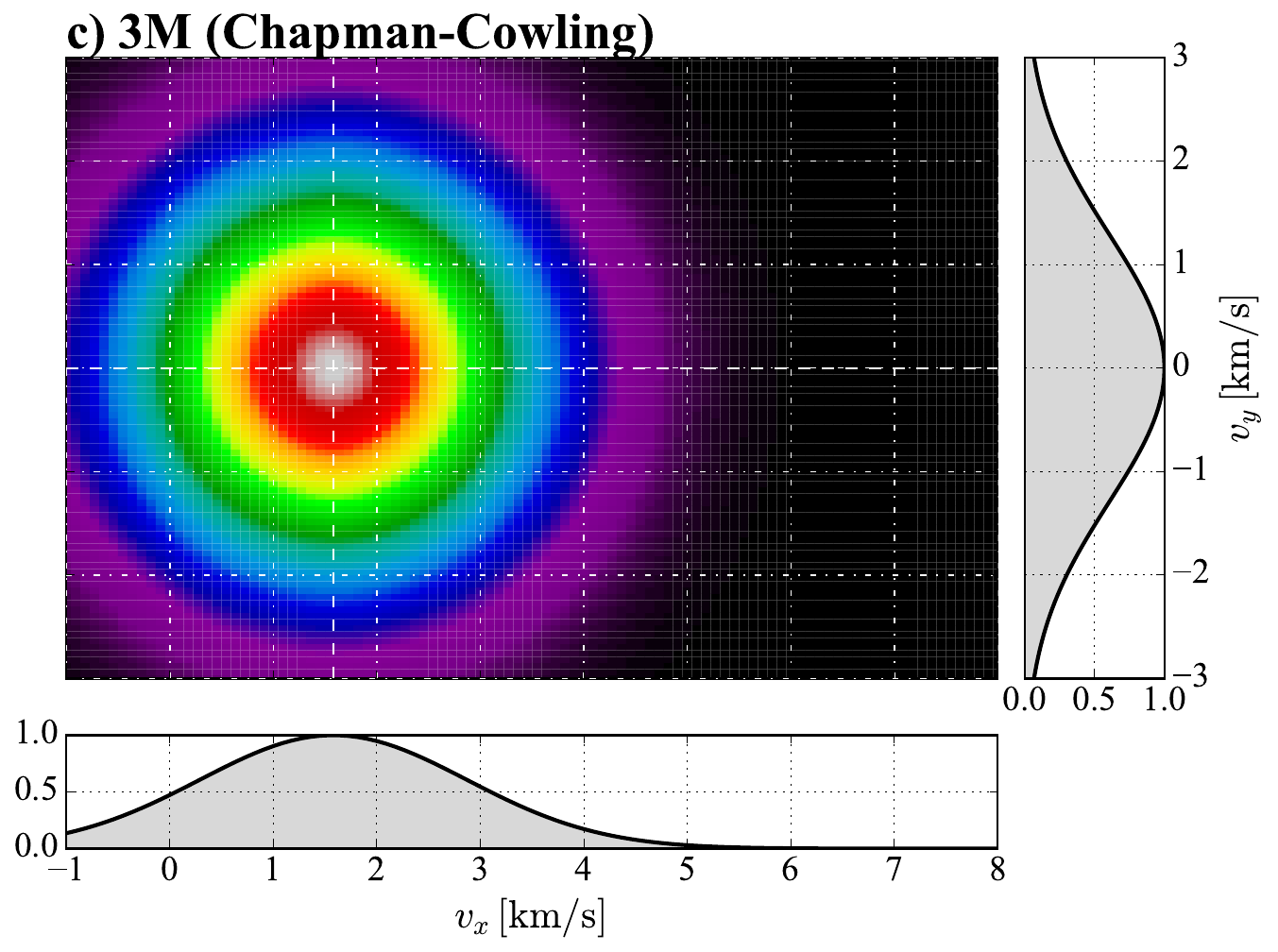}
    \end{minipage}

    \vspace{0.2cm}%

    \begin{minipage}{0.49\textwidth}
        \centering
        \includegraphics[width=\linewidth]{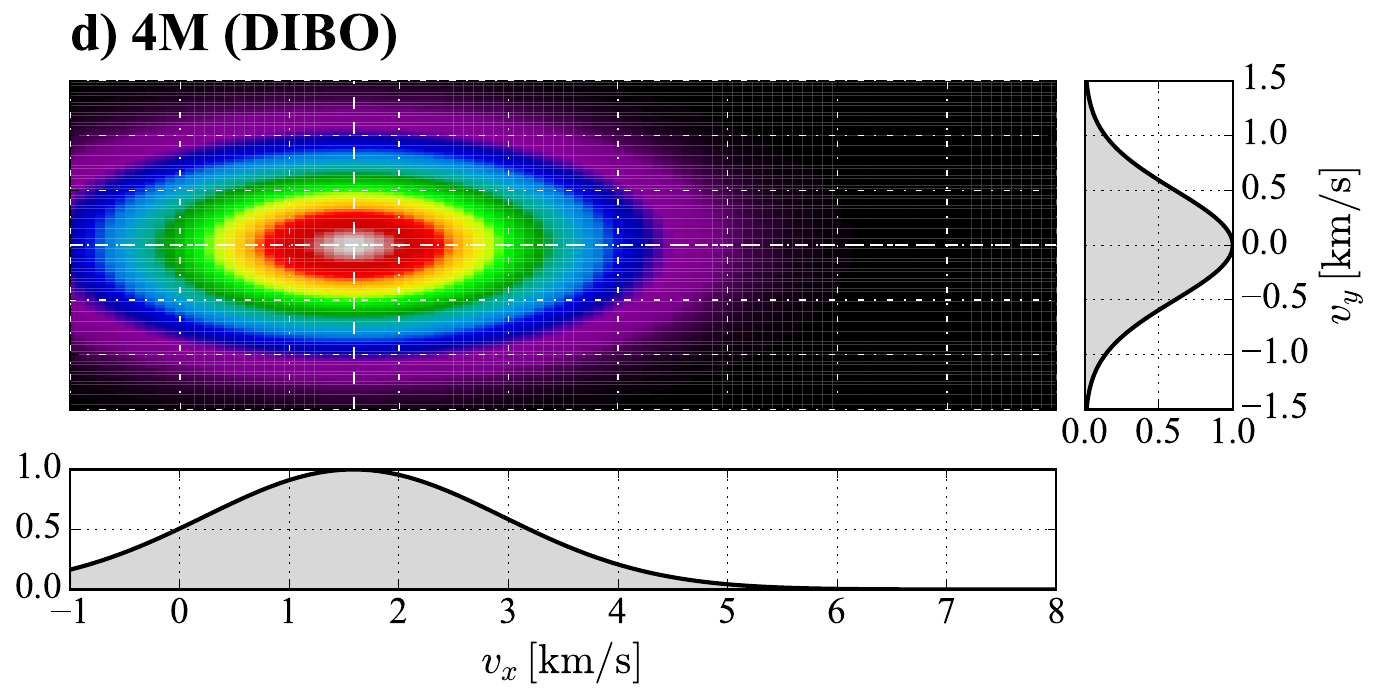}
    \end{minipage}
    \hfill%
    \begin{minipage}{0.49\textwidth}
        \centering
        \includegraphics[width=\linewidth]{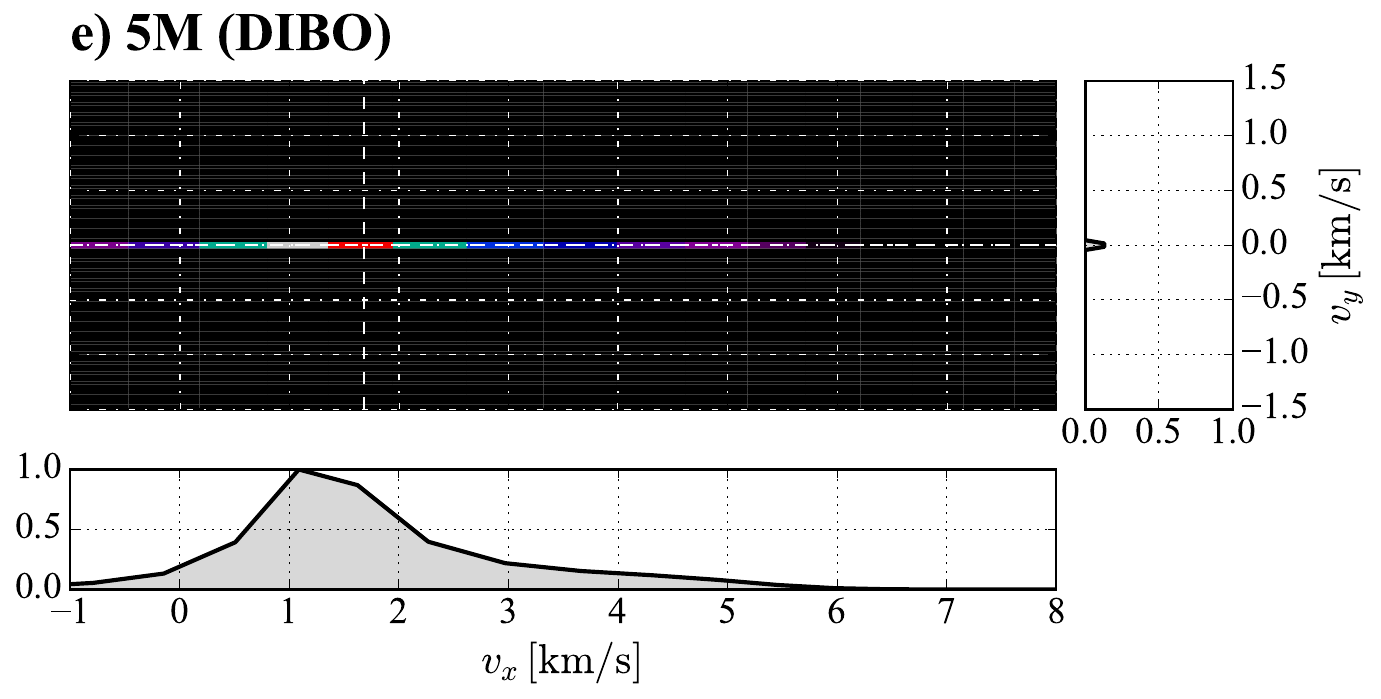}
    \end{minipage}

    \vspace{0.2cm}%

    \begin{minipage}{0.49\textwidth}
        \centering
        \includegraphics[width=\linewidth]{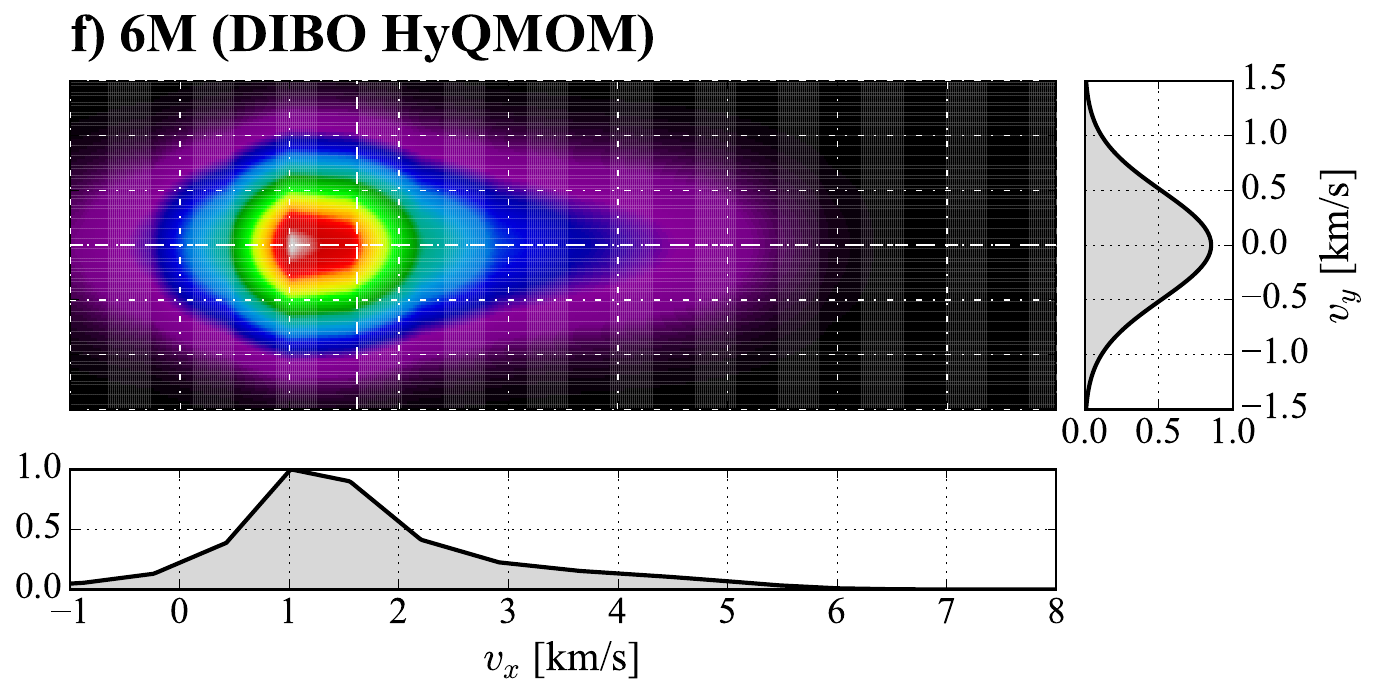}
    \end{minipage}
    \hfill%
    \begin{minipage}{0.49\textwidth}
        \centering
        \includegraphics[width=\linewidth]{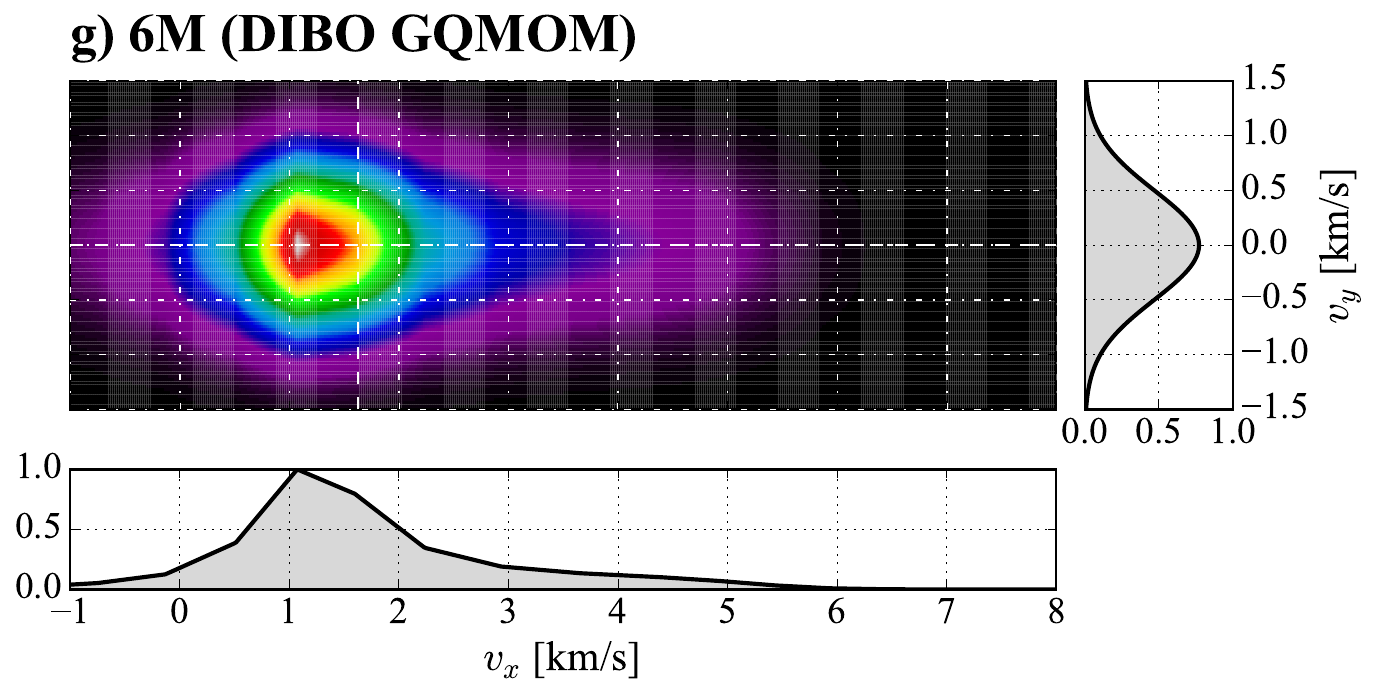}
    \end{minipage}

    \caption{Comparison of 2D distributions for different models in the converged solution at \(E/n_\text{gas} = 1000\) Td. The kinetic solution is found by averaging during $10\mu$s the solution with $10^5$ particles. The QMOM distributions are plotted as continuous functions, by reconstructing with the GQMOM technique with \(24\) Diracs. The continuous function is obtained by normalizing with the area and assuming linear distributions between the quadrature points.}
    \label{fig:comparison_2d}
\end{figure}

The results show that the $3M$ DIBO is able to predict the velocity but not the energy, however it largely improves the results from Chapman-Cowling and BGK approximations. The $4M$ DIBO model fails to find a solution for $E/n_g > 2000$ Td. A similar limitation is observed for the $3M$ DIBO at $E/n_g > 5000$ Td. Among the models, the $5M$ and $6M$ yielding the smallest relative errors across all studied moments. The largest error is found in the perpendicular temperature. However, as $T_\perp \ll T_x$, this error does not significantly affect the axial dynamics. It is also worth noting that the $6M$ GQMOM and HyQMOM produce very similar results, which motivates focusing solely on the HyQMOM in subsequent analyses.

Figure~\ref{fig:comparison_2d} shows the reconstructed 2D velocity distribution functions (VDFs) compared to the kinetic reference for $E/n_\gas = 1000$ Td. The $3M$ DIBO improves the Chapman--Cowling approximation, while the $4M$ DIBO captures the temperature anisotropy but fails to capture the skewness of the distribution. The QMOM distributions are plotted as continuous functions, computed using the GQMOM technique with \(24\) Diracs. The continuous function is obtained by normalizing with the area and assuming linear distributions between the quadrature points. The $5M$ DIBO successfully captures the skewness despite not capturing the perpendicular temperature, further confirming that the perpendicular temperature has limited impact on the axial moments due to $T_\perp \ll T_x$. Finally, the $6M$ model demonstrates remarkable fidelity in capturing the 2D VDF, though it slightly overestimates the perpendicular temperature. It is important to note that the hypothesis $f(v_x, v_\perp) = f_x(v_x)f_\perp(v_\perp)$ is not fully fulfilled by the reference solution and it is source of errors in the estimation of the perpendicular temperature. However, this assumption is well justified for distributions such as the one shown, as the large temperature anisotropy $T_x \ll T_\perp$ ensures that the dominant terms are in the axial quantities, while still providing a good approximation of the perpendicular distribution.}

{\subsection{Computational times}

Table~\ref{tab:wall_times} presents the computational times for the velocity relaxation problem. Times are compared to the MCC with comparable numerical error to the $6M$ solution in the axial moments, i.e., using $500$ particles. The computational gain is measured using an explicit Runge-Kutta scheme with the same time step as the MCC (which is just a first-order temporal scheme, as the dominant source of error is the statistical error) and also with an explicit--implicit LSODA scheme to highlight that using a set of equations allows the use of standard implicit discretizations, as opposed to MCC, where the implicitization is more complicated. We show the great gain in computational time of the DIBO approach as compared to MCC. Nevertheless, due to the interpolation tables, BGK and classical Chapman--Cowling are faster. The goal of the article is to illustrate this approach, and the computational performance was not a primary goal. Other approaches can be investigated to improve it, such as using interpolative analytical formulas or machine learning approaches for the interpolation.}

\begin{table}[H]
{\centering
\caption{Wall times and computational gain (speedup) relative to MCC with $500$ particles (Wall time $58.31$ s) for different models, measured on a 2.4 GHz 8-Core Intel Core i9 processor.}
\label{tab:wall_times}
\begin{tabular}{l S[table-format=1.2] S[table-format=5.2] S[table-format=1.2] S[table-format=5.2]}
\toprule
 & \multicolumn{2}{c}{RK2} & \multicolumn{2}{c}{LSODA} \\
\cmidrule(lr){2-3} \cmidrule(lr){4-5}
\textbf{Model} & {\textbf{Wall Time (s)}} & {\textbf{Gain}} & {\textbf{Wall Time (s)}} & {\textbf{Gain}} \\
\midrule
3M BGK         & 0.013 & 4445.35 & 0.00062 & 94195.48 \\
3M Chapman-Cowling     & 0.035 & 1675.50 & 0.0014 & 42460.29 \\
3M DIBO        & 0.33 & 175.39 & 0.028 & 2046.17 \\
4M DIBO        & 0.33 & 174.55 & 0.018 & 3154.50 \\
5M DIBO        & 0.82 & 70.72 & 0.014 & 4037.93 \\
6M DIBO HyQMOM & 1.88 & 31.07 & 0.13 & 448.49 \\
6M DIBO GQMOM ($5$ Diracs)  & 3.96 & 14.72 & 0.58 & 100.82 \\
\bottomrule
\end{tabular}
}
\end{table}

\section{Simulation results}
\label{sec:results}

\subsection{Numerical method}

The models presented in the previous section have been implemented in non-linear time-dependent simulations and compared to PIC simulations on the test case presented in Sec.~\ref{sec:set-ups}. This work focuses on the ion dynamics. For this reason, in the high-order moment simulations, we will only solve the ions equations, and the terms depending on the electron dynamics (namely the ionization rate and the electrical potential) will be imposed from the converged PIC simulations. {However, due to their stochastic nature, PIC-MCC simulations are intrinsically noisy, so the data are averaged over a large number of time steps (of the order of $\sim 10^7$) in order to have reasonably smooth data. This means an averaging window of $300~\mu$s for Case 1 and $150~\mu$s in Case 2.}

The simulations solve the time evolution of the moment systems for the 3M (Eqs.~\eqref{eq:system3M}), 4M (Eqs.~\eqref{eq:system4M}), 5M (Eqs.~\eqref{eq:system5M}) and 6M (Eqs.~\eqref{eq:system6M}) until steady state. The spatial discretization is done with a finite volume that uses the Rusanov numerical flux with a second order scheme that is obtained with a total variation diminishing (TVD) scheme, obtained by reconstructing the primitive variables with a slope limiter\cite{Cada09}.
{The boundary conditions are imposed with a ghost cell strategy with outflow conditions, where we assume that the ion flow is supersonic at the boundary.}
The time step of the forward-Euler time discretization is chosen by fixing CFL$=0.5$. We refer to Ref.~\cite{Berger25} for the details on the numerical scheme.

For Case 1, we will consider a non-uniform grid with a $\Delta x = 1$ mm in the bulk and a $\Delta x = 10^{-2}$ mm in the sheath (joined by a buffer zone for a smooth transition), resulting in $450$ cells in total; while in case 2, we consider a uniform grid with steps of $\Delta x = 0.25$~mm, i.e., $200$ cells.

The numerical schemes we consider are not preserving the realizability in the case of the 5M and 6M. These systems can leave the realizability region during the transient, which can produce numerical problems in the computation of the collision terms. To ensure that only realizable sets of moments are used in the computation of the collision terms, we modify any non-realizable set of moments into a realizable set. In practice, the most simple and systematic solution is to increase the kurtosis until the realizability condition $r_\star \geq 1 + q_\star^2$ is satisfied. We stress that this only happens during the (non-physical) transient and is not expected to have any effect on the steady-state solution that should be unique and realizable.

\subsection{Case 1: Bounded plasma between two floating walls at different pressures}

We recall that we solve for the ion equations, while the electric potential and ionization profile are the same as in the kinetic equation. The Case 1 represents a plasma between two floating walls separated by ${2L} = 10$ cm, as described in Section \ref{sec:set-ups}. We study four different pressures ($p_\gas = 5\cdot[10^{-2}, 10^{-1}, 1, 10]$ mTorr). These correspond to characteristic Knudsen numbers $\text{Kn} = \lambda_{\ion \gas}/{(2L)} = ({2}n_\gas\sigma_0L)^{-1} = 6.2\cdot[1,~10^{-1},~10^{-2},~10^{-3}]$, which goes from nearly collisionless to collisionally-dominated regimes.

\subsubsection{Comparison of moment profiles}
\label{sec:moment-profiles-coll}

We first compare the moment profiles of the moment closures and the kinetic simulations. We include solutions of the moment equations with BGK operators based on two different frequencies, \textit{i.e.} $\nu^{\mathrm{Schottky}}$ (Eq.~\eqref{eq:Schottky}) and $\nu^{\mathrm{Chabert}}$ (Eq.~\eqref{eq:Chabert}). The closures based on the direct integration of the Boltzmann operator that where presented in Section \ref{sec:DIBO-computation} are referred to as DIBO. We present only half of the domain as the simulation is symmetric.

\begin{figure}[t]%
    \centering
    \includegraphics[width=0.95\linewidth]{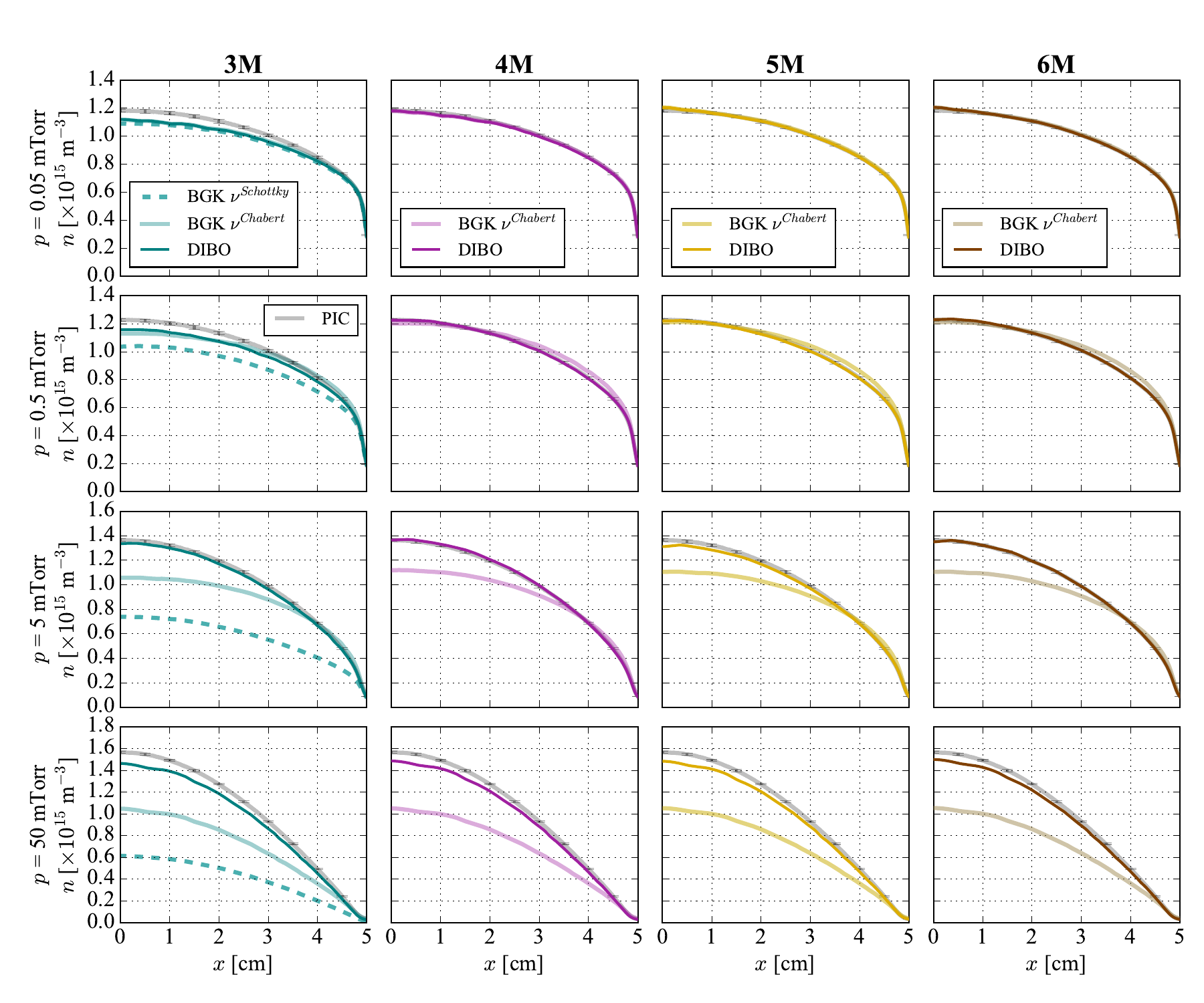}
    \caption{Density profiles of the different high-order moment closures compared to PIC simulations (in gray). We present the results in half of the simulated domain in four different pressures $0.05$ mTorr (top), $0.5$ mTorr (middle top), $5$ mTorr (middle bottom), and $50$ mTorr (bottom). We compare the results using the collisional source terms via the direct integration of the Boltzmann operator (DIBO) with different BGK operators. {The error bars indicate the standard deviation of the solution in the averaging window of $300\mu$s.}}
    \label{fig:profile_n_coll}
\end{figure}

The density profiles are shown in Fig.~\ref{fig:profile_n_coll}. In general, the high-order moment closures 4M, 5M and 6M using the DIBO approach have almost perfect agreement with the kinetic simulations. The 3M-DIBO solution is slightly less accurate, in particular at low pressures. We can also see that the BGK model with constant frequency is less accurate, confirming the need of taking into account the impact of the drift in the collision frequency (as naturally done in the DIBO model). Alternatively, the BGK model with $\nu^{\mathrm{Chabert}}$ has good accuracy, in particular at low pressure. However, it loses fidelity at high pressure. This is consistent with the results of Lafleur \cite{Lafleur15}, although in that reference the reason invoked was the non-homogeneous electron-impact ionization. In this paper, we show that it might be a combination of both effects.

\begin{figure}[t]%
    \centering
    \includegraphics[width=\linewidth]{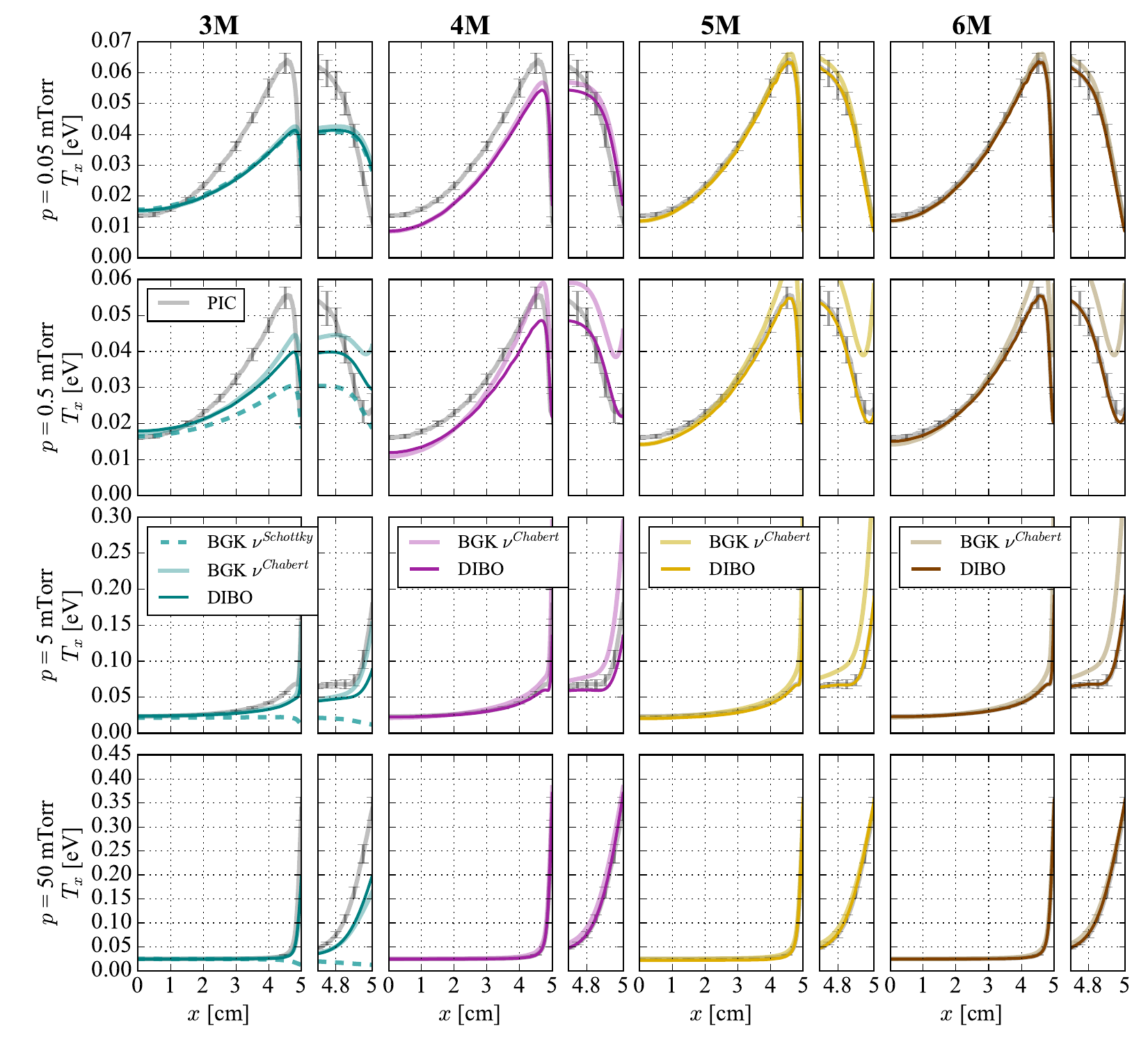}
    \caption{Axial temperature profiles of the different converged fluid simulations compared to PIC simulations (in gray). {The error bars indicate the standard deviation of the solution in the averaging window of $300\mu$s.}}
    \label{fig:profile_T_coll}
\end{figure}

\begin{figure}[t]%
    \centering
    \includegraphics[width=0.85\linewidth]{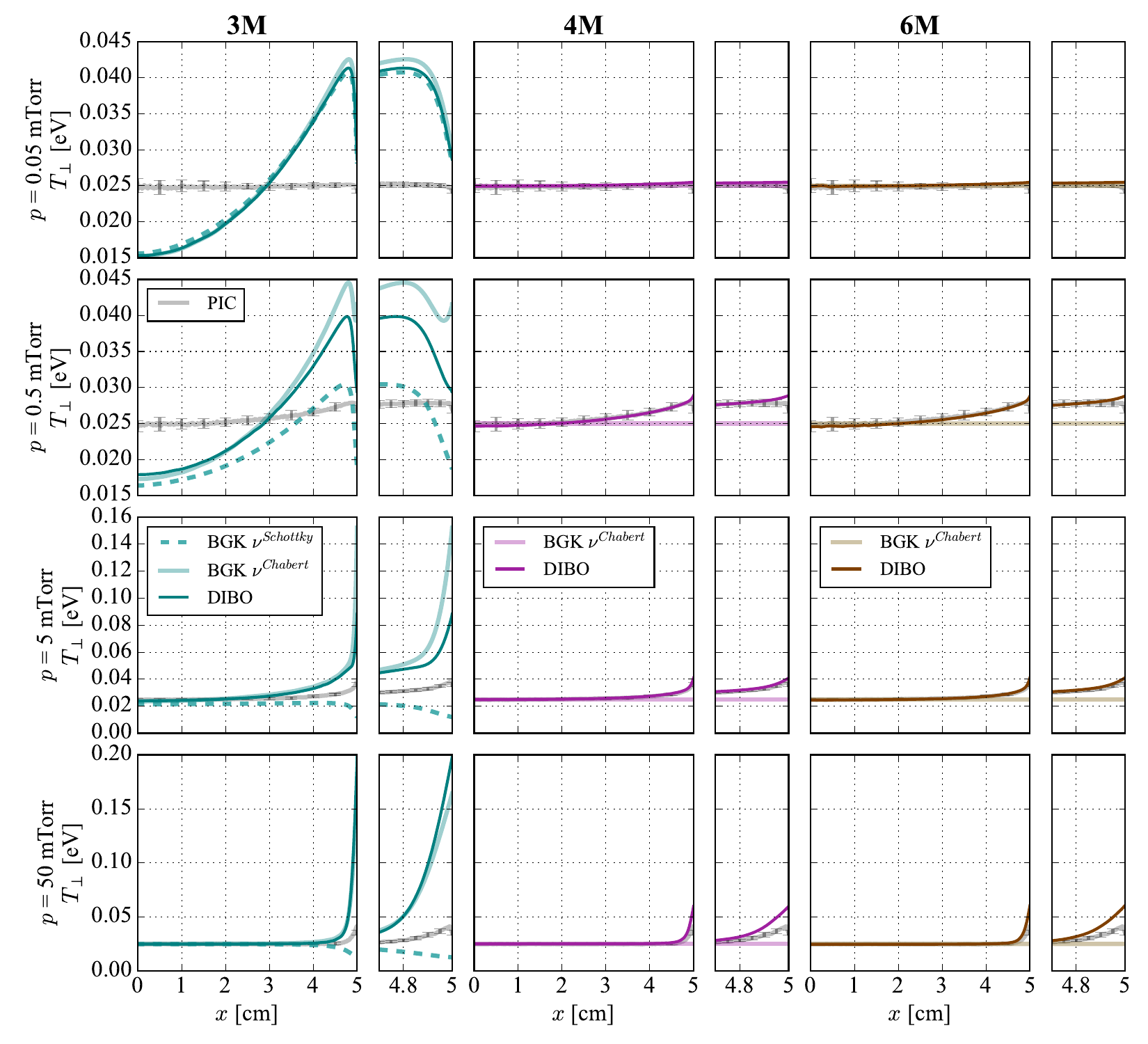}
    \caption{Perpendicular temperature profiles of the different high-order moment closures compared to PIC simulations (in gray). We present the results in half of the simulated domain in four different pressures $0.05$ mTorr (top), $0.5$ mTorr (middle top), $5$ mTorr (middle bottom), and $50$ mTorr (bottom). The 5M is not included as the perpendicular temperature is assumed to be zero. {The error bars indicate the standard deviation of the solution in the averaging window of $300\mu$s.}}
    \label{fig:profile_T_perp_coll}
\end{figure}

The axial temperature profile is shown in Fig.~\ref{fig:profile_T_coll}. The ion temperature in the kinetic profile increases toward the sheath, while it decreases within the sheath at low pressures (0.05 and 0.5 mTorr) due to strong advection. At higher pressures (5 and 50 {mTorr}), Joule heating dominates inside the sheath, causing the temperature to increase as well. Fig.~\ref{fig:profile_T_coll} also includes a zoom into the sheath region to highlight these features. The moment-based results show that the 3M-DIBO model fails to capture the low-pressure regime or the sheath behavior at high pressures. The 4M-DIBO model improves the results at high pressures but remains inaccurate at low pressures. In contrast, the 5M-DIBO and 6M-DIBO models successfully reproduce the temperature evolution across all pressures, as they correctly account for the axial heat flux, as demonstrated in the following. However, the BGK models exhibit accuracy issues, particularly inside the sheath.

The perpendicular temperature is shown in Fig.~\ref{fig:profile_T_perp_coll}. The kinetic profile shows a constant temperature at the gas temperature except in the sheath where the perpendicular temperature is increased as an effect of the drift and the elastic collisions (note that the charge exchange collisions do not heat the perpendicular direction). The 3M model presents the same temperature as in the axial profile and hence largely overestimates the effect of collisions. Alternatively, the 5M model is not presented as the perpendicular temperature is assumed to be zero (as the distribution is constructed as a sum of Dirac distributions with only axial velocity). In the comparison, we see that the 4M-DIBO and 6M-DIBO are able to correctly capture the perpendicular temperature. We note that the BGK models are not able to capture the temperature in the perpendicular direction. This is due to the anisotropy of the relaxation processes that result of the presence of charge exchange and elastic collisions.

\begin{figure}[t]%
    \centering
    \includegraphics[width=\linewidth]{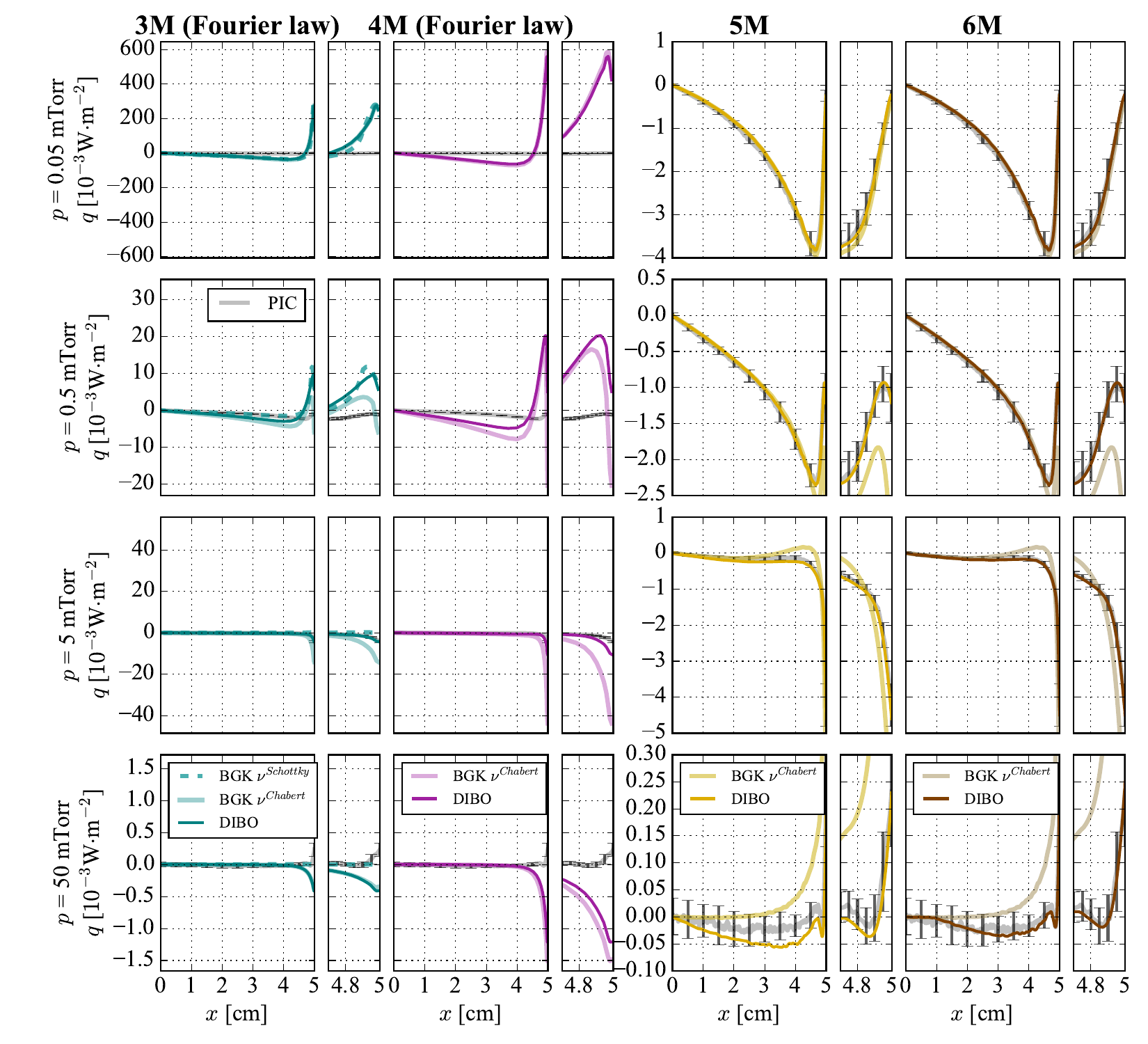}
    \caption{Heat flux profiles of the different high-order moment closures compared to PIC simulations (in gray). We present the results in half of the simulated domain in four different pressures $0.05$ mTorr (top), $0.5$ mTorr (middle top), $5$ mTorr (middle bottom), and $50$ mTorr (bottom). In the 3M and 4M case, we have included a Fourier law computed from the local fluid variables. {The error bars indicate the standard deviation of the solution in the averaging window of $300\mu$s.}}
    \label{fig:profile_q_coll}
\end{figure}

The heat flux profile is shown in Fig.~\ref{fig:profile_q_coll}. The heat flux is particularly large at low pressures everywhere in the domain (which is an impact of the ionization and charge exchange collisions that create a large tail in the distribution function). Alternatively, at larger pressures, the heat flux only changes inside the sheath (as a result of the charge exchange collisions and the presence of a large electric field). For the sake of comparison, we present the Fourier law with the 3M and 4M solutions. The Fourier approximation is shown to be not valid in this system, with either BGK of DIBO collisions and either at low and high pressures. We note that the 3M-DIBO Fourier law at 5 mTorr appears to have good fidelity inside the sheath, but this feature is attributed to a coincidence as the remaining pressures and cases have large discrepancies with this model. Alternatively, the 5M and 6M-DIBO models are able to capture the heat flux at all studied pressures. In particular, the 6M-DIBO model seems to slightly improve the high-pressure regime. This can be explained due to inclusion of the finite perpendicular pressure, which has an increasing impact in the computation of the collisional terms of the axial moments inside the sheath (as the thermal speed and the drift speed become increasingly comparable at large pressure).

\begin{figure}[t]%
    \centering
    \includegraphics[width=\linewidth]{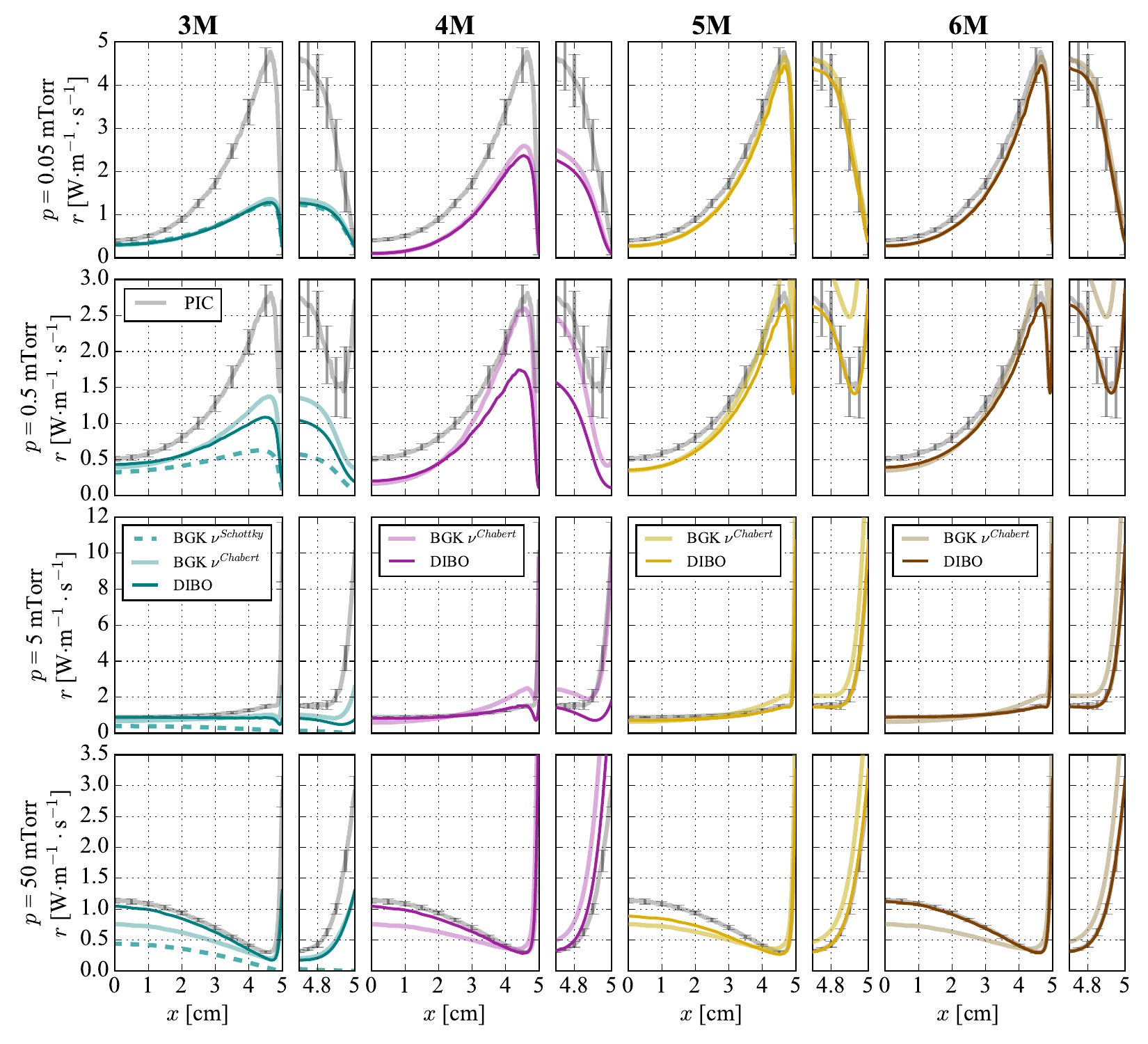}
    \caption{Kurtosis (fourth-order moment) profiles of the different high-order moment closures compared to PIC simulations (in gray). We present the results in half of the simulated domain in four different pressures $0.05$ mTorr (top), $0.5$ mTorr (middle top), $5$ mTorr (middle bottom), and $50$ mTorr (bottom). {The error bars indicate the standard deviation of the solution in the averaging window of $300\mu$s.}}
    \label{fig:profile_r_coll}
\end{figure}

Finally, the kurtosis (fourth-order moment) profile is shown in Fig.~\ref{fig:profile_r_coll}. The accuracy to capture the kurtosis of the different models is similar to that of the temperature. As in the previous moments, the best fidelity is provided by the 6M-DIBO model. Note that in the case of the kurtosis of 3M and 4M, it is this of a Maxwellian distribution. Alternatively, in the 5M and 6M models, this moment will be largely impacted by the closure as the fourth-order moment is the last of the moment hierarchy. These results, show that the HyQMOM closure is an accurate closure under the studied conditions. {We have added a quantification of the relative $L_2$ error norm for each model in the bulk and sheath in Appendix~\ref{sec:ErrorQuantification}, in Tables \ref{tb:tableiv}, \ref{tb:tablev}, and \ref{tb:tablevi}. Note that the reference solution has discretization errors and stochastic noise; thus, this quantification primarily serves to compare the models with each other, as the reference solution error can be comparable to that of the models for some moments (as shown in the error bars of the moment profiles).}

    \label{fig:profile_realizability_coll}

\subsubsection{Comparison of VDF reconstruction}
\label{sec:VDF_reconstruction_coll}

We will present here the ability of the different models to reconstruct a VDF from the moments. Note that this is particularly important for low-temperature plasma applications, as the energy of the ions at the wall plays an important role in processing applications. Similarly, the shape of the VDF can impact the collisional processes and the formation of the sheath\cite{Tsankov17}.

\begin{figure}%
    \centering
\includegraphics[width=0.9\linewidth,trim={0cm 6cm 0 0},clip]{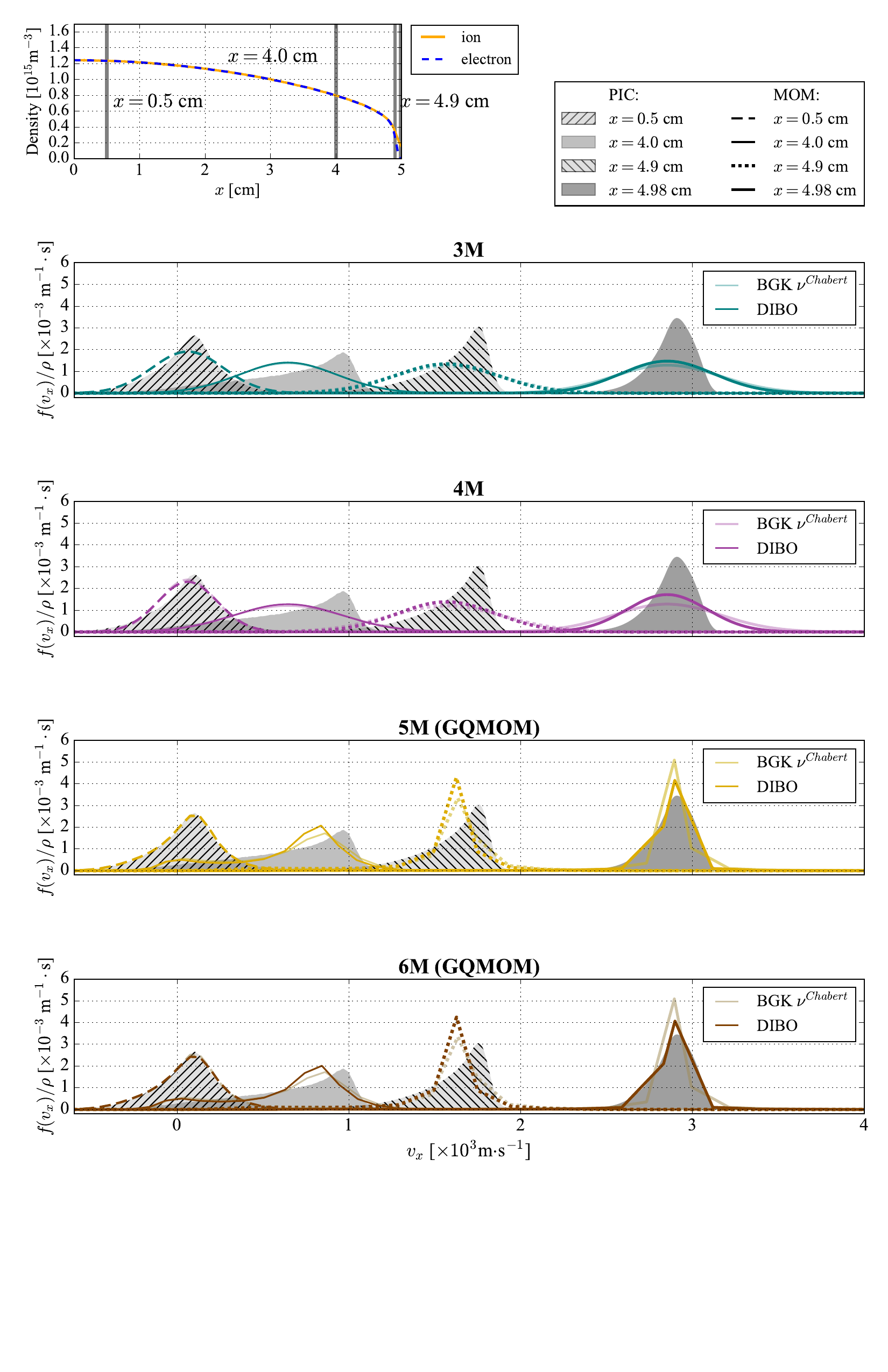}
    \caption{VDF from the kinetic simulations (in gray) and reconstructed from the MOMs simulation results for a pressure of 0.5~mTorr. The upper plot shows the positions at which each VDF is taken. The 5M and 6M are computed with the GQMOM method and presented as continuous distributions {through a direct linear interpolation between each Dirac using GQMOM}.}
    \label{fig:VDF_PIC&0_5M_IP_coll}
\end{figure}

\begin{figure}%
    \centering
    \includegraphics[width=0.9\linewidth,trim={0cm 6cm 0 0},clip]{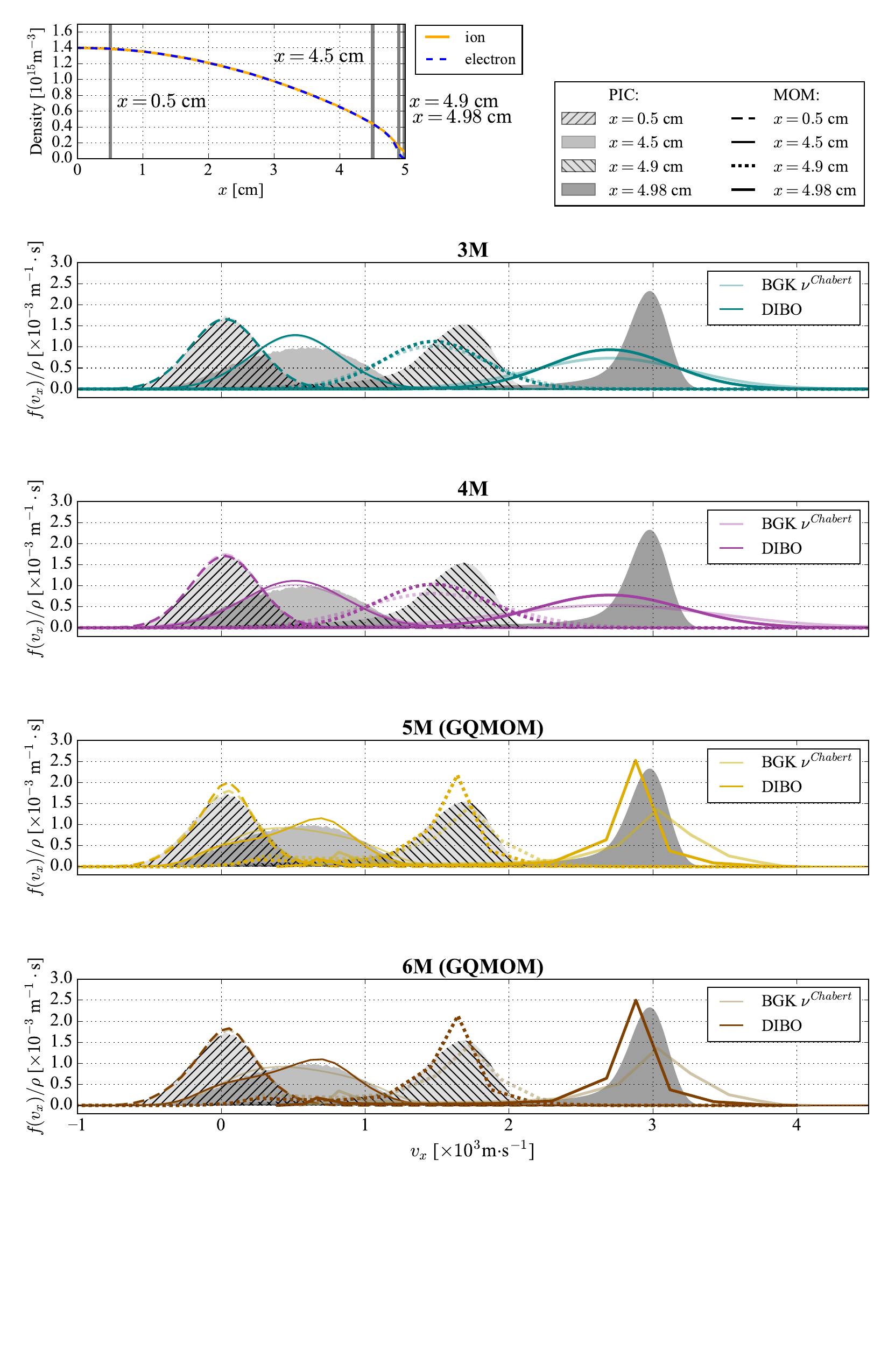}
    \caption{VDF from the kinetic simulations (in gray) and reconstructed from the MOMs simulation results for a pressure of 5~mTorr. The upper plot shows the positions at which each VDF is taken.}
    \label{fig:VDF_PIC&5M_IP_coll}
\end{figure}

\begin{figure}%
    \centering
    \includegraphics[width=0.9\linewidth,trim={0cm 6cm 0 0},clip]{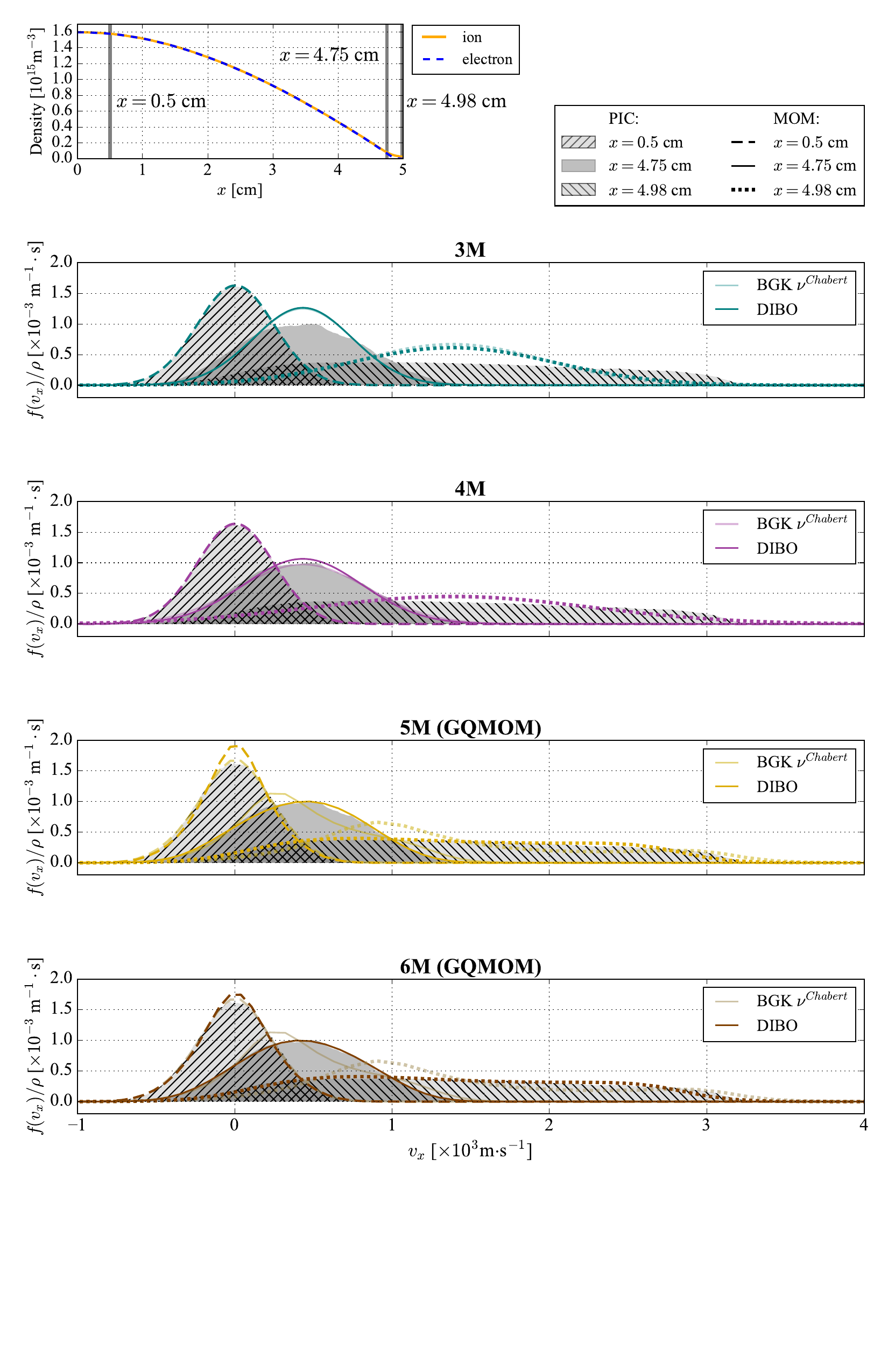}
    \caption{VDF from the kinetic simulations (in gray) and reconstructed from the MOMs simulation results for a pressure of 50~mTorr. The upper plot shows the positions at which each VDF is taken.}
    \label{fig:VDF_PIC&50M_HP_coll}
\end{figure}

In Figs.~\ref{fig:VDF_PIC&0_5M_IP_coll}, \ref{fig:VDF_PIC&5M_IP_coll} and \ref{fig:VDF_PIC&50M_HP_coll}, we show the VDF reconstruction 3M, 4M, 5M, and 6M considered with either BGK collisions (with Chabert's frequency, as it is the one that provides best results) and DIBO model at $p_\gas = 0.5,~5,$ and $50$ mTorr. We present the VDF in four different points of the discharge, at nearly the center, the presheath, near the sheath edge, and nearly the wall (x= 4.98 cm). In the high-pressure case, we do not present the presheath as the distribution is nearly the same as in the center (i.e., a Maxwellian). We note that the VDFs of the moment models are instantaneous values whereas the PIC results are averaged over long periods of time in order to remove the statistical noise. The 5M and 6M distributions are obtained with the generalized QMOM (GQMOM) method, as described by Ref.~\cite{Fox23}, in order to increase the number of Diracs for a given set of moments. {The number of Diracs varies from $30$ to $100$ depending on the pressure}.

The low pressure case of Fig.~\ref{fig:VDF_PIC&0_5M_IP_coll} shows that the distributions are far from a Maxwellian everywhere in the domain, with long tails in positions close to the sheath. These tails are a consequence of the charge exchange and ionization collisions. The 5M and 6M models show a very good agreement with the results. In particular, the distribution at the wall resulting from the 6M-DIBO solution. Alternatively, the 3M and 4M solutions over estimate the high-energy tails at the wall, which is consequence of a bad estimation of the temperature as well as the no inclusion of the heat flux and excess kurtosis in the model.

The intermediate pressure case of Fig.~\ref{fig:VDF_PIC&5M_IP_coll} shows that the distributions are closer to a Maxwellian everywhere in the domain, except in the vicinity of the sheath, where the electric field becomes large. The 5M and 6M models show a very good agreement with the results, largely improving the 3M and 4M solutions.

Finally, in the high-pressure case, (Fig.~\ref{fig:VDF_PIC&50M_HP_coll}) shows that the distributions are closer to a Maxwellian, except at the wall, where the distribution has a heavy tail. In this case, the asymmetry of the distribution is on the opposite side as compared to the low pressure case (as it can be seen by the change of sign in the heat flux). As in the previous cases, the 5M and 6M models show a very good agreement with the results, while the 4M shows that despite not capturing the asymmetry at the wall, it is still able to provide a very close agreement.

\subsection{Case 2: DC discharge at different pressures}

Case 2 (Fig.~\ref{fig:set-ups} b) models a 1D DC discharge in a domain $x \in [-L, L]$ ({with} $L = 2.5$~cm) at pressures $p_\gas = 200$ and $500$~mTorr ($T_\gas = 300$~K). The discharge is driven by a potential difference $\phi(x={-L}) = -300$~V (cathode) and $\phi(x=L) = 0$ (anode), sustained by self-consistent electron-impact ionization.

\subsubsection{Comparison of moment profiles}

\begin{figure}[t]%
    \centering
    \includegraphics[width=\linewidth]{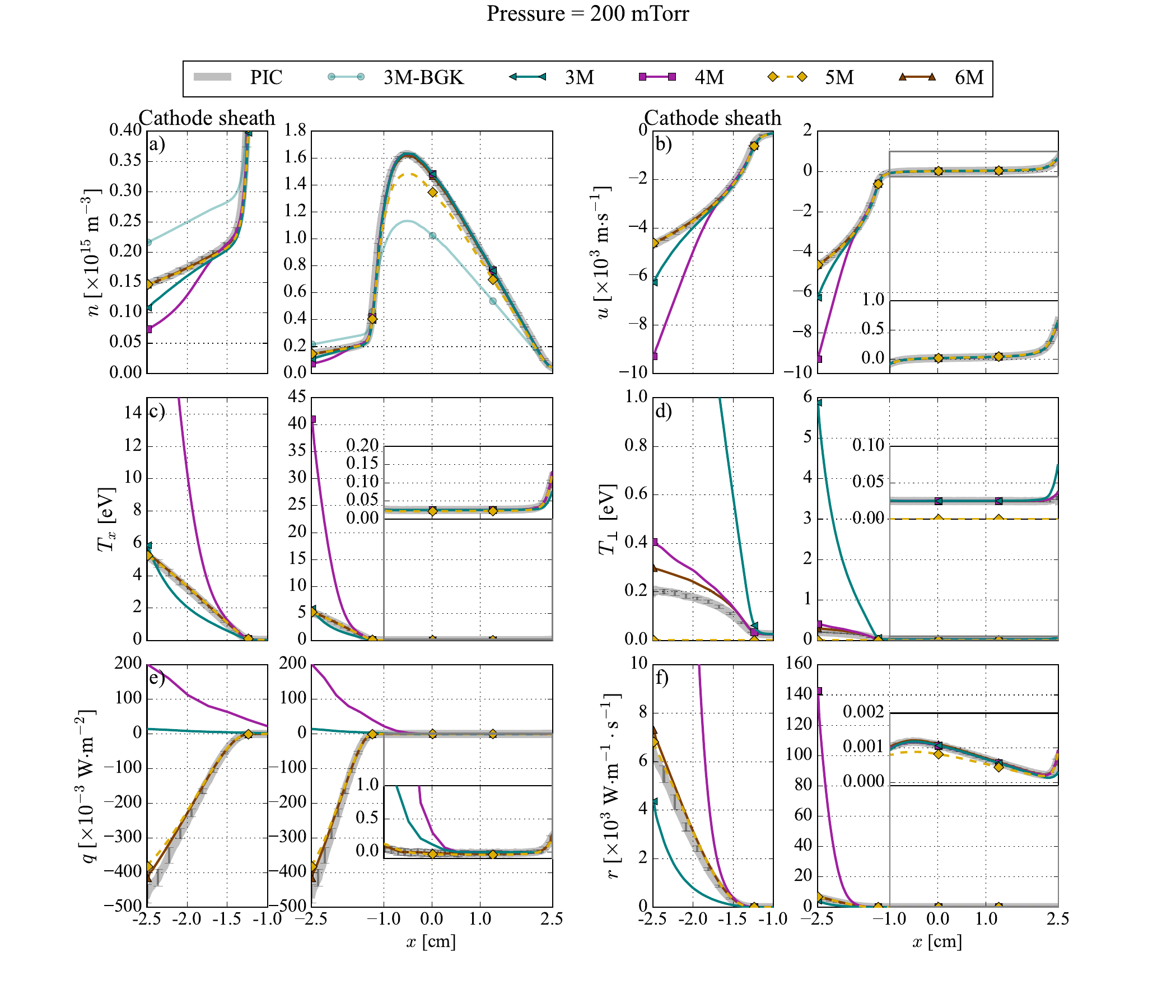}
    \caption{Moment profiles (first five axial moments and perpendicular temperature) from the DC discharge simulation at $200$ mTorr, compared to the reference kinetic solution (gray). We show 3M, 4M, 5M, and 6M models with collision source terms from direct Boltzmann integration, and include the 3M-BGK solution for density. Each profile features zoomed views of the cathode sheath and bulk region. Error bars indicate the standard deviation over a $150\,\mu$s averaging window.}
    \label{fig:profile_n_DC_200}
\end{figure}

\begin{figure}[t]%
    \centering
    \includegraphics[width=\linewidth]{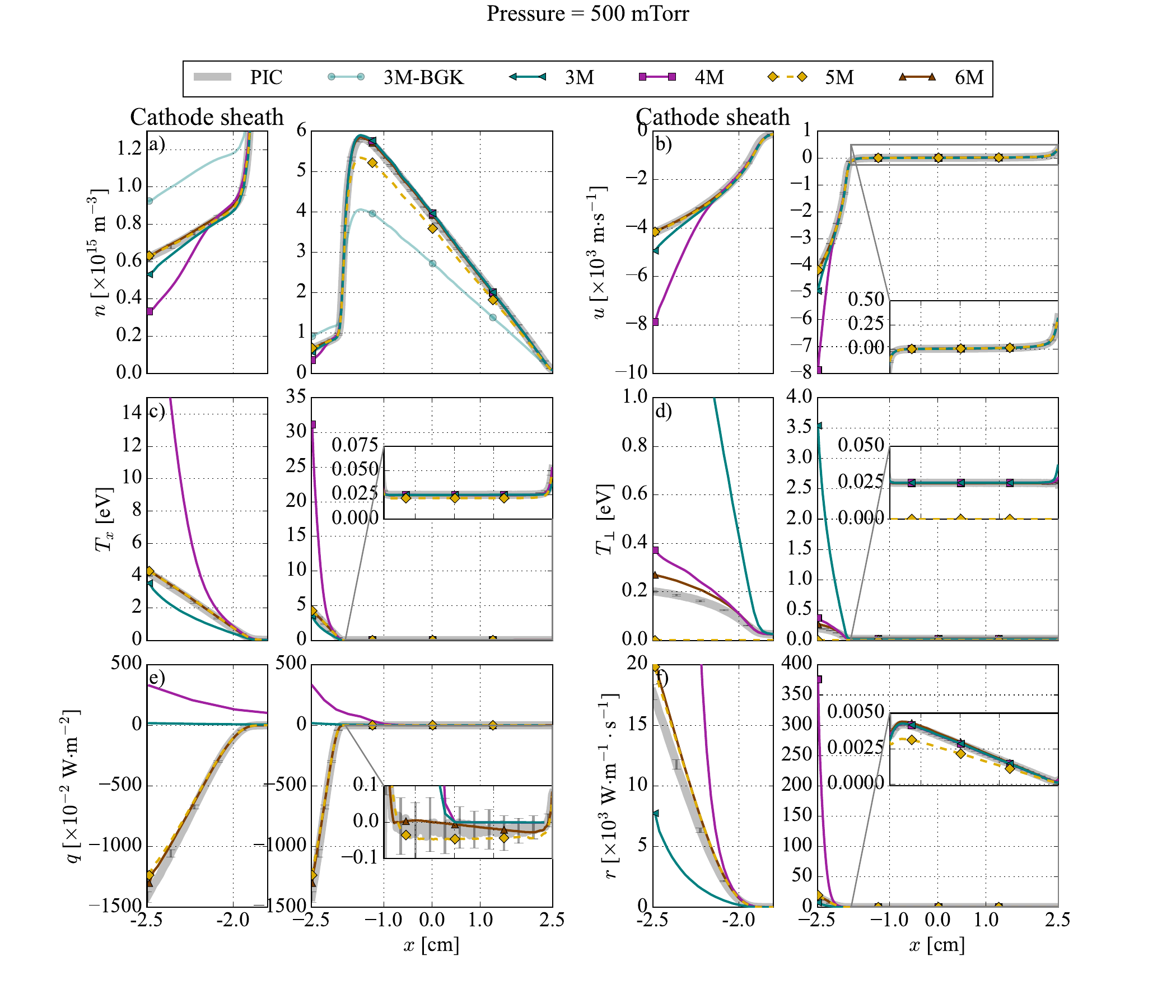}
    \caption{Moment profiles (first five axial moments and perpendicular temperature) from the DC discharge simulation at $500$ mTorr, compared to the reference kinetic solution (gray). We show 3M, 4M, 5M, and 6M models with collision source terms from direct Boltzmann integration, and include the 3M-BGK solution for density. Each profile features zoomed views of the cathode sheath and bulk region. Error bars indicate the standard deviation over a $150\,\mu$s averaging window.}
    \label{fig:profile_n_DC_500}
\end{figure}

In Figs.~\ref{fig:profile_n_DC_200} and \ref{fig:profile_n_DC_500}, we present the moments of the DC cases at $200$ and $500$\,mTorr, respectively, for the 3M, 4M, 5M, and 6M models using the DIBO collision model, and compare them to kinetic simulations. As a reference, we include only the 3M-BGK model using $\nu^{\text{Chabert}}$ in the density profile, as this model lacks accuracy for all other moments. As can be seen, the presence of a large potential drop (on the order of $300$\,V) at the cathode (left boundary) produces a large sheath, where ions are accelerated to high velocities. We provide a zoom into both the sheath and the bulk regions, as the behavior of the models differs significantly between these two regions. The behavior at the two studied pressures is very similar and summarized as follows.

In Figs.~\ref{fig:profile_n_DC_200}a and \ref{fig:profile_n_DC_500}a, we present the comparison of the density profiles. The 3M-DIBO, 4M-DIBO, and 6M-DIBO are able to capture correctly the density profile in the bulk. Alternatively, the 3M-BGK presents a large discrepancy (as noted in the high pressure cases of Case 1). The BGK presents discrepancies in all other moments and we choose to not present it, for the sake of clarity in the figure. Alternatively, the 5M-DIBO presents small errors in the maximum of the density. This is due to the approximation $T_{\ion,\gas} = 0$ that is done in the collision source terms (Eqs.~\eqref{eq:DiracVDF-simple}). Alternatively, inside the cathode sheath, we can note that the 3M-DIBO and 4M-DIBO present large errors in the cathode sheath. The reason of these errors can be explained by the behavior of the higher-order moments.

In Figs.~\ref{fig:profile_n_DC_200}b and \ref{fig:profile_n_DC_500}b, we present the comparison of the velocity, all the DIBO moment models are able to capture the bulk velocities. However, the 3M-DIBO and 4M-DIBO overestimate the sheath velocity, which explains the underestimation of the density drop at the sheath.

In Figs.~\ref{fig:profile_n_DC_200}c and \ref{fig:profile_n_DC_500}c, we present the comparison of $T_x$. Similarly, all the DIBO moment models are able to capture the bulk velocities. However, the 3M-DIBO and 4M-DIBO are not able to represent the sheath. In particular, the 4M presents a very large overestimation of the sheath temperature. This is, as it will be shown in the following, due to the absence of heath-flux in the axial direction. In the 3M, the fact of redistributing the energy in all direction works effectively as a flux of axial energy (to the other directions).

In Figs.~\ref{fig:profile_n_DC_200}d and \ref{fig:profile_n_DC_500}d, we present the comparison of $T_\perp$. The 4M-DIBO and 6M-DIBO are able to capture the temperature (5M considers this temperature to be zero). Alternatively, the 3M largely overestimates (by a factor $30$) the perpendicular temperature.

In Figs.~\ref{fig:profile_n_DC_200}e and \ref{fig:profile_n_DC_500}e, we present the comparison of the axial heat flux. In this case, the heat flux is very large, which is a result of the mutual effect of the charge exchange collisions and the electric field inside the sheath. We present the Fourier law for 3M and 4M models, which show that Fourier law predicts a heat flux in the opposite direction, which shows that the heat flux transport of these discharges is beyond the classical transport theories. Both the 5M and 6M-DIBO correctly capture this heat flux in a self-consistent manner, accounting for non-local effects.

Finally, in Figs.~\ref{fig:profile_n_DC_200}f and \ref{fig:profile_n_DC_500}f, we present the comparison of the axial kurtosis. The results show that the 6M-DIBO is able to both capture the profiles inside the sheath and the bulk, whereas the 3M and 4M are not able to capture the sheath (with a large overestimation of the 4M model of a factor $\sim 23$). {We have added a quantification of the relative L$_2$ error norm of each model in the bulk and in the sheath in Appendix \ref{sec:ErrorQuantification}, in Tables \ref{tb:tablevi} and \ref{tb:tablevii}. }

\subsubsection{Comparison of VDF reconstruction}

\begin{figure}[t]%
    \centering
    \includegraphics[width=\linewidth]{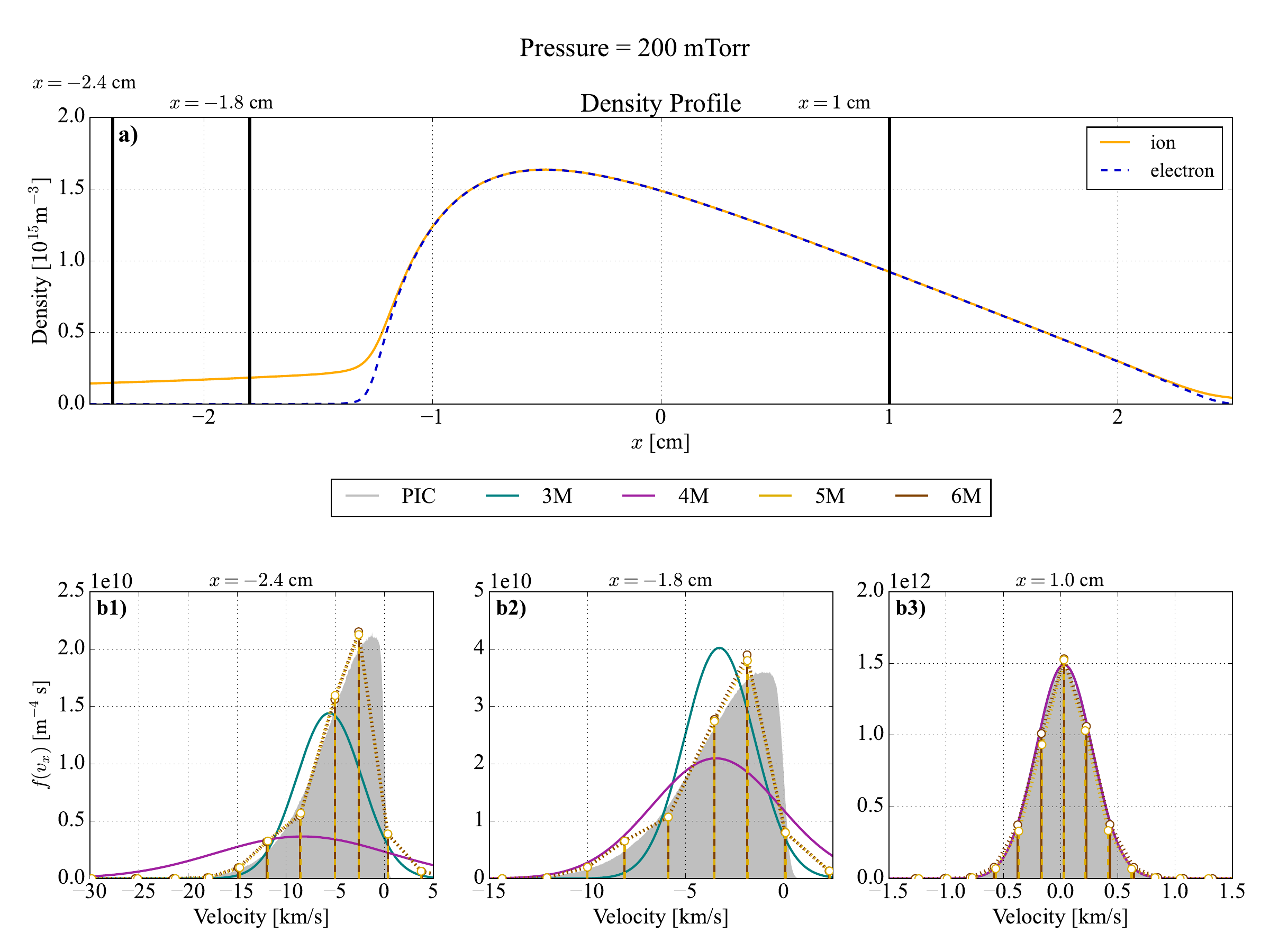}
    \caption{VDF from the kinetic simulations (in gray) and reconstructed from the moment simulation results for a pressure of $200$~mTorr. The upper plot shows the positions at which each VDF is taken on the density profile. The 5M and 6M distributions are computed with the GQMOM method {(with 15 nodes)} and the Dirac distributions are presented.}
    \label{fig:profile_VDF_DC_200}
\end{figure}

\begin{figure}[t]%
    \centering
    \includegraphics[width=\linewidth]{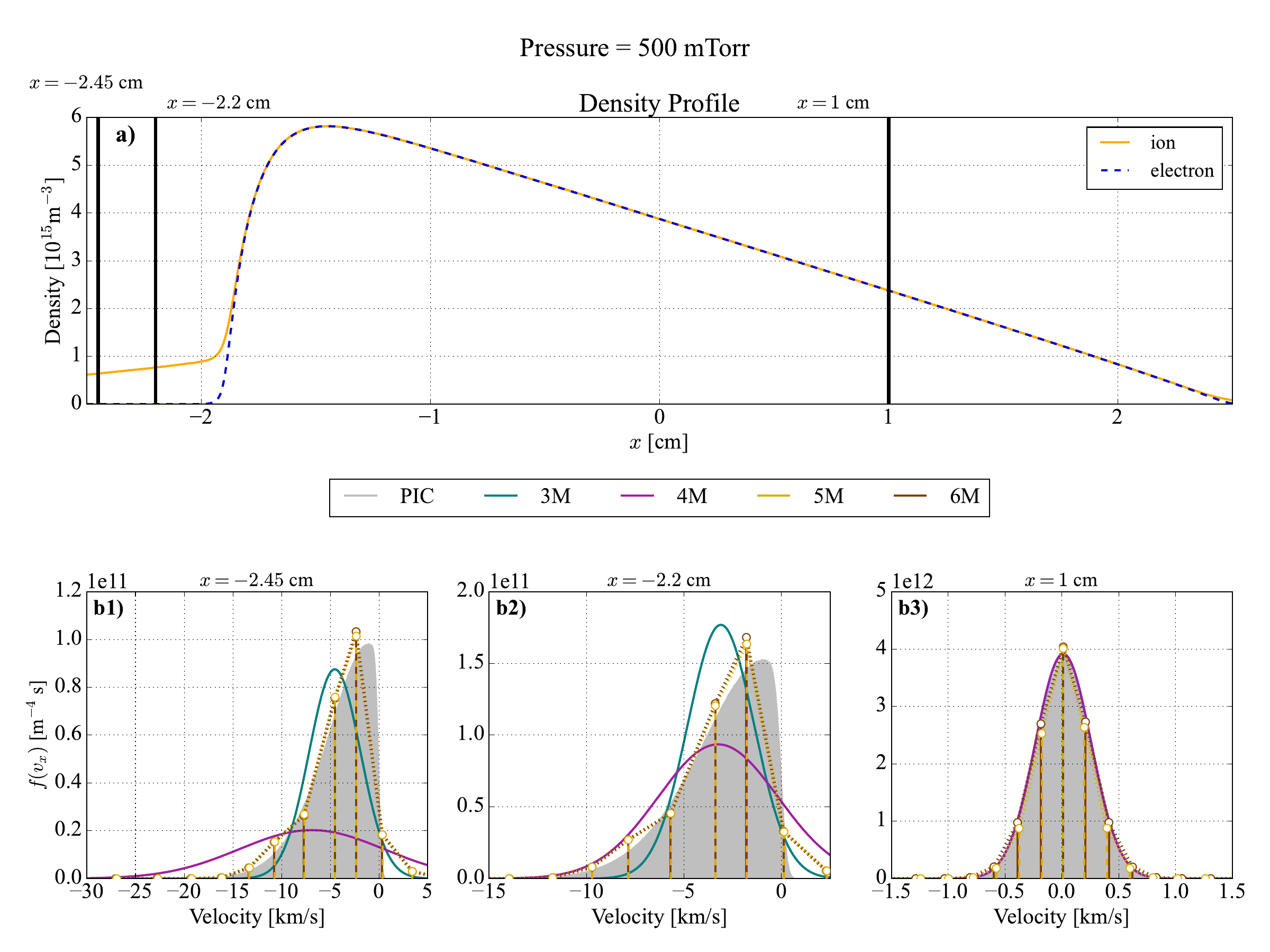}
    \caption{VDF from the kinetic simulations (in gray) and reconstructed from the moment simulation results for a pressure of $500$~mTorr. The upper plot shows the positions at which each VDF is taken on the density profile. The 5M and 6M distributions are computed with the GQMOM method {(with 15 nodes)} and the Dirac distributions are presented.}
    \label{fig:profile_VDF_DC_500}
\end{figure}

In Figs.~\ref{fig:profile_VDF_DC_200} and \ref{fig:profile_VDF_DC_500}, we present the reconstruction of the VDF of the different models as compared to the PIC-MCC solution at $200$ and $500$ mTorr, respectively. We present the ion VDF at three different positions: inside the cathode sheath close to the wall (Figs. \ref{fig:profile_VDF_DC_200}b1 and \ref{fig:profile_VDF_DC_500}b1, in the middle of the cathode sheath (Figs. \ref{fig:profile_VDF_DC_200}b2 and \ref{fig:profile_VDF_DC_500}b2) and in the bulk (Figs. \ref{fig:profile_VDF_DC_200}b3 and \ref{fig:profile_VDF_DC_500}b3). The kinetic VDF presents a shape that resembles the model proposed by Godyak, where the ion distribution is a half Maxwellian. Note that despite this resemblance, the BGK model does not provide good results as other non-local transport phenomena (pressure gradient, pressure anisotropy and heat flux) play a fundamental role. It is remarkable the accuracy of the 5M-DIBO and 6M-DIBO reconstruction. We present the representation using the GQMOM technique\cite{Fox23} (with $15$ Diracs). The 3M and 4M models are not able to capture this VDF, as it is beyond the capabilities of a Gaussian representation. It is worth noting that the ion VDF can potentially have a great influence in the plasma-material interactions and can potentially influence the bulk (by the production of secondary electrons, recombination of hot ions at the wall, etc). The proposed 6M model offers an alternative to kinetic solvers in order to quantify these effects.

\section{Conclusions}\label{sec:Conclusions}

In this paper, we have proposed different high-order moment models for the simulation of the nonequilibrium ion dynamics in one dimensional weakly ionized plasmas. The models are fully analytical and include the integration of the Boltzmann collision operator for the collisional source terms in the moment equations. Through several numerical simulations, we have benchmarked our models to kinetic simulations in a wide range of pressures (from 0.05 to 500 mTorr). In the following, we summarize our findings:
\begin{itemize}
    \item We have compared four moment models: 3M (considering mass, axial momentum and isotropic energy), 4M (mass, axial momentum and axial and perpendicular energies) , 5M (considering axial moments up to the fourth-order moment), and 6M (considering the 5M model with the perpendicular temperature). They increasingly incorporate different nonequilibrium phenomena. The most complete model is our 6M that captures arbitrary ion drifts, arbitrary pressure anisotropies, and arbitrary heat flux and kurtosis along the axial direction. The closure fluxes of all these models are analytical and represent distribution functions that are strictly positive (as opposed to other closures like Grad's that can represent distributions with negative tails).

    \item The collisional source terms are computed analytically for all the models with arbitrary collision scattering geometries and energy dependence. In particular, we propose a novel formulation that is a generalization of the classical Chapman-Cowling collisions for arbitrary drift velocities, temperature anisotropies and heat flux. In our formulation, the new collision integrals tend to the classical theory in the limit of small drift (as compared to the thermal speed) and zero temperature anisotropy. As shown in Fig.~\ref{fig:Omega111}, the drift velocity has a great impact in the rates. Similarly, large heat flux values can largely modify the collision rate at small drifts. {In addition, the proposed generalized model captures the net momentum exchange between a skewed ion distribution and stationary gas, as the fast tail collides more frequently than the slow bulk. This effect cannot be reproduced by neither BGK operators, with their velocity-independent rates, nor closures without a heat-flux equation.}
    \item In the simulations of a plasma between two floating walls (Case 1), the results show that the novel collision terms largely improve the BGK results (particularly at high pressures). The transport at low pressure requires considering an anisotropic pressure tensor. As a result, a rather simple 4M model is able to represent the density profile with high accuracy. The addition of the heat flux and kurtosis allow for a more accurate representation of the sheath as well as the reconstruction of the VDF close to the wall.
    \item In the simulation of a DC discharge (Case 2), the results show that the heat flux inside the cathode sheath is very large and in opposite direction to the classical Fourier flux. This heat flux is a consequence of the mutual action of a large potential drop ($300$ V) and the charge exchange collisions. Similarly, the pressure tensor is largely anisotropic in the sheath ($T_x/T_\perp\sim 20$). These two effects have a large impact on the transport of the ions, both in the energy and particle transport, being a challenge for classical fluid models. Despite this large nonequilibrium conditions, the 6M model is able to self-consistently capture the bulk and the sheath dynamics.
    \item The novel 6M model, based on a generalization of the 5M HyQMOM with anisotropic temperature, with the proposed analytical models of this paper for the collisional terms, demonstrates to be an efficient and robust model for the ion dynamics in one dimensional discharges. The generality of the model lies in its link to Gaussian quadratures, which allows to compute the collision terms with high accuracy. In addition, the 6M model contains all the other models (3M, 4M and 5M) as a particular case. As compared to the other models, the 6M model shows a great improvement at low pressure and in the presence of large electric fields, where the heat flux and the temperature anisotropy are very large.  As shown in the numerical results, the model is able to correctly reconstruct the distribution function in absence of noise and at a computational cost that is comparable to classical fluid models, as all the expressions are purely analytical and the collision terms can be read from tables computed with the cross sectional data. In addition, the model is extensible to other plasma mixtures , with {a self-consistent computation of the collision terms without} requiring the computation of the transport properties from 0D Boltzmann solvers or semi-empirical heuristic approximations.
    \item {Finally, we would like to highlight that the method can be extended to other species. In particular, it would be straightfoward to extend to the neutral species that are closer to equilibrium due to a smaller Knudsen number. The collisional terms with the ions presented in this paper would be easily adapted to the neutrals but self collisions and collisions with the electrons would be required. Note that in the conditions studied in this paper, the depletion of neutrals due to neutral heating might have a non-negligible impact on the dynamics of the charged species as discussed in Ref.~\cite{Liard07}.}
\end{itemize}
The model has been validated with a noble gas plasma. Nevertheless, the collision terms that are proposed in this paper are derived for arbitrary collisions and, therefore, it can be used for other plasma mixtures, including molecular plasmas. Similarly, the procedure to derive the collision terms in quadrature-based models can be extended to other moment closures including additional moments.

\section*{DATA AVAILABILITY}

The data that support the findings of this study are openly available in Zenodo at \doi{10.5281/zenodo.24197356}. Similarly, the code used to reproduce the results can be found in at
\url{https://github.com/alvarezlaguna/IonCollisionMomentModels_release}.

\section*{Acknowledgements}

This work is supported by the ERC Grant “HiMomPlas” under Grant Agreement Number 101222739 (PI AAL). AB's work is part of a PhD funded by the Institut Polytechnique de Paris (IPP) and EUR PlasmaScience.

\section*{Funding statement and disclaimer}
Funded by the European Union. Views and opinions expressed are however those of the author(s) only and do not necessarily reflect those of the European Union or the European Research Council Executive Agency. Neither the European Union nor the granting authority can be held responsible for them.

\appendix
\section{Computation of $\mathcal{I}_{k0}$}\label{app:IntegrationUnitSphere}

We summarize the derivation of the integral $\mathcal{I}_{k0}$ for the moments in the $x$ direction. By injecting Eq.~\eqref{eq:difference} into Eq.~\eqref{eq:angularI_ij}, we obtain,
\begin{equation}\label{eq:integralIk0}
    \mathcal{I}_{k0} = \int_{\mathbb{S}^2} \!\!\!\! m \left[(v'_{x})^k - v_{x}^k\right] \sigma(g, \Omega) \text{d}^2\Omega\ = \quad \mu \sum_{j=1}^k \binom{k}{j} G_x^{k-j} \left(\frac{\mu}{m}\right)^{j-1} \int_{\mathbb{S}^2} \!\!\!\!  \left[ (g'_x)^j - g_x^j\right] \sigma(g, \Omega) \text{d}^2\Omega.
\end{equation}
The $x$ component of $\vec{g}'$ reads $g'_x = g_x \cos\chi + g\sin\chi \left( \cos\varphi \,\eperpgOne \cdot \ex + \sin\varphi \,\eperpgTwo \cdot \ex \right)$, so the $j$-th power reads,
\begin{equation}
    (g'_x)^j = \sum_{i=0}^j \binom{j}{i} g_x^{j-i} \cos^{j-i}\chi g^i\sin^i\chi \left( \cos\varphi \,\vec{e}_{\perp g 1} \cdot \vec{e}_x + \sin\varphi \,\vec{e}_{\perp g 2} \cdot \vec{e}_x \right)^i.
\end{equation}

In order to perform the integration over the angles $\varphi$, we note first that
\begin{equation}
    \int_0^{2\pi} \left( \cos\varphi \,\vec{e}_{\perp g 1} \cdot \vec{e}_x + \sin\varphi \,\vec{e}_{\perp g 2} \cdot \vec{e}_x \right)^{2i+1} \,\text{d}\varphi = 0,
\end{equation}
while for the non-vanishing integrals, we will require the following identity,
\begin{equation}
    \int_0^{2\pi} \sin^{2p}\varphi \cos^{2q}\varphi \,\text{d}\varphi  = 2 B(p+\tfrac{1}{2}, q+\tfrac{1}{2})
\end{equation}
where $B$ is Euler's beta function, which is linked to the gamma function as
\begin{eqnarray}
    2 B(p + \tfrac{1}{2},q + \tfrac{1}{2}) &=& \frac{\Gamma(p + \tfrac{1}{2})\Gamma(q + \tfrac{1}{2})}{\Gamma(p+q + 1)} = \frac{2\pi}{4^{p+q}} \frac{\binom{2p}{p} \binom{2q}{q}}{\binom{p+q}{p}}.
\end{eqnarray}
This yields to the following integral over $\varphi$,
\begin{eqnarray}\label{eq:integralphi}
    \int_0^{2\pi} (g'_x)^j \,\text{d}\varphi &=& \sum_{i=0}^j \binom{j}{i} g_x^{j-i} \cos^{j-i}\chi g^i\sin^i\chi \int_0^{2\pi} \left( \cos\varphi \,\vec{e}_{\perp g 1} \cdot \vec{e}_x + \sin\varphi \,\vec{e}_{\perp g 2} \cdot \vec{e}_x \right)^i \,\text{d}\varphi \nonumber \\
    &=& \sum_{i=0}^{\lfloor j/2 \rfloor} \binom{j}{2i} g_x^{j-2i} \cos^{j-2i}\chi g^{2i}\sin^{2i}\chi \frac{2\pi}{4^{i}} \binom{2i}{i} \left( 1 - \frac{g_x^2}{g^2}\right)^{i} \nonumber \\
    &=& 2\pi \sum_{i=0}^{\lfloor j/2 \rfloor} \binom{2i}{i} \binom{j}{2i} \left(g_x \cos\chi\right)^{j-2i} \left(\left(g^2 - g_x^2\right)\frac{\sin^2\chi}{4}\right)^{i}.
\end{eqnarray}
The integration over the scattering angle $\chi$ needs to be expressed as function of the transport cross sections\cite{Chapman70}, defined as,
\begin{equation}
    Q^{(l)}(|\vec{g}|) = 2\pi\int_0^{\pi} (1 - \cos^l\chi)\sigma(|\vec{g}|,\,\chi)\sin\chi\,d\chi.
\end{equation}
With this definition, we can integrate the part that depends on $\chi$ of Eq.~\eqref{eq:integralphi}, as follows,
\begin{eqnarray}\label{eq:integrationchi}
    2\pi \int_0^{\pi} \!\!\!\!   \cos^{j-2i}\chi \sin^{2i}\chi \sigma(g, \chi) \sin\chi \text{d}\chi &=& \int_0^{\pi} \!\!\!\!\sum_{l=0}^i \binom{i}{l} (-1)^{i-l+1} (1-\cos^{j-2l}\chi) 2\pi \sigma(|\vec{g}|,\,\chi) \sin\chi \text{d}\chi  \nonumber \\
    &=& \sum_{l=0}^i \binom{i}{l} (-1)^{i-l+1} Q^{(j-2l)}(|\vec{g}|).
\end{eqnarray}
By introducing Eq.~\eqref{eq:integralphi} and Eq.~\eqref{eq:integrationchi} into Eq.~\eqref{eq:integralIk0}, this yields
\begin{multline}
    \mathcal{I}_{k0} = \mu \sum_{j=1}^k \binom{k}{j} G_x^{k-j} \left(\frac{\mu}{m}\right)^{j-1} \sum_{i=0}^{\lfloor j/2 \rfloor} \binom{2i}{i} \binom{j}{2i} g_x^{j-2i} \left(\tfrac{1}{4}\left(g^2 - g_x^2\right)\right)^{i} \sum_{l=0}^i \binom{i}{l} (-1)^{i-l+1} Q^{(j-2l)}(g).
\end{multline}

\section[Collision term with anisotropic Maxwellian]{Derivation of the collision terms }
\subsection{Anisotropic and isotropic Maxwellians (4M and 3M)}
\label{app:3M-4Mmodel}
In this section, we will detail the derivation of the 4M collision integrals with a two-temperature anisotropic Maxwellian with arbitrary axial drift. It is to be noted, that the 3M model is a particular case of the 4M where the temperature of the ions become isotropic. We refer to Ref.\cite{Benilov97} for a full derivation of the 3M model. Here, we will treat the 3M as a particular case of the 4M. As seen in Section \ref{sec:DIBO-computation}, the results of the 3M are strictly equivalent as these of Ref.~\cite{Benilov97}.

We will solve the integral of Eq.~\eqref{eq:integral1} with the angular integrals \eqref{eq:integralIk0Final} and \eqref{eq:integralI02Final}, with the following distribution functions
\begin{subequations}
\begin{align}
    f^\mathrm{(4M)}(v_x,\,v_\perp) &= n \frac{\gamma^{1/2}_x\gamma_\perp}{(2\pi)^{3/2}} e^{-\frac{\gamma_x}{2}\left(v_x - u \right)^2-\frac{\gamma_\perp}{2}v^2_\perp}\, ,  \\
    f_\gas(\vec{v}_\gas) &= n_\gas \left(\frac{\gamma_\gas}{2\pi}\right)^{3/2} \exp\left( -\tfrac{\gamma_\gas}{2}v_\gas^2 \right) \,.
\end{align}
\end{subequations}
We define the following velocity variable, similar to the center-of-mass velocity while accounting for the temperatures of ions (axial and perpendicular) and neutrals, as
\begin{equation}
X_x = \frac{\gamma_x (v_x - u) + \gamma_\gas v_{\gas,x}}{\gamma_x + \gamma_\gas},~~~~ \vec{X}_\perp = \frac{\gamma_\perp \vec{v}_{\perp} + \gamma_\gas \vec{v}_{\gas,\perp}}{\gamma_\perp + \gamma_\gas},
\end{equation}
so that
\begin{subequations}
\begin{align}
v_{x} - u &= X_x + \tfrac{\gamma_\gas}{\gamma_x + \gamma_{\gas}} (g_x - u), & \vec{v}_{\perp} &= \vec{X}_\perp + \tfrac{\gamma_\gas}{\gamma_\perp + \gamma_{\gas}} \vec{g}_\perp, \\
v_{\gas,x} &= X_x - \tfrac{\gamma_x}{\gamma_x + \gamma_{\gas}} (g_x - u), & \vec{v}_{\gas,\perp} &= \vec{X}_\perp - \tfrac{\gamma_\perp}{\gamma_\perp + \gamma_{\gas}} \vec{g}_\perp.
\end{align}
\end{subequations}
The absolute value of the determinant of the determinant of the Jacobian of the transformation $(v_{x},\vec{v}_\perp, x_{\gas,x},\vec{v}_{\gas,\perp})\rightarrow(\vec{X},\vec{g})$ is unity. The product of VDFs can be written as
\begin{multline}
    f_\ion(\vec{v}_\ion) f_\gas(\vec{v}_\gas) = \\
    n_\ion n_\gas \frac{\sqrt{\gamma_x} \gamma_\perp \gamma_\gas^{3/2}}{(2\pi)^{3}} \,\exp\left(-\frac{\gamma_x + \gamma_{\gas}}{2} X_x^2 - \frac{\gamma_{x \gas}}{2} (g_x - u)^2 - \frac{\gamma_\perp + \gamma_{\gas}}{2} X_\perp^2 - \frac{\gamma_{\perp \gas}}{2} g_\perp^2 \right) \,.
\end{multline}
Since the collision integrals of the axial moments in our problem do not depend on $\vec{X}_\perp$, we can integrate the product of VDF over $\vec{X}_\perp$ to get
\begin{multline}\label{eq:integralX}
    \int_{\mathbb{R}^2} \, f_\ion(\vec{v}_\ion) f_\gas(\vec{v}_\gas) \text{d}^2\vec{X}_\perp
    = \\
    n_\ion n_\gas \gamma_{\perp,\gas} \frac{\sqrt{\gamma_x \gamma_\gas}}{(2\pi)^2} \,\exp\left(-\frac{\gamma_x + \gamma_{\gas}}{2} X_x^2 - \frac{\gamma_{x \gas}}{2} (g_x - u)^2 - \frac{\gamma_{\perp \gas}}{2} g_\perp^2 \right)\,.
\end{multline}

\paragraph{Axial momentum exchange:}
From Eq.~\eqref{eq:integralIk0Final},
\begin{equation}
    \mathcal{I}_{10} = \int_{\mathbb{S}^2} \!\!\!\! \, m (v'_{x} - v_{x}) \sigma(g, \Omega) \text{d}^2\Omega\,
    = -\mu g_x Q^{(1)}(g) ,
\end{equation}
so after integrating over $\vec{X}$,
\begin{equation}
    \mathcal{C}_{10} = -\mu n_\ion n_\gas \frac{\gamma_{\perp,\gas} \sqrt{\gamma_{x,\gas}}}{(2\pi)^{3/2}} \int_{\mathbb{R}^3} g_x g Q^{(1)}(g) \exp\left( - \frac{\gamma_{x \gas}}{2} (g_x - u)^2 - \frac{\gamma_{\perp \gas}}{2} g_\perp^2 \right) \mathrm{d}^3\vec{g} \,.
\end{equation}
By integrating $\vec{g}$ in spherical coordinates $(g, \theta, \phi)$, with the polar axis directed parallel to $\vec{u}$ (in this case the x-direction), \textit{i.e.} $\vec{g} = g(\cos{\theta}\,, \sin{\theta}\cos{\phi}\,, \sin{\theta}\sin{\phi})$, where $\theta$ represents the angle between $\vec{u}$ and $\vec{g}$. Performing the change of variable $\zeta = \cos\theta$, we obtain
\begin{multline}
    \mathcal{C}_{10}
    = - \mu n_\ion n_\gas \gamma_{\perp,\gas} \sqrt{\frac{\gamma_{x,\gas}}{2\pi}} \int_0^\infty \!\!\!\!\!\! \text{d}g \, g^4 Q^{(1)}(g)\times \\
    \int_{-1}^1 \!\! \text{d}\zeta \, \zeta \,\exp\left( -\frac{1}{2} \left( \gamma_{x,\gas} u^2 + \gamma_{\perp,\gas} g^2 - 2 \gamma_{x,\gas} u g \zeta + (\gamma_{x,\gas} - \gamma_{\perp,g}) g^2 \zeta^2 \right) \right)
\end{multline}
that can be rewritten in non-dimensional form, as
\begin{multline}
    \mathcal{C}_{10} = - \mu n_{\ion} n_{\gas} \kappa_{x,\perp} \frac{1}{\gamma_{x,\gas}^{3/2}}\sqrt{\frac{\gamma_{x,\gas}}{2\pi}} \int_0^\infty \!\!\!\!\!\! \text{d}\bar{g} \, \bar{g}^4 Q^{(1)}\left(\frac{\bar{g}}{\sqrt{\gamma_{x,\gas}}}\right) \times \\
    \int_{-1}^1 \!\! \text{d}\zeta \, \zeta \,\exp\left( -\frac{1}{2} \left( \bar{u}^2 + \kappa_{x,\perp} \bar{g}^2 - 2\bar{u} \bar{g} \zeta + (1 - \kappa_{x,\perp}) \bar{g}^2 \zeta^2 \right) \right)
\end{multline}
with
\begin{align}\label{eq:normalization}
\bar{u} &= \sqrt{\gamma_{x,\gas}} u, & \bar{g} &= \sqrt{\gamma_{x,\gas}} g, \\
{\kappa}_{x,\perp} &= \frac{\gamma_{\perp,\gas}}{\gamma_{x,\gas}} = \frac{m_\gas T_x + m_\ion T_\gas}{m_\gas T_\perp + m_\ion T_\gas}, & \kappa_{\perp,\gas} &= \frac{\gamma_{\perp,\gas}}{\gamma_\gas},~~~~~\kappa_{x,\gas} = \frac{\gamma_{x,\gas}}{\gamma_\gas}
\end{align}
We define the generalized integral of Eq.~\eqref{eq:OmegaAn} and, substituting, we obtain Eq.~\eqref{eq:collTerm1_AnisoMaxw1}. By assuming a isotropic Maxwellian, is equivalent to $T_x=T_\perp = T$, which means $\kappa_{x,\perp} = 1$ and $\gamma_{x,\gas} = \gamma_{\ion,\gas}$ and therefore, it reads Eq.~\eqref{eq:collTerm1_isoMaxw1}.

\paragraph{Axial energy exchange:}
From Eq.~\eqref{eq:integralIk0Final},
\begin{equation}
    \mathcal{I}_{20} = \int_{\mathbb{S}^2} \!\!\!\! \, m ((v'_{x})^2 - v_{x}^2) \sigma(g, \Omega) \text{d}^2\Omega
    = -\mu \left( 2 G_{x} g_{x} Q^{(1)} - \frac{1}{2}\frac{\mu}{m} \left(g^{2} - 3 g_{x}^{2}\right) Q^{(2)}
    \right) \,.
\end{equation}
We write the axial velocity of the center of mass in the new variables,
\begin{equation}
G_x = X_x + \frac{\mu}{(\gamma_{\gas} + \gamma_x)} \left( \frac{\gamma_\gas}{m_{\gas}} - \frac{\gamma_{x}}{m} \right) g_x + \frac{\gamma_{x,\gas}}{\gamma_{\gas}} u.
\end{equation}
We define,
\begin{equation}
    \hat{\kappa}_x = \frac{\mu}{(\gamma_{\gas} + \gamma_x)} \left( \frac{\gamma_\gas}{m_{\gas}} - \frac{\gamma_{x}}{m} \right) = \frac{\boltz(T_x - T_\gas)\gamma_{x,\gas}}{(m + m_\gas)}.
\end{equation}
After integration over $\vec{X}$ (injecting the result of Eq.~\eqref{eq:integralX}), we obtain
\begin{equation}
\begin{aligned}
&\mathcal{C}_{02} = -\mu n_{\ion} n_{\gas} \frac{\gamma_{\perp,\gas}}{2\pi} \sqrt{\frac{\gamma_{x,\gas}}{2\pi}} \int_{\mathbb{R}^3}\exp \left( -\frac{\gamma_{x,\gas}}{2} (g_x - u)^2 - \frac{\gamma_{\perp,\gas}}{2} g_{\perp}^2 \right) \\
&\quad\quad \times \left[\hat{\kappa}_x g_x^2 Q^{(1)} + 2 g_x \left( \frac{\gamma_{x,\gas}}{\gamma_{\gas}} u \right) Q^{(1)} - \frac{\mu}{m} \left( \frac{g^2}{2} - \frac{3g_x^2}{2} \right) Q^{(2)} \right]\mathrm{d}^3\vec{g}.
\end{aligned}
\end{equation}
As done before, we integrate with spherical coordinates and we use the previous normalization of Eq.~\eqref{eq:normalization}, which yields,
\begin{equation}
\begin{aligned}
&\mathcal{C}_{02} = - \mu n_\ion n_\gas \frac{{\kappa}_{\perp,\gas}}{\sqrt{2\pi} \gamma_{x,\gas}^{3/2}} \int_0^\infty \!\!\!\!\!\! \text{d}\bar{g} \bar{g}^3 \int_{-1}^1 \!\! \text{d}\zeta \, \left( \left( 2 \hat{\kappa}_x \bar{g}^2 \zeta^2 \frac{\gamma_{x,\gas}}{\gamma_\gas} \bar{u} \bar{g} \zeta \right) Q^{(1)} - \frac{\mu}{m} \frac{\bar{g}^{2}}{2} \left(1 - 3 \zeta^2\right) Q^{(2)} \right) \times \nonumber \\
&\quad\quad \exp\left( -\frac{1}{2} \left( \bar{u}^2 + \kappa_{x,\,\perp} \bar{g}^2 - 2 \bar{u} \bar{g} \zeta + (1 - \kappa_{x,\,\perp}) \bar{g}^2 \zeta^2 \right) \right)
\end{aligned}
\end{equation}
By introducing the generalized integral of Eq.~\eqref{eq:OmegaAn} and, substituting, we obtain Eq.~\eqref{eq:collTerm1_AnisoMaxw2} and Eq.~\eqref{eq:collTerm1_isoMaxw2} in the isotropic limit.

\paragraph{Perpendicular energy exchange:}
The angular integral $\mathcal{I}_{02}$ is given in Eq.~\eqref{eq:integralI02Final}. We introduce the new velocity variable
\begin{equation}
\vec{G}_\perp = \vec{X}_\perp + \frac{\mu}{(\gamma_{\gas} + \gamma_\perp)} \left( \frac{\gamma_\gas}{m_{\gas}} - \frac{\gamma_{\perp}}{m_{\ion}} \right) \vec{g}_\perp = \vec{X}_\perp + \frac{\boltz(T_\perp - T_\gas)\kappa_{x,\perp}\gamma_{x,\gas}}{(m_\ion + m_\gas)} \vec{g}_\perp = \vec{X}_\perp + \hat{\kappa}_\perp \vec{g}_\perp
\end{equation}
The integration over the $\vec{X}$ velocity yields,
\begin{equation}
    \begin{aligned}
        &\mathcal{C}_{02} = -\mu n_\text{i} n_\text{g} \frac{\sqrt{\gamma_{x,\text{g}}} \gamma_{\perp,\text{g}}}{(2\pi)^{3/2}} \int_{\mathbb{R}^3} \!\!\!\!\! \text{d}^3\! \vec{g}\, g \left( 2 \hat{\kappa}_\perp g_y^2 Q^{(1)}(g) - \tfrac{1}{2}\tfrac{\mu}{m} \left( g^2 - 3 g_y^2\right) Q^{(2)}(g) \right)\,\times\nonumber\\
        &\quad\quad \exp\left(- \frac{\gamma_{x \text{g}}}{2} (g_x - u)^2 - \frac{\gamma_{\perp \text{g}}}{2} g_\perp^2 \right)
    \end{aligned}
\end{equation}
We integrate using spherical coordinates, introducing the following relations, $g_\perp^2 = g^2 \sin^2\theta$ and the integration over the azimuthal angle $\phi$ can be done easily and gives that $\int_0^{2\pi} g_y^2 \text{d}\phi = \tfrac{1}{2} g^2 \sin^2\theta$. With the previous normalization of Eq.~\eqref{eq:normalization}, this yields the following integral
\begin{equation}
    \begin{aligned}
        &\mathcal{C}_{02} = -\mu n_\text{i} n_\text{g} \frac{{\kappa}_{\perp,\text{g}}}{\sqrt{2\pi} \gamma_{x,\text{g}}^{3/2}} \int_{0}^\infty \!\! \text{d}\! \bar{g}\, \bar{g}^5 \!\!\int_{-1}^1 \!\!\text{d}\zeta \left( \hat{\kappa}_\perp (1-\zeta^2) Q^{(1)}(g) - \tfrac{1}{4}\tfrac{\mu}{m} \left(- 1 + 3\zeta^2\right) Q^{(2)}(g) \right) \times \nonumber \\
        &\quad \quad \,\exp\left(- \tfrac{1}{2}(1 - \kappa_{x,\,\perp}) \bar{g}^2 \zeta^2 -  \bar{u} \bar{g}\zeta + \tfrac{1}{2} \bar{u}^2 - \frac{\kappa_{x,\,\perp}}{2} \bar{g}^2 \right)
    \end{aligned}
\end{equation}
By introducing the generalized integral of Eq.~\eqref{eq:OmegaAn} and, substituting, we obtain Eq.~\eqref{eq:collTerm1_AnisoMaxw3} and Eq.~\eqref{eq:collTerm1_isoMaxw3} in the isotropic limit.
\subsection{Dirac distributions for 5M model}\label{app:5Mmodel}
We will solve the integral of Eq.~\eqref{eq:integral1} with the angular integrals \eqref{eq:integralIk0Final}, with the following distribution functions,
\begin{equation}\label{eq:DiracsDistr}
    f^\mathrm{(5M)}(v_x,\,v_\perp) = n \sum_{i=0}^2 w_i\delta(\vec{v}-\vec{u}_i)~~~\text{and}~~~f_\gas(\vec{v}_\gas) = n_\gas \delta(\vec{v}_\gas).
\end{equation}
We can perform the integration of all the axial moments in a very straight forward manner. In the following, we consider a single Dirac distribution, while the total contribution will be the sum of the three, as explained in Eq.~\eqref{eq:sumDiracs}. As a result, we obtain
\begin{equation}
    \begin{aligned}\label{eq:DiracVDF-simple_integral}
        \mathcal{C}^{(i)}_{k0} ={} & \mu n n_\gas\int_{\mathbb{R}^3}\int_{\mathbb{R}^3}\mathrm{d}^3\vec{v}\mathrm{d}^3\vec{v}_\gas \delta(\vec{v}-u\hat{\vec{e}}_x) \delta(\vec{v}_\gas) \sum_{j=1}^k \binom{k}{j} G_x^{k-j} \left(\frac{\mu}{m}\right)^{j-1} \\
        & \sum_{i=0}^{\lfloor j/2 \rfloor} \binom{2i}{i} \binom{j}{2i} g_x^{j-2i} \left(\tfrac{1}{4}\left(g^2 - g_x^2\right)\right)^{i} \sum_{l=0}^i \binom{i}{l} (-1)^{i-l+1} Q^{j-2l}(g).
    \end{aligned}
\end{equation}
We can largely simplify in the case that the Diracs are in the $v_x$ direction, as
\begin{eqnarray}
    \int_\mathbb{R} \delta(v_x - u)  g_x^{j-2i} \left(g^2 - g_x^2\right)^i\mathrm{d}v_x
    &=& \left\{ \begin{array}{l}
        u \quad \text{if } i = 0 \\
        0 \quad \text{otherwise.}
    \end{array} \right.
\end{eqnarray}
{If we consider the collision between a Dirac of ions at $\vec v_\ion = u_\ion \hat{\vec{e}}_x$ and density $n_\ion$ and a Dirac of gas species at $\vec v_\gas = u_\gas \hat{\vec{e}}_x$ and density $n_\gas$, so the relative velocity reads $g_x = u_\ion - u_\gas$ and $G_x = \tfrac{\mu}{m_\gas}u_\ion +  \tfrac{\mu}{m_\ion}u_\gas$. Then, the collision integral reads:
\begin{equation}
\mathcal{C}^{(D-D)}_{k0} = -\mu n_\ion n_\gas \sum_{j=1}^{k} \binom{k}{j} \left( \frac{\mu}{m_{\gas}} u_{\ion} + \frac{\mu}{m_{\ion}} u_{\gas} \right)^{k-j} \left( \frac{\mu}{m_{\ion}} \right)^{j-1} (u_{\ion} - u_{\gas})^j Q^{(j)}(|u_{\ion} - u_{\gas}|) |u_{\ion} - u_{\gas}|
\end{equation}
}
{As a result, considering $u_\ion = u$ and $u_\gas = 0$ as in Eq.~\eqref{eq:DiracsDistr}, the previous expression yields,
\begin{equation}\label{eq:DiracVDF-simple-app}
    \mathcal{C}_{k0}
    = -\mu n n_\gas \lvert u\rvert \sum_{j=1}^k \binom{k}{j} \frac{\mu^{k-1}}{m_\gas^{k-j}m^{j-1}} u_i^{k} Q^{(j)}\left(\lvert u\rvert\right) \,.
\end{equation}}
{By abuse of notation, we have dropped the ion subindex in the latter.}
\subsection{Dirac distributions with anisotropic temperature (6M)}\label{app:6Mmodel}
We solve Eq.~\eqref{eq:integral1} with the angular integrals \eqref{eq:integralIk0Final} and the following distribution functions
\begin{equation}
    f(\vec{v}_\text{i}) = \sum_{i = 0}^2 n w_i\delta\left(v_x - u_i\right) \frac{\gamma_\perp}{2\pi} \,\text{e}^{-\frac{1}{2} \gamma_\perp v_\perp^2}
    \quad\text{and}\quad f_\text{g}(\vec{v}_\text{g}) = n_\text{g} \left(\frac{\gamma_\text{g}}{2\pi}\right)^{3/2} \,\text{e}^{-\frac{1}{2} \gamma_\text{g} v_\text{g}^2}.
\end{equation}
In the following, we will consider just one of the Diracs as the collision operator is bilinear and the total result is just a linear combinantion of the individual Diracs.

We will use the following change of variables, as defined previously, $(v_x, \vec{v}_\perp, \vec{v}_\gas) \rightarrow (G_x, \vec{X}_\perp, \vec{g})$. Note that as the temperature in the $x$ of the Dirac is zero, we do not need to change the $x$ component of the center of mass velocity. As a result, the multiplication of the Diracs reads
\begin{multline}
    f_\text{i}(\vec{v}_\text{i}) f_\text{g}(\vec{v}_\text{g}) = n_\text{i} n_\text{g} \frac{\gamma_\perp \gamma_\text{g}^{3/2}}{(2\pi)^{5/2}} \,\delta\left(G_x + \tfrac{\mu}{m} g_x - u\right) \times\\
    \,\exp\left(-\frac{1}{2} \gamma_\text{g} (u - g_x)^2 - \frac{1}{2} \left((\gamma_\perp + \gamma_\text{g}) X_\perp^2 + \gamma_{\perp,\text{g}} g_\perp^2 \right) \right)
\end{multline}
The procedure is very similar to the previous derivations. In particular, the axial momentum, axial energy, and perpendicular energy exchanges of the Dirac with anisotropic temperature is a particular solution of the 4M moment for $T_x = 0$. As a result, we will just outline the derivation of the axial heat flux and kurtorsis exchanges.

The integration over the $G_x$ of the heat flux exchange reads,
\begin{eqnarray}
    &\int_{\mathbb{R}}  \delta\left(G_x + \tfrac{\mu}{m_1} g_x - u\right)\mathcal{I}_{30}\,\text{d}G_x
        = -\mu \left(
        \left( 3 u^2 g_{x} - 6\tfrac{\mu}{m}u g_x^2 + \tfrac{3}{2}\tfrac{\mu^2}{m^2} g_x^3 + \tfrac{3}{2}\tfrac{\mu^2}{m^2} g_x g^2 \right) Q^{(1)} \right. \nonumber \\
    &\quad\quad\quad\left. + \left( \tfrac{3}{2}\tfrac{\mu}{m} \left(u - \tfrac{\mu}{m} g_x\right) (3g_x^2 - g^2) \right) Q^{(2)}
        + \tfrac{1}{2}\tfrac{\mu^2}{m^2} g_x \left( 5 g_x^2 - 3 g^2\right) Q^{(3)}
    \right).
\end{eqnarray}
The integration over the $\vec{g}$ is following the same procedure as in the 4M model explained above, leading to Eq.~\eqref{eq:collTerm1_4_AniHyQMOM_3}.

Finally, we specify the integration over the $G_x$ of the kurtosis exchange reads
\begin{eqnarray}
    &&\int_{\mathbb{R}}  \delta\left(G_x + \tfrac{\mu}{m_1} g_x - u\right)\mathcal{I}_{40}\,\text{d}G_x= \nonumber \\
    && -\mu \left(
        \left( 4 u^3 g_x  - 12\tfrac{\mu}{m} u^2 g_x^2 + 6\left(\tfrac{\mu}{m}\right)^2 u g_x (g_x^2 + g^2) + 2\left(\tfrac{\mu}{m}\right)^3 g_x^2 (g_x^2 - 3 g^2) \right) Q^{(1)} \right. \nonumber \\
    &&+ \left( 3\tfrac{\mu}{m} u^2 (3 g_x^2 - g^2) - 6\left(\tfrac{\mu}{m}\right)^2 u g_x (3 g_x^2 - g^2) + \tfrac{3}{4}\left(\tfrac{\mu}{m}\right)^3 \left( 7 g_x^4 + 2 g^2 g_x^2 - g^4\right) \right) Q^{(2)} \nonumber \\
    &&\left. + \left( 10\left(\tfrac{\mu}{m}\right)^2 u g_x^3 - 6\left(\tfrac{\mu}{m}\right)^2 u g^2 g_x - 10\left(\tfrac{\mu}{m}\right)^3 g_x^4 + 6\left(\tfrac{\mu}{m}\right)^3 g^2 g_x^2 \right) Q^{(3)} \right.\nonumber\\
    &&+\left. \tfrac{1}{8} \left(\tfrac{\mu}{m}\right)^3\left( 3 g^4 - 30 g^2 g_x^2 + 35 g_x^4 \right) Q^{(4)}.
    \right)
\end{eqnarray}
The integration over the $\vec{g}$ is following the same procedure as in the 4M model explained above. This leads to Eq.~\eqref{eq:collTerm1_4_AniHyQMOM_4}.

\newpage
\section{Quantification of the error in the one-dimensional cases}
\label{sec:ErrorQuantification}
\begin{table*}[h]
\centering
\resizebox{\textwidth}{!}{%
\begin{tabular}{c c | c c c c c c c c c}
\hline
Pressure [mTorr] & Moment & 3M BGK $\nu^{Schottky}$ & 3M BGK $\nu^{Chabert}$ & 3M DIBO & 4M BGK $\nu^{Chabert}$ & 4M DIBO & 5M BGK $\nu^{Chabert}$ & 5M DIBO & 6M BGK $\nu^{Chabert}$ & 6M DIBO \\
\hline
\multirow{6}{*}{0.05} & $n$ & $6.52\times 10^{-2}$ & $4.98\times 10^{-2}$ & $5.13\times 10^{-2}$ & $7.88\times 10^{-3}$ & $9.34\times 10^{-3}$ & $7.65\times 10^{-3}$ & $5.95\times 10^{-3}$ & $7.65\times 10^{-3}$ & $6.28\times 10^{-3}$ \\
 & $u$ & $4.26\times 10^{-2}$ & $3.01\times 10^{-2}$ & $3.56\times 10^{-2}$ & $6.93\times 10^{-3}$ & $6.75\times 10^{-3}$ & $8.71\times 10^{-3}$ & $2.42\times 10^{-3}$ & $8.71\times 10^{-3}$ & $2.29\times 10^{-3}$ \\
 & $T_x$ & $3.36\times 10^{-1}$ & $3.29\times 10^{-1}$ & $3.38\times 10^{-1}$ & $1.86\times 10^{-1}$ & $2.03\times 10^{-1}$ & $3.28\times 10^{-2}$ & $3.13\times 10^{-2}$ & $3.28\times 10^{-2}$ & $2.92\times 10^{-2}$ \\
 & $T_\perp$ & $3.04\times 10^{-1}$ & $3.26\times 10^{-1}$ & $3.15\times 10^{-1}$ & $6.47\times 10^{-3}$ & $8.13\times 10^{-3}$ & $1.00\times 10^{0}$ & $1.00\times 10^{0}$ & $6.60\times 10^{-3}$ & $7.87\times 10^{-3}$ \\
 & $q$ & $1.04\times 10^{1}$ & $1.15\times 10^{1}$ & $1.06\times 10^{1}$ & $2.37\times 10^{1}$ & $2.18\times 10^{1}$ & $4.28\times 10^{-2}$ & $4.25\times 10^{-2}$ & $4.28\times 10^{-2}$ & $4.22\times 10^{-2}$ \\
 & $r$ & $6.74\times 10^{-1}$ & $6.59\times 10^{-1}$ & $6.72\times 10^{-1}$ & $4.28\times 10^{-1}$ & $4.66\times 10^{-1}$ & $1.07\times 10^{-1}$ & $1.24\times 10^{-1}$ & $1.07\times 10^{-1}$ & $1.22\times 10^{-1}$ \\
\hline
\multirow{6}{*}{0.5} & $n$ & $1.40\times 10^{-1}$ & $5.54\times 10^{-2}$ & $4.93\times 10^{-2}$ & $2.77\times 10^{-2}$ & $3.62\times 10^{-3}$ & $2.79\times 10^{-2}$ & $4.76\times 10^{-3}$ & $2.79\times 10^{-2}$ & $4.27\times 10^{-3}$ \\
 & $u$ & $1.25\times 10^{-1}$ & $3.95\times 10^{-2}$ & $3.48\times 10^{-2}$ & $5.09\times 10^{-2}$ & $5.96\times 10^{-3}$ & $5.21\times 10^{-2}$ & $8.06\times 10^{-3}$ & $5.21\times 10^{-2}$ & $2.56\times 10^{-3}$ \\
 & $T_x$ & $3.70\times 10^{-1}$ & $2.16\times 10^{-1}$ & $2.52\times 10^{-1}$ & $1.25\times 10^{-1}$ & $1.63\times 10^{-1}$ & $1.06\times 10^{-1}$ & $4.51\times 10^{-2}$ & $1.06\times 10^{-1}$ & $2.29\times 10^{-2}$ \\
 & $T_\perp$ & $2.17\times 10^{-1}$ & $2.72\times 10^{-1}$ & $2.27\times 10^{-1}$ & $3.91\times 10^{-2}$ & $6.13\times 10^{-3}$ & $1.00\times 10^{0}$ & $1.00\times 10^{0}$ & $3.90\times 10^{-2}$ & $8.14\times 10^{-3}$ \\
 & $q$ & $4.76\times 10^{-1}$ & $1.24\times 10^{0}$ & $8.50\times 10^{-1}$ & $3.48\times 10^{0}$ & $2.50\times 10^{0}$ & $6.25\times 10^{-2}$ & $2.93\times 10^{-2}$ & $6.25\times 10^{-2}$ & $3.09\times 10^{-2}$ \\
 & $r$ & $6.90\times 10^{-1}$ & $4.40\times 10^{-1}$ & $5.15\times 10^{-1}$ & $2.07\times 10^{-1}$ & $3.57\times 10^{-1}$ & $1.21\times 10^{-1}$ & $1.42\times 10^{-1}$ & $1.21\times 10^{-1}$ & $1.16\times 10^{-1}$ \\
\hline
\multirow{6}{*}{5} & $n$ & $4.47\times 10^{-1}$ & $1.79\times 10^{-1}$ & $2.02\times 10^{-2}$ & $1.40\times 10^{-1}$ & $8.40\times 10^{-3}$ & $1.47\times 10^{-1}$ & $2.46\times 10^{-2}$ & $1.47\times 10^{-1}$ & $3.25\times 10^{-3}$ \\
 & $u$ & $6.21\times 10^{-1}$ & $9.03\times 10^{-2}$ & $2.48\times 10^{-2}$ & $7.43\times 10^{-2}$ & $1.32\times 10^{-2}$ & $8.28\times 10^{-2}$ & $3.84\times 10^{-2}$ & $8.28\times 10^{-2}$ & $5.68\times 10^{-3}$ \\
 & $T_x$ & $4.29\times 10^{-1}$ & $1.73\times 10^{-1}$ & $1.97\times 10^{-1}$ & $1.06\times 10^{-1}$ & $4.69\times 10^{-2}$ & $1.02\times 10^{-1}$ & $7.71\times 10^{-2}$ & $1.02\times 10^{-1}$ & $1.36\times 10^{-2}$ \\
 & $T_\perp$ & $1.60\times 10^{-1}$ & $2.02\times 10^{-1}$ & $1.70\times 10^{-1}$ & $6.14\times 10^{-2}$ & $8.92\times 10^{-3}$ & $1.00\times 10^{0}$ & $1.00\times 10^{0}$ & $6.14\times 10^{-2}$ & $1.02\times 10^{-2}$ \\
 & $q$ & $1.07\times 10^{0}$ & $1.13\times 10^{0}$ & $4.99\times 10^{-1}$ & $3.67\times 10^{0}$ & $1.28\times 10^{0}$ & $1.01\times 10^{0}$ & $4.19\times 10^{-1}$ & $1.01\times 10^{0}$ & $1.60\times 10^{-1}$ \\
 & $r$ & $7.25\times 10^{-1}$ & $2.48\times 10^{-1}$ & $2.57\times 10^{-1}$ & $3.77\times 10^{-1}$ & $5.79\times 10^{-2}$ & $2.24\times 10^{-1}$ & $1.53\times 10^{-1}$ & $2.24\times 10^{-1}$ & $2.09\times 10^{-2}$ \\
\hline
\multirow{6}{*}{50} & $n$ & $6.05\times 10^{-1}$ & $3.29\times 10^{-1}$ & $7.08\times 10^{-2}$ & $3.27\times 10^{-1}$ & $5.36\times 10^{-2}$ & $3.25\times 10^{-1}$ & $5.55\times 10^{-2}$ & $3.24\times 10^{-1}$ & $4.52\times 10^{-2}$ \\
 & $u$ & $1.12\times 10^{0}$ & $1.92\times 10^{-1}$ & $3.74\times 10^{-2}$ & $1.85\times 10^{-1}$ & $1.97\times 10^{-2}$ & $1.85\times 10^{-1}$ & $7.30\times 10^{-2}$ & $1.85\times 10^{-1}$ & $1.81\times 10^{-2}$ \\
 & $T_x$ & $2.04\times 10^{-1}$ & $7.01\times 10^{-2}$ & $8.80\times 10^{-2}$ & $9.13\times 10^{-2}$ & $2.12\times 10^{-2}$ & $8.25\times 10^{-2}$ & $1.23\times 10^{-1}$ & $8.25\times 10^{-2}$ & $1.89\times 10^{-2}$ \\
 & $T_\perp$ & $5.61\times 10^{-2}$ & $9.08\times 10^{-2}$ & $7.07\times 10^{-2}$ & $1.22\times 10^{-2}$ & $1.68\times 10^{-2}$ & $1.00\times 10^{0}$ & $1.00\times 10^{0}$ & $1.22\times 10^{-2}$ & $1.36\times 10^{-2}$ \\
 & $q$ & $1.03\times 10^{0}$ & $1.50\times 10^{0}$ & $1.49\times 10^{0}$ & $3.96\times 10^{0}$ & $2.98\times 10^{0}$ & $2.38\times 10^{0}$ & $1.40\times 10^{0}$ & $2.38\times 10^{0}$ & $4.87\times 10^{-1}$ \\
 & $r$ & $6.22\times 10^{-1}$ & $3.33\times 10^{-1}$ & $1.05\times 10^{-1}$ & $3.21\times 10^{-1}$ & $9.43\times 10^{-2}$ & $3.19\times 10^{-1}$ & $2.17\times 10^{-1}$ & $3.18\times 10^{-1}$ & $1.95\times 10^{-2}$ \\
\hline
\end{tabular}}
\caption{Relative $L_2$ errors for Case I in the bulk region $x\in(0.0,\,4.75)$ cm.}\label{tb:tableiv}
\end{table*}

\begin{table*}[h]
\centering
\resizebox{\textwidth}{!}{%
\begin{tabular}{c c | c c c c c c c c c}
\hline
Pressure [mTorr] & Moment & 3M BGK $\nu^{Schottky}$ & 3M BGK $\nu^{Chabert}$ & 3M DIBO & 4M BGK $\nu^{Chabert}$ & 4M DIBO & 5M BGK $\nu^{Chabert}$ & 5M DIBO & 6M BGK $\nu^{Chabert}$ & 6M DIBO \\
\hline
\multirow{6}{*}{0.05} & $n$ & $8.07\times 10^{-3}$ & $1.42\times 10^{-2}$ & $8.26\times 10^{-3}$ & $1.43\times 10^{-2}$ & $9.14\times 10^{-3}$ & $1.10\times 10^{-2}$ & $1.47\times 10^{-2}$ & $1.10\times 10^{-2}$ & $1.46\times 10^{-2}$ \\
 & $u$ & $1.28\times 10^{-2}$ & $1.55\times 10^{-2}$ & $1.28\times 10^{-2}$ & $1.74\times 10^{-2}$ & $1.65\times 10^{-2}$ & $1.97\times 10^{-2}$ & $2.30\times 10^{-2}$ & $1.97\times 10^{-2}$ & $2.30\times 10^{-2}$ \\
 & $T_x$ & $2.64\times 10^{-1}$ & $2.54\times 10^{-1}$ & $2.60\times 10^{-1}$ & $1.42\times 10^{-1}$ & $1.15\times 10^{-1}$ & $5.51\times 10^{-2}$ & $3.52\times 10^{-2}$ & $5.51\times 10^{-2}$ & $3.47\times 10^{-2}$ \\
 & $T_\perp$ & $5.51\times 10^{-1}$ & $6.25\times 10^{-1}$ & $5.76\times 10^{-1}$ & $4.80\times 10^{-3}$ & $1.47\times 10^{-2}$ & $1.00\times 10^{0}$ & $1.00\times 10^{0}$ & $4.81\times 10^{-3}$ & $1.42\times 10^{-2}$ \\
 & $q$ & $5.73\times 10^{1}$ & $5.52\times 10^{1}$ & $5.50\times 10^{1}$ & $1.50\times 10^{2}$ & $1.43\times 10^{2}$ & $9.12\times 10^{-2}$ & $3.45\times 10^{-2}$ & $9.12\times 10^{-2}$ & $3.49\times 10^{-2}$ \\
 & $r$ & $6.99\times 10^{-1}$ & $6.66\times 10^{-1}$ & $6.88\times 10^{-1}$ & $4.77\times 10^{-1}$ & $5.30\times 10^{-1}$ & $7.00\times 10^{-2}$ & $2.25\times 10^{-2}$ & $7.00\times 10^{-2}$ & $2.20\times 10^{-2}$ \\
\hline
\multirow{6}{*}{0.5} & $n$ & $4.40\times 10^{-2}$ & $6.05\times 10^{-2}$ & $1.34\times 10^{-2}$ & $5.24\times 10^{-2}$ & $1.20\times 10^{-2}$ & $4.20\times 10^{-2}$ & $1.45\times 10^{-2}$ & $4.20\times 10^{-2}$ & $1.53\times 10^{-2}$ \\
 & $u$ & $4.00\times 10^{-2}$ & $3.86\times 10^{-2}$ & $1.96\times 10^{-2}$ & $3.27\times 10^{-2}$ & $2.32\times 10^{-2}$ & $2.65\times 10^{-2}$ & $2.79\times 10^{-2}$ & $2.65\times 10^{-2}$ & $2.73\times 10^{-2}$ \\
 & $T_x$ & $3.16\times 10^{-1}$ & $2.72\times 10^{-1}$ & $1.96\times 10^{-1}$ & $3.44\times 10^{-1}$ & $7.10\times 10^{-2}$ & $3.85\times 10^{-1}$ & $4.39\times 10^{-2}$ & $3.85\times 10^{-1}$ & $3.64\times 10^{-2}$ \\
 & $T_\perp$ & $1.19\times 10^{-1}$ & $5.37\times 10^{-1}$ & $3.45\times 10^{-1}$ & $1.01\times 10^{-1}$ & $1.36\times 10^{-2}$ & $1.00\times 10^{0}$ & $1.00\times 10^{0}$ & $1.01\times 10^{-1}$ & $1.20\times 10^{-2}$ \\
 & $q$ & $5.54\times 10^{0}$ & $2.51\times 10^{0}$ & $5.12\times 10^{0}$ & $9.61\times 10^{0}$ & $1.15\times 10^{1}$ & $5.25\times 10^{-1}$ & $2.87\times 10^{-2}$ & $5.25\times 10^{-1}$ & $2.96\times 10^{-2}$ \\
 & $r$ & $8.19\times 10^{-1}$ & $5.46\times 10^{-1}$ & $6.65\times 10^{-1}$ & $3.94\times 10^{-1}$ & $6.18\times 10^{-1}$ & $7.05\times 10^{-1}$ & $4.34\times 10^{-2}$ & $7.05\times 10^{-1}$ & $3.86\times 10^{-2}$ \\
\hline
\multirow{6}{*}{5} & $n$ & $2.24\times 10^{-1}$ & $1.08\times 10^{-1}$ & $1.42\times 10^{-2}$ & $8.67\times 10^{-2}$ & $2.77\times 10^{-2}$ & $1.00\times 10^{-1}$ & $6.06\times 10^{-3}$ & $1.00\times 10^{-1}$ & $2.41\times 10^{-2}$ \\
 & $u$ & $1.86\times 10^{-1}$ & $7.22\times 10^{-2}$ & $3.14\times 10^{-2}$ & $6.89\times 10^{-2}$ & $3.72\times 10^{-2}$ & $7.81\times 10^{-2}$ & $3.72\times 10^{-2}$ & $7.81\times 10^{-2}$ & $3.32\times 10^{-2}$ \\
 & $T_x$ & $8.11\times 10^{-1}$ & $1.74\times 10^{-1}$ & $3.76\times 10^{-1}$ & $4.32\times 10^{-1}$ & $1.87\times 10^{-1}$ & $7.68\times 10^{-1}$ & $2.25\times 10^{-2}$ & $7.68\times 10^{-1}$ & $2.60\times 10^{-2}$ \\
 & $T_\perp$ & $4.81\times 10^{-1}$ & $1.21\times 10^{0}$ & $6.55\times 10^{-1}$ & $2.46\times 10^{-1}$ & $3.37\times 10^{-2}$ & $1.00\times 10^{0}$ & $1.00\times 10^{0}$ & $2.46\times 10^{-1}$ & $3.64\times 10^{-2}$ \\
 & $q$ & $1.10\times 10^{0}$ & $2.80\times 10^{0}$ & $1.80\times 10^{-1}$ & $1.01\times 10^{1}$ & $1.62\times 10^{0}$ & $6.35\times 10^{-1}$ & $6.45\times 10^{-2}$ & $6.35\times 10^{-1}$ & $4.05\times 10^{-2}$ \\
 & $r$ & $9.87\times 10^{-1}$ & $6.87\times 10^{-1}$ & $8.51\times 10^{-1}$ & $1.76\times 10^{-1}$ & $7.50\times 10^{-1}$ & $7.15\times 10^{-1}$ & $3.72\times 10^{-2}$ & $7.15\times 10^{-1}$ & $2.57\times 10^{-2}$ \\
\hline
\multirow{6}{*}{50} & $n$ & $4.88\times 10^{-1}$ & $7.18\times 10^{-2}$ & $1.33\times 10^{-1}$ & $9.32\times 10^{-2}$ & $9.12\times 10^{-2}$ & $9.00\times 10^{-2}$ & $3.08\times 10^{-2}$ & $9.11\times 10^{-2}$ & $7.07\times 10^{-2}$ \\
 & $u$ & $8.27\times 10^{-1}$ & $1.56\times 10^{-1}$ & $5.96\times 10^{-2}$ & $1.80\times 10^{-1}$ & $1.98\times 10^{-2}$ & $1.74\times 10^{-1}$ & $3.93\times 10^{-2}$ & $1.74\times 10^{-1}$ & $1.88\times 10^{-2}$ \\
 & $T_x$ & $9.23\times 10^{-1}$ & $4.68\times 10^{-1}$ & $4.06\times 10^{-1}$ & $1.49\times 10^{-1}$ & $5.02\times 10^{-2}$ & $7.71\times 10^{-2}$ & $4.01\times 10^{-2}$ & $7.71\times 10^{-2}$ & $4.01\times 10^{-2}$ \\
 & $T_\perp$ & $5.37\times 10^{-1}$ & $1.84\times 10^{0}$ & $2.17\times 10^{0}$ & $2.54\times 10^{-1}$ & $2.53\times 10^{-1}$ & $1.00\times 10^{0}$ & $1.00\times 10^{0}$ & $2.54\times 10^{-1}$ & $2.64\times 10^{-1}$ \\
 & $q$ & $9.92\times 10^{-1}$ & $4.34\times 10^{0}$ & $4.64\times 10^{0}$ & $1.50\times 10^{1}$ & $1.17\times 10^{1}$ & $7.32\times 10^{0}$ & $4.35\times 10^{-1}$ & $7.33\times 10^{0}$ & $1.86\times 10^{-1}$ \\
 & $r$ & $9.95\times 10^{-1}$ & $5.44\times 10^{-1}$ & $5.60\times 10^{-1}$ & $1.26\times 10^{0}$ & $4.97\times 10^{-1}$ & $5.68\times 10^{-1}$ & $5.86\times 10^{-2}$ & $5.70\times 10^{-1}$ & $1.78\times 10^{-2}$ \\
\hline
\end{tabular}}
\caption{Relative $L_2$ errors for Case I in the sheath region $x\in(4.75,\,5)$ cm.}\label{tb:tablev}
\end{table*}

\begin{table*}[h]
\centering
\resizebox{\textwidth}{!}{%
\begin{tabular}{c c | c c c c c}
\hline
Pressure [mTorr] & Moment & 3M (BGK) & 3M (DIBO) & 4M (DIBO) & 5M (DIBO) & 6M (DIBO) \\
\hline
\multirow{6}{*}{200} & $n$ & $2.92\times 10^{-1}$ & $2.32\times 10^{-2}$ & $1.78\times 10^{-2}$ & $8.42\times 10^{-2}$ & $1.54\times 10^{-2}$ \\
 & $u$ & $2.65\times 10^{-1}$ & $5.02\times 10^{-2}$ & $4.81\times 10^{-2}$ & $3.41\times 10^{-2}$ & $2.37\times 10^{-2}$ \\
 & $T_x$ & $6.72\times 10^{-1}$ & $4.24\times 10^{-1}$ & $6.08\times 10^{-1}$ & $6.21\times 10^{-2}$ & $5.03\times 10^{-2}$ \\
 & $T_\perp$ & $2.62\times 10^{0}$ & $5.23\times 10^{0}$ & $3.52\times 10^{-1}$ & $1.00\times 10^{0}$ & $2.70\times 10^{-1}$ \\
 & $q$ & $1.01\times 10^{0}$ & $1.06\times 10^{0}$ & $1.82\times 10^{0}$ & $8.30\times 10^{-2}$ & $6.50\times 10^{-2}$ \\
 & $r$ & $8.85\times 10^{-1}$ & $7.40\times 10^{-1}$ & $1.23\times 10^{0}$ & $1.57\times 10^{-1}$ & $1.60\times 10^{-1}$ \\
\hline
\multirow{6}{*}{500} & $n$ & $2.99\times 10^{-1}$ & $1.74\times 10^{-2}$ & $1.76\times 10^{-2}$ & $7.87\times 10^{-2}$ & $9.53\times 10^{-3}$ \\
 & $u$ & $4.06\times 10^{-1}$ & $3.32\times 10^{-2}$ & $2.81\times 10^{-2}$ & $1.06\times 10^{-1}$ & $3.00\times 10^{-2}$ \\
 & $T_x$ & $2.91\times 10^{-3}$ & $8.59\times 10^{-3}$ & $7.75\times 10^{-3}$ & $1.37\times 10^{-1}$ & $1.72\times 10^{-2}$ \\
 & $T_\perp$ & $1.19\times 10^{-2}$ & $5.98\times 10^{-3}$ & $4.92\times 10^{-3}$ & $1.00\times 10^{0}$ & $1.17\times 10^{-2}$ \\
 & $q$ & $1.61\times 10^{1}$ & $5.51\times 10^{1}$ & $8.60\times 10^{2}$ & $1.39\times 10^{0}$ & $6.55\times 10^{-1}$ \\
 & $r$ & $2.99\times 10^{-1}$ & $1.62\times 10^{-2}$ & $8.81\times 10^{-3}$ & $2.36\times 10^{-1}$ & $4.79\times 10^{-2}$ \\
\hline
\end{tabular}}
\caption{Relative $L_2$ errors for Case II in the bulk region $x\in(-1.8,2.25)$ cm.}\label{tb:tablevi}
\end{table*}

\begin{table*}[h]
\centering
\resizebox{\textwidth}{!}{%
\begin{tabular}{c c | c c c c c}
\hline
Pressure [mTorr] & Moment & 3M (BGK) & 3M (DIBO) & 4M (DIBO) & 5M (DIBO) & 6M (DIBO) \\
\hline
\multirow{6}{*}{200} & $n$ & $4.40\times 10^{-1}$ & $1.35\times 10^{-1}$ & $3.41\times 10^{-1}$ & $9.64\times 10^{-3}$ & $3.78\times 10^{-3}$ \\
 & $u$ & $3.18\times 10^{-1}$ & $1.77\times 10^{-1}$ & $6.39\times 10^{-1}$ & $7.30\times 10^{-3}$ & $9.71\times 10^{-3}$ \\
 & $T_x$ & $7.26\times 10^{-1}$ & $2.37\times 10^{-1}$ & $4.53\times 10^{0}$ & $2.36\times 10^{-2}$ & $2.83\times 10^{-2}$ \\
 & $T_\perp$ & $4.74\times 10^{0}$ & $1.65\times 10^{1}$ & $7.34\times 10^{-1}$ & $1.00\times 10^{0}$ & $3.88\times 10^{-1}$ \\
 & $q$ & $1.00\times 10^{0}$ & $1.03\times 10^{0}$ & $1.45\times 10^{0}$ & $1.43\times 10^{-1}$ & $9.72\times 10^{-2}$ \\
 & $r$ & $9.20\times 10^{-1}$ & $5.54\times 10^{-1}$ & $1.40\times 10^{1}$ & $9.76\times 10^{-2}$ & $1.37\times 10^{-1}$ \\
\hline
\multirow{6}{*}{500} & $n$ & $2.56\times 10^{-1}$ & $4.60\times 10^{-2}$ & $1.11\times 10^{-1}$ & $6.78\times 10^{-2}$ & $2.91\times 10^{-2}$ \\
 & $u$ & $2.98\times 10^{-1}$ & $8.96\times 10^{-2}$ & $4.69\times 10^{-1}$ & $1.38\times 10^{-2}$ & $9.40\times 10^{-3}$ \\
 & $T_x$ & $7.12\times 10^{-1}$ & $3.36\times 10^{-1}$ & $3.80\times 10^{0}$ & $4.14\times 10^{-2}$ & $3.69\times 10^{-2}$ \\
 & $T_\perp$ & $3.66\times 10^{0}$ & $1.00\times 10^{1}$ & $5.77\times 10^{-1}$ & $1.00\times 10^{0}$ & $2.95\times 10^{-1}$ \\
 & $q$ & $1.00\times 10^{0}$ & $1.02\times 10^{0}$ & $1.27\times 10^{0}$ & $1.02\times 10^{-1}$ & $7.24\times 10^{-2}$ \\
 & $r$ & $9.14\times 10^{-1}$ & $6.68\times 10^{-1}$ & $1.24\times 10^{1}$ & $1.53\times 10^{-1}$ & $1.73\times 10^{-1}$ \\
\hline
\end{tabular}}
\caption{Relative $L_2$ errors for Case II in the sheath region $x\in(-2.5,-1.8)$ cm.}\label{tb:tablevii}
\end{table*}

\clearpage
\bibliography{references}%

\end{document}